\documentclass[prl,twocolumn,floatfix,nopacs,nofootinbib,superscriptaddress]{revtex4-2}

\usepackage{graphicx}
\usepackage{dcolumn}
\usepackage{placeins}
\usepackage{color}

\usepackage[dvips]{epsfig}
\usepackage{amssymb}
\usepackage{soul}
\usepackage{amsmath}
\usepackage{mathtools}
\usepackage{multirow}
\usepackage{hyperref}
\hypersetup{
    colorlinks=true,
    linkcolor=black,
    citecolor=black,
    urlcolor=black
}

\usepackage{eucal}
\newcommand{\rev}[1]{{#1}}
\newcommand{\revred}[1]{{#1}}
\definecolor{BLACK}{rgb}{0,0,0}
\newcommand{\revtc}[1]{\textcolor{black}{#1}}
\AtBeginDocument{\setcitestyle{round}}
\makeatletter
\def\@bibdataout@aps{\immediate\write\@bibdataout{@CONTROL{apsrev42Control,author="08",editor="1",pages="1",title="0",year="1"}}\if@filesw\immediate\write\@auxout{\string\citation{apsrev42Control}}\fi}
\AtBeginDocument{\def\place@bibnumber#1{#1.}}
\makeatother
\usepackage{amsfonts}
\usepackage{latexsym}

\makeatletter
\renewcommand\section{\@startsection{section}{1}{\z@}%
  {-2ex \@plus -1ex \@minus -.2ex}%
  {0.8ex \@plus .2ex}%
  {\normalfont\large\bfseries\raggedright}}
\renewcommand\subsection{\@startsection{subsection}{2}{\z@}%
  {-1.5ex \@plus -1ex \@minus -.2ex}%
  {0.5ex \@plus .2ex}%
  {\normalfont\normalsize\bfseries\raggedright}}
\makeatother

\newcounter{movie}

\begin{document}
\makeatletter
\let\oldaddcontentsline\addcontentsline
\renewcommand{\addcontentsline}[3]{}
\makeatother
 \title{\rev{Squid funnels time their elastic recoil to amplify pulsed jets}}

\author{Daehyun Choi}
\affiliation{\revred{BioFrontiers Institute and Department of Chemical and Biological Engineering, CU Boulder, Colorado 80303, USA}}
\affiliation{School of Chemical and Biomolecular Engineering, Georgia Institute of Technology, Atlanta, GA 30318}

\author{Paras Singh}
\affiliation{Daniel Guggenheim School of Aerospace Engineering, Georgia Institute of Technology, Atlanta, GA 30332}

\author{Ian Bergerson}
\affiliation{\revred{BioFrontiers Institute and Department of Chemical and Biological Engineering, CU Boulder, Colorado 80303, USA}}
\affiliation{School of Chemical and Biomolecular Engineering, Georgia Institute of Technology, Atlanta, GA 30318}
\affiliation{School of Physics, Georgia Institute of Technology, Atlanta, GA 30318}

\author{Minho Kim}
\affiliation{Department of Mechanical Engineering, Ajou University, Suwon 16499, Republic of Korea}

\author{Jieun Park}
\affiliation{Department of Mechanical Engineering, Pohang University of Science and Technology, Pohang 37673, Republic of Korea}

\author{Halley J. Wallace}
\affiliation{School of Chemical and Biomolecular Engineering, Georgia Institute of Technology, Atlanta, GA 30318}

\author{Kenny Zhang}
\affiliation{School of Chemical and Biomolecular Engineering, Georgia Institute of Technology, Atlanta, GA 30318}

\author{Sandy Y. Hsieh}
\affiliation{Optical Microscopy Core, Parker H. Petit Institute for Bioengineering and Biosciences, Georgia Institute of Technology, Atlanta, GA 30332}

\author{Aqua T. Asberry}
\affiliation{Histology Core, Parker H. Petit Institute for Bioengineering and Biosciences, Georgia Institute of Technology, Atlanta, GA 30332}

\author{Theodore A. Uyeno}
\affiliation{Department of Biology, Valdosta State University, Valdosta, GA 31698}

\author{William F. Gilly}
\affiliation{Hopkins Marine Station, Stanford University, Pacific Grove, CA 93950, USA}

\author{Hyungmin Park}
\affiliation{Department of Mechanical Engineering, Seoul National University, Seoul 08826, Republic of Korea}

\author{Daeshik Kang}
\affiliation{Department of Mechanical Engineering, Pohang University of Science and Technology, Pohang 37673, Republic of Korea}

\author{Chandan Bose}
\affiliation{Aerospace Engineering, School of Metallurgy and Materials, University of Birmingham, Birmingham B15 2TT, United Kingdom}

\author{Saad Bhamla}
\email{saad.bhamla@colorado.edu}
\affiliation{\revred{BioFrontiers Institute and Department of Chemical and Biological Engineering, CU Boulder, Colorado 80303, USA}}
\affiliation{School of Chemical and Biomolecular Engineering, Georgia Institute of Technology, Atlanta, GA 30318}

\date{\today}

\begin{abstract}
Pulsed jets propel squid across nearly four orders of magnitude in body size. \rev{Each pulse leaves through the funnel, a soft nozzle, but whether its compliance acts within a pulse is not known.} We tracked chromatophores on two squid species and found that \rev{the funnel dilates and recoils, with a repeatable phase lag, while the mantle is still contracting\rev{, and that the lag persists after denervation}.} \rev{For inertia-dominated jets, silicone} nozzles, 3D fluid--structure simulations and a reduced-order model \rev{agree that the impulse reaches about 4$\times$ the rigid value} when the nozzle response time is 0.2 to 0.4 of the jet-acceleration time, a range that overlaps the \rev{\textit{in vivo}} timing. \rev{The wall stores strain energy during acceleration and returns it during expulsion (akin to an elastic capacitor), which we term ``superpropulsion''.} Synthetic nozzles tuned \rev{for superpropulsion} launch jets farther in air, disperse dye faster, and drive a boat faster at a lower cost of transport.
\end{abstract}

\maketitle

\begin{figure*}[t!]
	\centering
	\includegraphics[width=\textwidth]{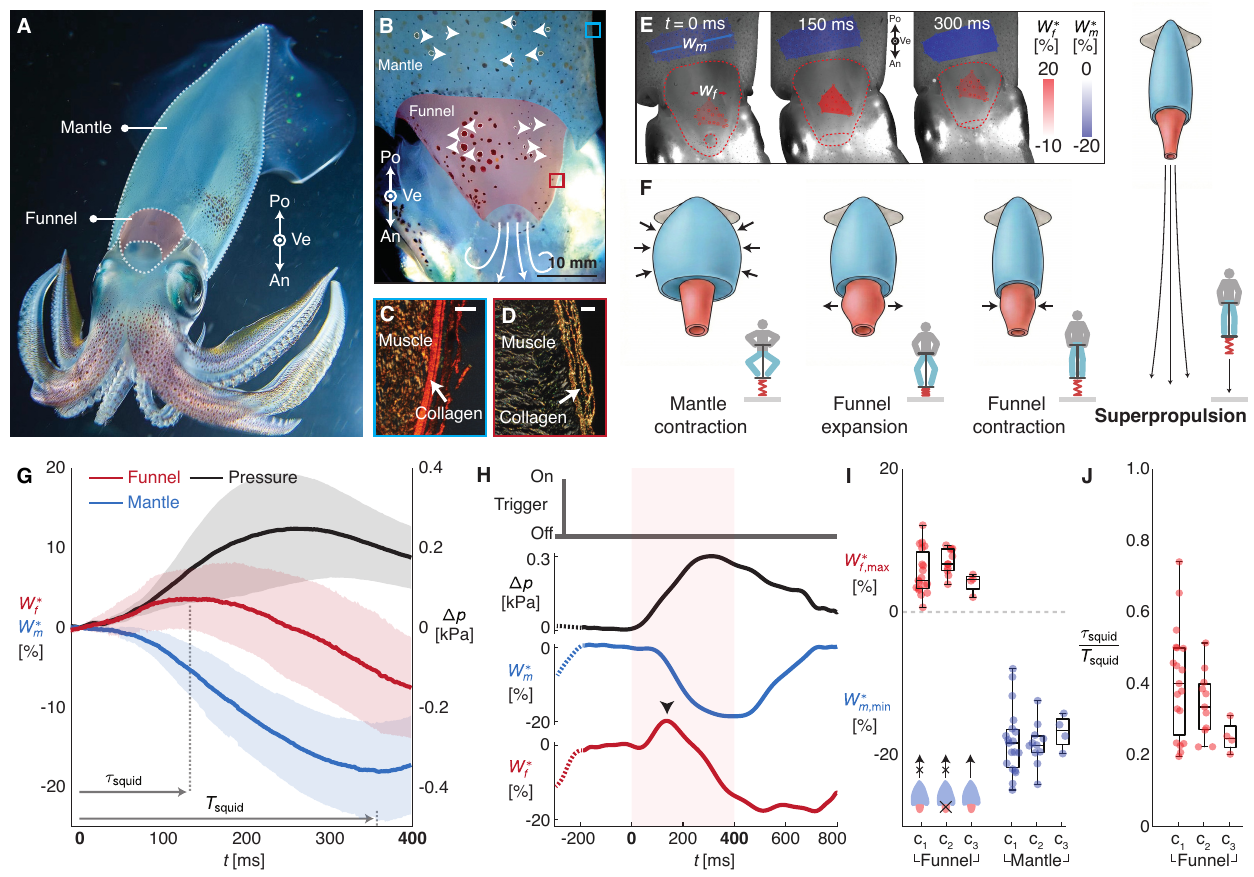}
	\caption{\textbf{Squid funnel collagen architecture and phase-lagged funnel recoil during jetting.}
	  (\textbf{A})~ Squid mantle cavity and funnel (siphon) of \rev{\textit{Sepioteuthis sepioidea}\revred{, identified from fin morphology and locality (Palm Beach County, Florida)}}.
    (\textbf{B})~Image \rev{of \textit{Doryteuthis opalescens}} illustrating mantle contraction and funnel expansion quantified by tracking chromatophore features.
    (\textbf{C--D})~Picrosirius Red staining \revred{of \textit{D.~opalescens}} under polarized light highlighting a  birefringent  collagen-rich layer in the (\textbf{C}) mantle and (\textbf{D}) funnel (arrow); scale bar, 50\,$\mu$m.
    (\textbf{E})~Snapshots \rev{of \textit{D.~opalescens}} at three representative times showing the initial state, peak funnel widening, and contraction\revred{--}recoil of the funnel, as the mantle decreases in size; $W_f$ and $W_m$ denote   funnel and mantle widths. 
    (\textbf{F})~Conceptual schematic summarizing the observed out-of-phase expansion--recoil sequence: as mantle-cavity pressure rises during expulsion, the funnel widens and subsequently recoils during the same jet pulse. We refer to this store--release sequence of an elastic nozzle boundary condition as superpropulsion\rev{.}   A  spring-mass  analogy (pogo-stick) is shown for intuition.
    (\textbf{G})~ Normalized funnel width ($W_f^*$), mantle width ($W_m^*$), and mantle-cavity pressure ($\Delta p$) \revred{(mean and one standard deviation across the jets)} under the normal immobilized condition ($c_1$). Widths are normalized by their initial values: $W_i^* = W_i / W_{i,0}$, where $i\in\{f,m\}$.
    Data represent $N=19$ jets collected from $n=5$ individuals.
    (\textbf{H})~Time history of the stimulus trigger, mantle-cavity pressure ($\Delta p$), mantle width ($W_m^*$), and funnel width ($W_f^*$).
    (\textbf{I})~Maximum funnel widening ($W_{f,\text{max}}^*$) and minimum mantle width ($W_{m,\text{min}}^*$) for \rev{\emph{Doryteuthis opalescens}} under \revred{intact} immobilized ($c_1$) and funnel-\rev{denervated} immobilized ($c_2$) conditions, and for freely \revred{swimming} \rev{\emph{Sepioteuthis lessoniana}} ($c_3$).
    (\textbf{J})~Funnel response-time ratio ($\tau_{\text{squid}}/T_{\text{squid}}$), where $\tau_{\mathrm{squid}}$ is the delay from the onset of mantle contraction to peak funnel expansion and $T_{\mathrm{squid}}$ is the mantle contraction duration (illustrated in \textbf{G}). In (\textbf{I--J}), $N$ ($n$) = 19 (5), 11 (3), and 4 (3) for $c_1$, $c_2$, and $c_3$, respectively.
    Box plots show the median (center line), interquartile range (box), and full data range (whiskers). Image credit for panel A: Michael Patrick O'Neill. \rev{Axis labels in (\textbf{A}), (\textbf{B}) and (\textbf{E}) give the animal's orientation, where An is anterior, Po is posterior and Ve is ventral, the last pointing out of the image plane toward the viewer.}}
	\label{f_squid}
\end{figure*}

\section*{\revtc{Summary}}
A squid’s funnel springs back while the jet is still leaving, and soft nozzles timed the same way quadruple the impulse.

\section*{\revtc{Introduction}}
\subsection{\textbf{Squid jetting across scales and habitats}}
\rev{Since emerging in the Cambrian ($\sim$500~Ma), cephalopods have relied on pulsed jetting for rapid locomotion, and extant squids retain this mode of propulsion across environments}~\cite{kroger2011cephalopod, staaf2017squid}. 
Jetting spans nearly four orders of magnitude in body length, from millimeter-scale hatchlings to the largest species approaching 10~m~\cite{staaf2014aperture, york2020squids,rosa2017biology}.
 Squids typically cruise using fin undulation~\cite{odor1988forces, bartol2001swimming}, but switch to  jetting for high-performance behaviors, including fast escapes  (up to 10 body lengths per second in \textit{Doryteuthis pealeii}) and long-range movements such as Humboldt squid migrations ($1000$~km)~\cite{bartol2009hydrodynamics, bartol2008swimming, anderson2000mechanics, gilly2006vertical}. They occupy habitats from coastal waters to ocean depths exceeding 2000~m~\cite{hoving2014deep, roper2010cephalopods}, and some species undertake brief aerial excursions by jetting at the surface~\cite{muramatsu2013oceanic}. Escape jetting is mediated  by giant axons that synchronize  mantle activation on millisecond timescales~\cite{young1938functioning, hodgkin1952quantitative, gilly1984threshold, otis1990concerted,pumphrey1938rates, bullock1948properties}. Mechanically, jetting expels  discrete pulses of water as the mantle contracts. \rev{The funnel sets the nozzle boundary condition through its outlet diameter and the curvature of its wall, both of which vary through the pulse}, so performance depends on the unsteady flow development within each pulse.

\subsection{\textbf{Vortex pinch-off and timescale matching in pulsed jets}}

Each jet pulse forms a starting jet and vortex ring, and circulation growth is limited by vortex-ring pinch-off, which defines a characteristic stroke ratio (formation number)~\cite{gharib1998universal, krueger2003significance, gao2020jfm}.
Operating near this pinch-off limit is often associated with improved thrust-to-power \revred{(by up to 20--50\%)} compared with steady jets~\cite{dabiri2009optimal, dabiri2005starting, krueger2006formation,whittlesey2013optimal,christianson2020cephalopod,moslemi2010propulsive}, and related vortex-formation constraints appear across jetting swimmers and cardiac filling flows~\cite{hove2003intracardiac}.
In shape-changing jet-propelled bodies, rapid 
contraction (e.g., of the squid mantle) can reduce added-mass penalties and recover fluid kinetic energy, enhancing acceleration and efficiency~\cite{weymouth2013ultrafast, weymouth2015octopus}. Elastic structures can further couple internal and external fluid inertia, yielding oscillator-like dynamics in pulsed-jet propulsion~\cite{bujard2021resonant}. These studies emphasize that pulsed-jet performance is set by the acceleration history and the time-dependent boundary condition, not only by total expelled volume. 

However, most prior work on squid focuses on mantle kinematics and downstream vortex formation. The deformable funnel that sets the  nozzle boundary condition has received comparatively little attention. Recent work on engineered compliant nozzles shows that wall-wave dynamics can strengthen vortex-ring jets and increase thrust and entrainment~\cite{choi2022flow, choi2024mechanism, morris2025optimal}. Still, a  theory that predicts impulse and power amplification and connects the optimal regime to \rev{\textit{in vivo}} funnel timing and practical propulsor-relevant design remains to be established.

\subsection{\textbf{Funnel compliance and jet coupling}}
\rev{In the loliginid squid studied here, the} funnel (siphon) is a cone-shaped muscular nozzle that forms a sealed outlet with the mantle via the mantle--funnel locking cartilages, so exhalant water from the mantle cavity is directed through the funnel~\cite{anderson2000mechanics, bartol2009hydrodynamics, gosline1983patterns, Ward1972} (Fig.~\ref{f_squid}A). Squid vector thrust by reorienting the funnel, which also serves as the outlet for   waste and ink discharge. Funnel musculature can further modulate the outlet diameter during ejection, altering the effective stroke ratio and circulation accumulated before pinch-off~\cite{staaf2014aperture, dabiri2005starting, dabiri2009optimal}. Bioinspired platforms have incorporated deformable siphons for thrust vectoring and aperture control~\cite{zhang2021siphon, flores2025robonautilus}, but most analyses prescribe nozzle kinematics  rather than treating the nozzle as a compliant structure mechanically loaded by the accelerating jet.  As a result, it remains unclear how funnel compliance couples to  unsteady jet acceleration, whether  elastic energy is stored and returned within a pulse, and how this coupling alters jet impulse and power. Here we quantify funnel structure and deformation in live squid and, using  engineered compliant nozzles and theory, test when matching of nozzle response to jet acceleration enhances pulsed-jet performance.

\section*{Results}

\subsection{\textbf{Funnel collagen sheath and phase-lagged recoil during jetting}}

Histological sections of \textit{Doryteuthis opalescens}\rev{, a locally available species with a strong and well-characterized escape jet~\cite{otis1990concerted, neumeister2000temperature, li2019hypoxia} \revred{(Fig.~\ref{sif_squid})},} show a collagen-rich outer sheath surrounding the funnel musculature, consistent with the muscular-hydrostat organization of cephalopod soft tissues~\cite{Kier1985} (Fig.~\ref{f_squid}C,D and Fig.~\ref{sif_histology}).
Picrosirius Red staining under polarized illumination reveals a birefringent, interwoven collagen network. The collagen layer is $72 \pm 12\,\mu$m thick in the funnel, compared with $24 \pm 6\,\mu$m in the mantle. This collagen-rich sheath is consistent with an elastic contribution to funnel mechanics.

To relate this structure to \rev{\textit{in vivo}} deformation, we quantify mantle and funnel widths from high-speed video while simultaneously recording mantle-cavity pressure (Fig.~\ref{f_squid}B,H; squid acquisition in Fig.~\ref{sif_squid}\rev{,} measurement in Fig.~\ref{sif_chromatophore} \rev{and validation in Figs.~\ref{sif_membrane_dic}--\ref{sif_balloon_dic}}). Both mantle and funnel widths increase before the main jet-associated pressure rise  in the mantle cavity (Fig.~\ref{f_squid}H), consistent with a pre-jet mantle hyperinflation phase reported previously~\cite{gosline1983patterns}.

After the onset of mantle contraction ($t=0$), mantle width decreases monotonically, whereas the funnel exhibits an additional expansion--recoil event during the same jet pulse (Fig.~\ref{f_squid}E,G,H). Across intact, restrained \textit{D. opalescens} ($c_1$; $N=19$ jets from $n=5$ individuals), funnel width increases beyond its initial value (up to \rev{$12.1\%$}\rev{, median $4.4\%$}) and then recoils while mantle contraction continues (Fig.~\ref{f_squid}G,H; Supplementary Movie~\ref{mv_squid}). 
To quantify the timing of this sequence \rev{\textit{in vivo}}, we define a funnel response time $\tau_{\mathrm{squid}}$ as the delay from the onset of mantle contraction ($t=0$) to peak funnel width, and normalize it by the mantle contraction duration $T_{\mathrm{squid}}$ (Fig.~\ref{f_squid}G). Across conditions, $\tau_{\mathrm{squid}}/T_{\mathrm{squid}}$ ranges from \rev{$\sim$0.25--0.40} (Fig.~\ref{f_squid}J). 

To probe whether this phase-lagged deformation requires precisely timed active funnel control, we repeat measurements after  disrupting funnel innervation (method in Fig.~\ref{sif_paralyze}; $c_2$; $N=11$, $n=3$). The expansion--recoil signature persists, and the response-time ratio remains of similar order. \rev{The pulse-scale timing therefore survives} perturbed neural input\rev{, consistent} with a substantial passive mechanical contribution (Fig.~\ref{f_squid}I,J). We also observe comparable behavior in freely swimming \textit{Sepioteuthis lessoniana} ($c_3$; $N=4$, $n=3$), indicating that the funnel expansion--recoil sequence is not unique to the restrained preparation (Fig.~\ref{f_squid}I,J).

These \rev{\textit{in vivo}} measurements motivate controlled tests of how a compliant nozzle boundary condition, decoupled from active muscular actuation,  modifies jet impulse and power. We explore this next using engineered nozzles, 3D-CFD simulations, and theory.

\begin{figure*}[p]
	\centering
	\includegraphics[width=\textwidth]{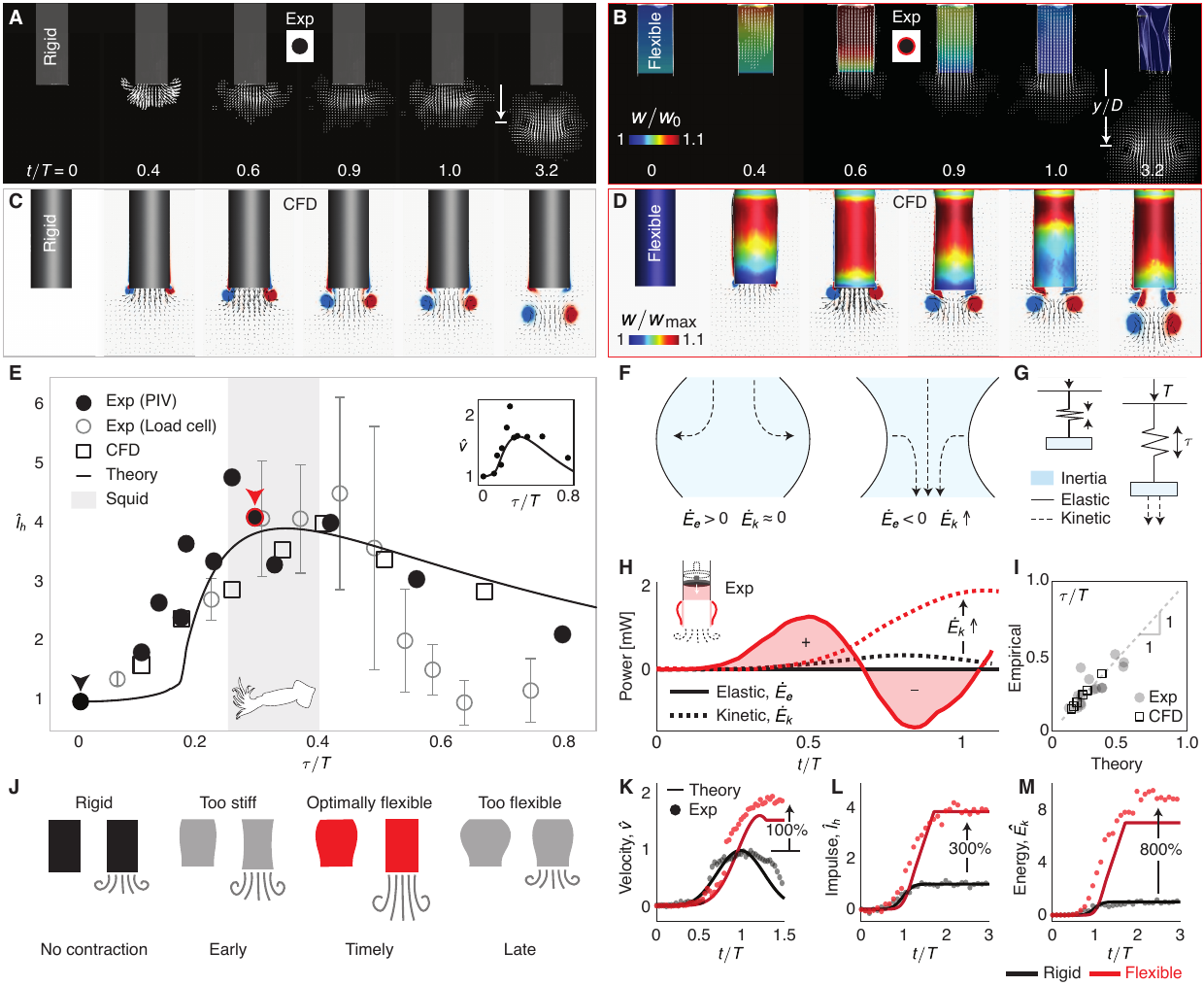}
    \caption{\textbf{Superpropulsion amplifies jet impulse with an optimum at $\boldsymbol{\tau/T\sim}\;$\rev{$\boldsymbol{0.2}$}--$\boldsymbol{0.4}$.}
	  (\textbf{A--D})~Development of a single-pulse jet in experiments  (\textbf{A,B})  and fluid--structure simulations (3D-CFD; \textbf{C,D}) for rigid (\textbf{A,C}) and compliant nozzles (\textbf{B,D}). Velocity vectors show the evolving jet. Colors indicate normalized nozzle width for visualization (\textbf{B}: $W/W_0$; \textbf{D}: $W/W_{\max}$). Time is normalized by the imposed acceleration time $T$.
    (\textbf{E})~Hydrodynamic impulse ratio, $\hat{I}_h = I_h^{\mathrm{flex}}/I_h^{\mathrm{rigid}}$,  versus the response-time ratio $\tau/T$ from \rev{PIV} experiments (filled circles), simulations (open squares), and the superpropulsion model (solid line). 
    \rev{Gray open circles show the flexible-to-rigid ratio of the impulse obtained from load-cell force histories (Figs.~\ref{sif_force_rig} and~\ref{sif_force_histories}). The error bars give the standard error of the mean (SEM) of the trials combined with the uncertainty of the correction for the water column that moves with the piston.}
    \revred{The shaded band marks the \textit{in vivo} range $\tau_{\mathrm{squid}}/T_{\mathrm{squid}} = 0.25$--0.40, spanned by the three condition means in Fig.~\ref{f_squid}J.} Inset: maximum exit-velocity ratio $\hat{v}=v_{\max}^{\mathrm{flex}}/v_{\max}^{\mathrm{rigid}}$  versus $\tau/T$ showing good agreement between theory (solid line) and experiments (filled circles).
    \rev{The PIV} data points include cases reprocessed from~\cite{choi2024mechanism} and new cases acquired to refine the $\tau/T$ range (Table~\ref{tab:S1}).
    (\textbf{F})~Schematic of the superpropulsion mechanism: \revred{during} jet acceleration, the nozzle dilates and absorbs elastic strain energy (elastic strain power $\dot{E}_e > 0$) while the  jet kinetic-power delivery is still small ($\dot{E}_k \approx 0$). During recoil, stored elastic energy is released ($\dot{E}_e < 0$) and transferred to the jet ($\dot{E}_k \uparrow$).
    (\textbf{G})~Analogy to a spring-mass resonator: the actuation time $T$ and nozzle response time $\tau$ govern the interplay between inertia, elastic storage, and kinetic output.
    (\textbf{H})~Time traces of elastic strain power in the nozzle wall ($\dot{E}_e$) and jet kinetic power ($\dot{E}_k$) \revred{for the flexible (red) and the rigid nozzle (black)}, estimated from measured deformation and PIV. Positive $\dot{E}_e$ indicates elastic-energy storage during dilation and negative $\dot{E}_e$ indicates energy return during recoil, coincident with increased $\dot{E}_k$.
    (\textbf{I})~Measured  $\tau/T$ compared with theoretical estimates for experiments and simulations.
    (\textbf{J})~Schematic: impulse amplification is maximized when nozzle recoil timing matches the jet-acceleration timescale. Nozzles that are too stiff or too compliant recoil too early or too late, respectively.
    (\textbf{K--M})~Time evolution for the optimally compliant case (red) versus the rigid nozzle (black), from experiments (symbols) and theory (lines):  (\textbf{K})~$\hat{v}$ (peak $\approx 2\times$, $\sim$100\% increase), (\textbf{L})~$\hat{I}_h$ (\rev{terminal value} $\approx 4\times$, $\sim$300\% increase), and (\textbf{M})~$\hat{E}_k$ (peak $\approx 9\times$, $\sim$800\% increase).
  }
	\label{f_mechanism}
\end{figure*}

\subsection{\textbf{Superpropulsion in compliant nozzles amplifies vortex-ring jets}}
To isolate the hydrodynamic consequences of squid-like jet--funnel coupling, we use a piston-driven single-pulse jet facility~\cite{choi2022flow, choi2024mechanism} and compare an effectively rigid \revred{aluminum} nozzle  with thin-walled silicone nozzles of matched diameter and tunable compliance (Fig.~\ref{f_mechanism}A,B; Fig\rev{s}.~\ref{sif_nozzlefabrication}\rev{--\ref{sif_silicone_stiffness}}). We quantify the unsteady jet using particle-image velocimetry (Fig.~\ref{sif_piv}) and measure nozzle-wall kinematics from synchronized high-speed imaging (Supplementary Movie~\ref{mv_squid}). \revred{The jet timescales of all experiments are compared with those of squid in Fig.~\ref{sif_timescales}.}

    For the representative pulse in Fig.~\ref{f_mechanism}A,B, the rigid nozzle ($Eh \sim 10^7~\mathrm{N/m}$) generates a canonical vortex ring that pinches off and translates downstream. Under the same imposed inlet waveform, a compliant nozzle ($Eh = 14.4~\mathrm{N/m}$) produces a primary vortex ring that convects substantially faster. At matched normalized time the vortex-core displacement  increases by \revred{80--90}\% relative to the rigid case (see arrows at $t/T=3.2$ in Fig.~\ref{f_mechanism}A,B and trajectories in Fig.~\ref{sif_piv}).  PIV-derived integrated metrics show  \rev{a 4$\times$ ($\sim$300\%) increase in the terminal}  hydrodynamic impulse \rev{(Fig.~\ref{sif_common_window}\revred{, matched by a direct load-cell measurement, Figs.~\ref{sif_force_rig} and~\ref{sif_force_histories}})}, and a $\sim$800\% increase in jet kinetic energy (as reported previously for flexible nozzles~\cite{choi2024mechanism}), along with \rev{66\% greater cumulative entrainment (Fig.~\ref{sif_piv}, H)}~\cite{krueger2003significance, morris2025optimal}.

Nozzle kinematics from high-speed imaging show a within-pulse store--release sequence (color contour in Fig.~\ref{f_mechanism}B). During the acceleration phase, the compliant wall dilates under the rising internal pressure, with the dilation initiating near the support and propagating toward the exit (Fig.~\ref{f_mechanism}B,D; $t/T=0.4$). 
The nozzle then recoils (contracts) during the same jet pulse ($t/T \gtrsim 0.6$), while expulsion continues. Power accounting based on measured deformation and PIV shows that the nozzle absorbs  elastic power during dilation and returns it during recoil, coincident with an increase in jet kinetic-power delivery (Fig.~\ref{f_mechanism}F,H). With this process, the flexible nozzle \rev{delivers more power than the rigid nozzle} under identical actuation by \rev{redistributing} energy delivery \rev{in time} (Fig.~\ref{f_squid}F).

We refer to this within-pulse, phase-lagged store--release mechanism, in which the compliant boundary condition returns elastic energy during expulsion out of phase with the imposed inlet waveform, as \emph{superpropulsion}, borrowing terminology from droplet superpropulsion~\cite{prl119108001, challita2023superpropulsion}. \rev{In droplet superpropulsion, a moving surface (the actuator) launches a droplet, which deforms elastically during the launch. When the recoil of the droplet is timed with the motion of the surface, the droplet takes off faster than the maximum speed of the surface. The prefix \emph{super} refers to this speed above the limit of the actuator rather than to a large gain in general. The same holds here, since the jet leaves the flexible nozzle faster than the same actuator drives it through a rigid nozzle.} 

Finally, 3D-CFD reproduces the same dilation--recoil sequence and vortex-ring strengthening for compliant nozzles ($Eh$ = \revtc{75}~$\mathrm{N/m}$) under controlled actuation (Fig.~\ref{f_mechanism}C,D; methodology in Fig.~\ref{sif_cfd_method} and 3D vortex dynamics in Fig.~\ref{sif_cfd_3dflow})~\revred{\cite{choi2026computational}}. \revred{A load cell on a separate rig measures the thrust directly, and once the force is integrated in time and the momentum of the moving solid and water column is subtracted, the flexible nozzle delivers about four times the impulse of the rigid nozzle over $\tau/T \approx 0.3$--$0.45$ (gray circles in Fig.~\ref{f_mechanism}E, rig and histories in Figs.~\ref{sif_force_rig} and~\ref{sif_force_histories}).} These \revred{PIV and load-cell} experiments and simulations indicate that impulse amplification \rev{in inertia-dominated jets} \revred{($Re \approx 700$--$30{,}000$, Table~\ref{tab:S1})} arises from phase-lagged nozzle compliance and is controlled primarily by the response-time ratio $\tau/T$, with a maximum near $\tau/T\approx\rev{0.2}$--0.4 (Fig.~\ref{f_mechanism}E). \rev{Three independent impulse estimates are consistent with this optimum}: hydrodynamic impulse from PIV vorticity fields, a control-volume impulse from 3D-CFD (pressure and momentum-flux contributions), and direct force-time integration from a load cell (\rev{Figs.~\ref{sif_force_rig} and~\ref{sif_force_histories}}). 

\subsection{\textbf{Superpropulsion theory predicts compliant-nozzle jet amplification}}
To interpret the observed $\tau/T$ optimum and provide a predictive design rule, we model the nozzle as a thin-walled compliant tube driven by an imposed inlet flow rate $q_{\mathrm{in}}(t)$. A one-dimensional long-wave formulation for pulsatile flow in compliant tubes~\cite{li2004dynamics, vandevosse2011pulse, hughes1973onedimensional} yields a tractable input--output relation between $q_{\mathrm{in}}$ and the  exit flow rate $q_{\mathrm{out}}$ (Figs.~\ref{f_mechanism}K--M, ~\ref{sif_mathmodel_setting}). 
We nondimensionalize time by the acceleration time $T$ (dots denote derivatives with respect to $t/T$),  giving the leading-order response

\begin{equation}
\ddot{q}_{\mathrm{out}} + \omega_0^2\,q_{\mathrm{out}} = \omega_0^2\,q_{\mathrm{in}},\qquad \omega_0=\frac{\pi}{2}\frac{T}{\tau},
\label{eq:lumped_io_main}
\end{equation}
where $\tau = L/c$ is the nozzle response time (wave-transit time over length $L$) and $c = \sqrt{Eh/(\rho D)}$ is the Moens--Korteweg wave speed for a tube of diameter $D$ (wall stiffness $Eh$ and fluid density $\rho$). \rev{Since $\tau \propto 1/\sqrt{Eh}$, a stiffer nozzle has a faster wall wave and a smaller $\tau/T$, so the ratio measures nozzle stiffness relative to the jet acceleration.} Following the experiments, we prescribe $q_{\mathrm{in}}$ from the rigid-nozzle case under identical upstream actuation~\cite{choi2022flow, choi2024mechanism}.

Equation~\ref{eq:lumped_io_main} is a forced\rev{, undamped} second-order oscillator (spring--mass/\rev{LC} analogue, Figs.~\ref{f_mechanism}G and \ref{sif_mathmodel_lumped}), with dynamics governed by the response-time ratio $\tau/T$. This forced-oscillator structure is mathematically analogous to reduced-order descriptions used in droplet and elastic-projectile superpropulsion, where transient storage and out-of-phase return of internal energy can amplify the output~\cite{prl119108001, challita2023superpropulsion, giombini2022use, celestini2020contactlayer}. When $\tau/T$ is $O(1)$ (here \revred{$\sim 0.2$--$0.4$}), the nozzle boundary condition becomes phase-lagged relative to the imposed inlet waveform, allowing elastic strain energy accumulated during early pressurization to be returned during expulsion (Fig.~\ref{f_mechanism}F). 

Although Eq.~\ref{eq:lumped_io_main} captures the early phase shift and amplitude modulation, the linear response can predict late-time suction (negative $q_{\mathrm{out}}$), which we do not observe experimentally. High-speed imaging instead shows buckling/collapse of the thin-walled nozzle (Fig.~\ref{f_mechanism}B). \rev{This collapse adds the fluid inside the nozzle to the jet, so that about twice the mass of the rigid-nozzle jet leaves at about twice the velocity (Fig.~\ref{sif_slug_budget})\rev{, which accounts for the fourfold impulse and the roughly eightfold kinetic energy of Figs.~\ref{f_mechanism}L,M}.} We therefore augment the model with a minimal collapse closure that prevents suction by expelling the stored nozzle volume after a collapse-onset time defined from a cumulative inflow--outflow criterion (Figs.~\ref{f_mechanism}K--M, \ref{sif_mathmodel_collapsemodel}). 

To compare directly with experiments and 3D-CFD, we compute $\hat{v}$, $\hat{I}_h$, and $\hat{E}_k$ from $q_{\mathrm{out}}$ using the same metric definitions\revred{, restrict integrals to intervals with $q_{\mathrm{out}}>0$, and evaluate them at $t = 4T$, the same time as the PIV and 3D-CFD values} (\revred{Figs.~\ref{sif_common_window} and}~\ref{sif_mathmodel_setting}). Despite its minimal form, the model reproduces the measured time histories of exit-velocity gain, hydrodynamic impulse, and jet kinetic energy for the compliant nozzle (Figs.~\ref{f_mechanism}E, K--M,~\ref{sif_mathmodel_prediction}). It also predicts a pronounced  impulse maximum at $\tau/T \approx 0.2$--$0.4$ consistent with experiments and 3D-CFD. Notably, the \rev{\textit{in vivo}} timing ratio $\tau_{\mathrm{squid}}/T_{\mathrm{squid}}$ lies in the same regime (Fig.~\ref{f_mechanism}E), linking funnel-scale phase lag to the superpropulsion mechanism identified here. 

The forced-oscillator form frames this optimum as an impedance-matching and timing problem. If the nozzle is too stiff ($\tau/T\ll 1$), it completes dilation and recoil before the acceleration-driven pressure rise peaks, so it returns energy prematurely. If the nozzle is too compliant ($\tau/T\gg 1$), recoil arrives after the pulse and fails to overlap the expulsion window (Fig.~\ref{f_mechanism}J). 
For purely sinusoidal forcing, this timing optimum  would occur at $\tau/T=0.5$, but the piston-driven inlet waveform is more sharply peaked than a sinusoid, shifting the optimum to lower $\tau/T$ (Fig.~\ref{sif_mathmodel_collapsemodel}). 
Since \revred{$\tau/T = \sqrt{\rho D L^2/(E h T^2)}$} depends only on nozzle geometry, material properties and the programmed acceleration time, independently measured $\tau$ and programmed $T$ identify the impulse-amplifying regime without additional fitting (Fig.~\ref{f_mechanism}I), a rule we apply next to jet-driven functions in air and water. 

\begin{figure*}[p]
	\centering
    \includegraphics[width=0.92\textwidth]{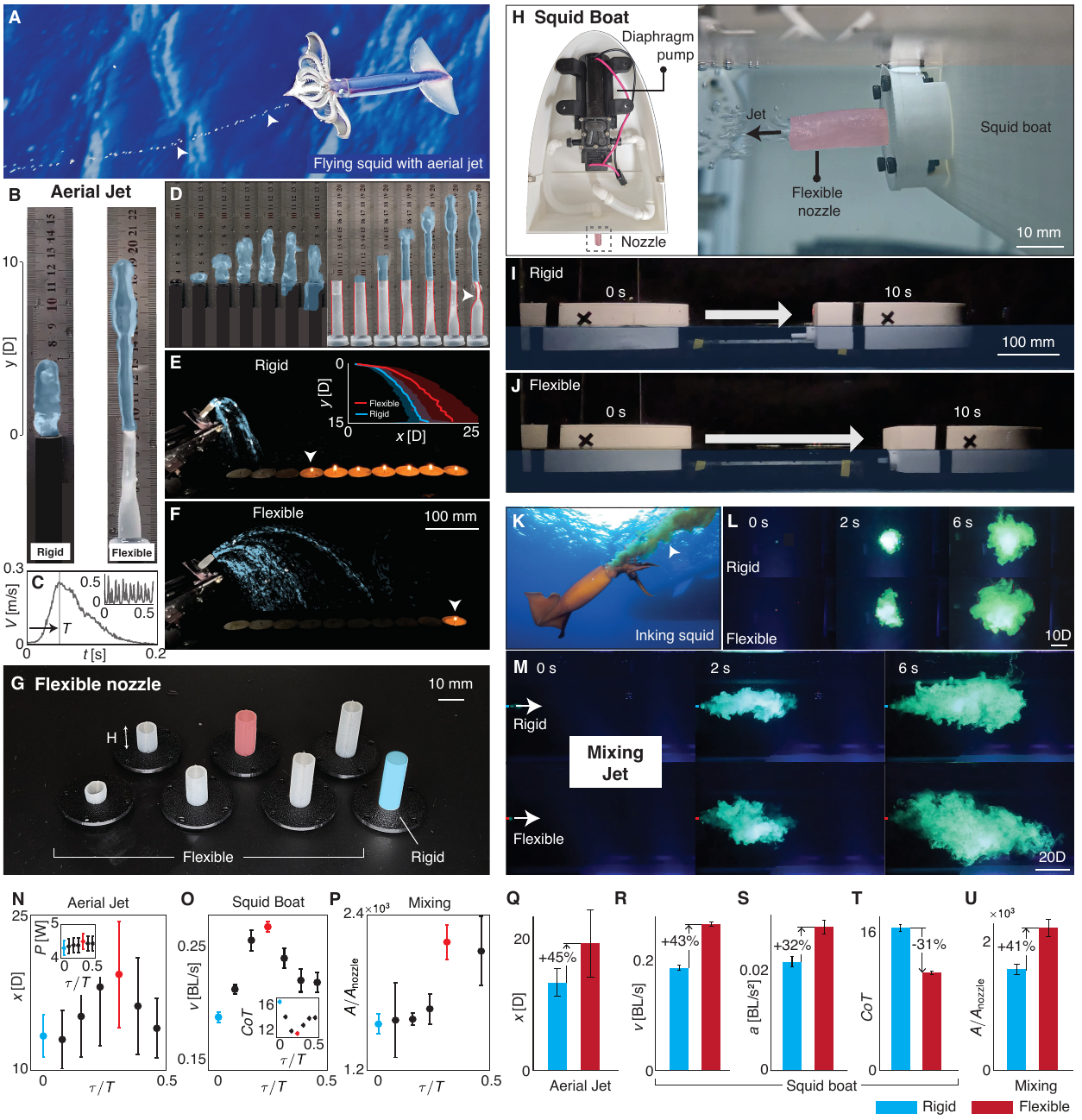}
    \caption{\textbf{Superpropulsion-tuned compliant nozzles improve jet reach, boat propulsion and mixing.}
    (\textbf{A--F})~Aerial-jet demonstrations inspired by flying squid.
    (\textbf{A})~Aerial jetting in a \rev{neon flying squid (\textit{Ommastrephes bartramii})}.
    (\textbf{B})~Single-pulse aerial jet: a  compliant nozzle produces a peak jet height 110\% larger than a rigid nozzle. 
    (\textbf{C})~Measured input waveforms for single-pulse actuation and repeated pulsing (inset).
    (\textbf{D})~Time sequence of jet development and nozzle deformation for the single-pulse case (33~ms between frames).
    (\textbf{E, F})~Candle-extinguishing range test\rev{, where the standoff distance is the farthest distance at which the jet still extinguishes the flame}. Under identical pumping, the compliant nozzle increases the maximum extinguishing standoff distance  by up to $\sim$3$\times$ relative to the rigid nozzle. Inset,  averaged jet trajectories.
    (\textbf{G})~Nozzle set used to tune response time: compliant nozzles with heights 5--30~mm (\revred{$\tau/T = 0.08$--0.46}) and a rigid nozzle. Shading marks the optimally compliant and rigid cases used for comparison.
    (\textbf{H--J})~Jet-propelled ``squid-boat'' demonstration. 
    (\textbf{H})~ Squid boat using a compliant nozzle.
    (\textbf{I, J})~Representative  translations under identical pumping. The compliant nozzle increases the  mean boat speed by \revred{43\%} relative to the rigid nozzle.
    (\textbf{K--M})~Mixing demonstration inspired by squid inking. 
    (\textbf{K})~Inking in \rev{\textit{Dosidicus gigas}}. (\textbf{L,M})~Dye visualization (front and side views): the compliant nozzle produces a more dispersed dye cloud than the rigid nozzle.
    (\textbf{N--P})~Dependence of application-level performance on $\tau/T$: 
    (\textbf{N}) normalized horizontal jet distance $x/D$ \revred{with the power consumption $P$ in the inset}, (\textbf{O}) boat speed \revred{with the cost of transport (CoT) in the inset}, and (\textbf{P}) normalized dye dispersion area $A/A_\text{nozzle}$.
    \rev{Performance} metrics peak near $\tau/T \approx 0.3$.
    (\textbf{Q--U})~Performance gains at the optimal condition relative to the rigid nozzle: (\textbf{Q})~mean horizontal jet distance +44.6\%, (\textbf{R})~boat speed \revred{+42.6\%}, (\textbf{S})~acceleration \revred{+32.0\%}, (\textbf{T})~CoT \revred{$-31.4\%$}, (\textbf{U})~normalized dye dispersion +40.5\%. 
    In (\textbf{N}), error bars represent temporal standard deviations over $t/T = 0$--$8{,}000$; $N$ $(n)$ = 6 (1), where $n$ is \revred{the number of} replicates per condition and $N$ \revred{the total number of} measurements.
    \revred{In (\textbf{O}) and (\textbf{P}), values are taken at a representative time, with $N$ $(n)$ = 35 (5) in (\textbf{O}) and 18 (3) in (\textbf{P}).}
    Image credits: panel A, Anthony Pierce; panel K, \rev{Blue Planet Archive / Richard Herrmann}.
    }
\label{f_application}
\end{figure*}

\subsection{\textbf{Aerial jets}}
As a first demonstration, we discharge liquid jets into air, motivated by \rev{the aerial jetting of flying squid~\cite{odor2013rocket, muramatsu2013oceanic, marshall2012squid}} (Fig.~\ref{f_application}A).
For a fixed single-pulse actuation waveform, the compliant nozzle increases the peak jet height by 110\% relative to a  rigid nozzle (Figs.~\ref{f_application}B,C;~\ref{sif_aerialjet}).
Time-resolved imaging shows that the  emerging water column reaches its peak about \revred{70}~ms later for the compliant nozzle than for the rigid nozzle, consistent with within-pulse elastic storage and recoil (Fig.~\ref{f_application}D). Under repeated pulsing driven by  a commercial diaphragm pump, the compliant nozzle similarly increases  jet strength across successive pulses, as reflected in nozzle-exit velocity histories (inset of Fig.~\ref{f_application}C; Fig.~\ref{sif_repeated_pulse}). 

\rev{As an operational measure of working reach, we use the candle-extinguishing standoff distance, the farthest distance at which the jet still arrives with enough momentum to extinguish a candle flame. This measure gives us a pass-or-fail threshold that an exit-velocity measurement alone does not provide.} In a range test at a fixed launch angle of 45$^\circ$, the compliant nozzle increases the maximum candle-extinguishing standoff distance by $\sim$3$\times$ (Fig.~\ref{f_application}E,F; \revred{Supplementary} Movie~\ref{mv_squid}).  Trajectory tracking shows a mean increase of $\sim$40\% in jet displacement at $y/D=15$ (inset of Fig.~\ref{f_application}E), with larger shot-to-shot scatter consistent with steeper velocity gradients measured in the compliant case (Fig.~\ref{sif_aerialjet}). \revred{We also examine how the gain changes with the launch angle relative to gravity (Fig.~\ref{sif_downwardjet}). When the hydrostatic pressure is not negligible relative to the elastic pressure of the wall, a downward jet leaves the flexible nozzle about twice as fast as it leaves the rigid one.}

\subsection{\textbf{Squid-boat jet propulsion}}
We next test whether tuning nozzle response time improves vehicle-scale propulsion using a jet-driven ``squid-boat'' that travels at the water surface while expelling a submerged pulsed jet through interchangeable nozzles (Fig.~\ref{f_application}H; Fig.~\ref{sif_boat}).
We track  boat motion and  electrical power draw to quantify speed and energetic cost. We also define cost of transport as $\text{CoT} = E / (m g d)$, where $E$ is the  electrical  energy consumed, $m$ is  vehicle mass, $g$ is gravitational acceleration, and $d$ is  distance traveled.  With a compliant nozzle tuned to the superpropulsion regime ($\tau/T = 0.23$), the boat moves \revred{42.6\%} faster than with the rigid nozzle \revred{(Fig.~\ref{f_application}I, J, R; Supplementary Movie~\ref{mv_squid})} and reduces cost of transport  by \revred{31.4\%} under the same pump setting \revred{(Fig.~\ref{f_application}T)}. 

To assess portability across scale, we repeat the experiment in a miniaturized platform with both  vehicle mass and jet pulse volume  reduced by a factor of $\sim40$ (Fig.~\ref{sif_smallboat}). Using a miniature diaphragm pump and a \revred{dip-coated compliant nozzle 2~mm in diameter}  (Fig.~\ref{sif_nozzlefabrication}), the compliant-nozzle mini-boat again outperforms its rigid counterpart, reaching speeds  $>50\%$ higher under the same pumping conditions. 

\subsection{\textbf{Inking-inspired dispersion and mixing}}
Squid  disperse ink to form large clouds that act as visual  decoys (Fig.~\ref{f_application}K). \rev{The animal may steer or move the funnel during inking, so here we focus only on the contribution of a passive flexible nozzle.}  To test whether compliant-nozzle superpropulsion enhances mixing, we inject  fluorescent dye into a quiescent tank and record the evolving plume from orthogonal views (Figs.~\ref{f_application}L,M). We quantify dispersion by segmenting the largest connected dye region and tracking its projected  area over time, normalized by the nozzle cross-sectional area (Fig.~\ref{sif_mixing}).

Compared with the rigid nozzle, the compliant nozzle produces faster axial penetration and stronger lateral spreading, consistent with enhanced entrainment (Fig.~\ref{f_application}L,M; Fig.~\ref{sif_piv}; \revred{Supplementary} Movie~\ref{mv_squid}). At the superpropulsion-tuned condition, the normalized dispersion metric increases by 40.5\% relative to the rigid nozzle (Fig.~\ref{f_application}U). The spreading resembles ``blooming\revred{''} jets, in which broken axisymmetry and self-induced vortex dynamics promote plume widening~\cite{reynolds2003bifurcating, gohil2015simulation}.
Consistent with this interpretation, high-speed imaging shows self-excited motion at the compliant nozzle tip (Fig.~\ref{f_application}D), which can provide an intrinsic symmetry-breaking perturbation without external forcing.

\section*{\revtc{Discussion}}
We examine an understudied component of squid jet propulsion, the soft funnel. Using histology and \rev{\textit{in vivo}}  kinematics (Fig.~\ref{f_squid}I--J, \ref{sif_chromatophore}) in two species ($n=11$ animals; $N>30$ jet pulses), we identify a reproducible, phase-lagged expansion--recoil during jetting. We then isolate its hydrodynamic consequences with engineered soft nozzles, supported by 3D fluid--structure simulations and reduced-order theory, and we test the resulting timescale-matching prediction across $>\revred{300}$ nozzle and application trials. Across platforms, performance peaks near $\tau/T\approx0.3$, consistent with the impulse-optimal superpropulsion regime (Figs.~\ref{f_mechanism}E,~\ref{f_application}G, N--P). Because power consumption varies only modestly across the sweep (Fig.~\ref{f_application}N, inset), tuning $\tau/T$ primarily  shifts when stored elastic energy returns to the flow, translating within-pulse elastic recoil into gains in jet reach, vehicle transport, and plume dispersion (Fig.~\ref{f_application}Q--U). That  these outcomes agree across air and water, and across single-pulse and repeated-pulse operation, supports $\tau/T$ as a practical design parameter for pulsed-jet systems. 

At first glance, the reduced CoT in the compliant-nozzle boat can seem counterintuitive because the compliant case also delivers higher jet kinetic energy per pulse (Figs.~\ref{f_mechanism}M, \ref{f_application}T). Superpropulsion does not create energy. Under prescribed kinematics, nozzle compliance reshapes the pressure history seen by the actuator and allows an underloaded drive to transfer additional work to the fluid. The elastic nozzle stores part of that work as strain during dilation and returns it on recoil within the same pulse, enhancing \rev{the terminal} impulse (Figs.~\ref{f_mechanism}L,M)~\cite{prl119108001,challita2023superpropulsion,choi2024mechanism}. 

\rev{The lower  cost of transport becomes clearer from the power budget. For a self-propelled body of weight $W$ cruising at steady speed $v$, the power splits into a speed-independent overhead and the power needed to overcome drag,
\begin{equation}
  P_{\mathrm{total}} = P_{\mathrm{fixed}} + P_{\mathrm{prop}},
  \label{eq:ptotal}
\end{equation}
where $P_{\mathrm{fixed}}$  comes from pump friction, valve and cam losses, and $P_{\mathrm{prop}} = \tfrac{1}{2}\rho C_D \revred{A_{\mathrm{f}}}\,v^{3} \equiv \revred{k}\,v^{3}$ is the power needed to overcome drag \revred{(drag coefficient $C_D$, frontal area $A_{\mathrm{f}}$)}. Writing the cost of transport as energy per unit weight per unit distance, which is the definition used above expressed as a power ratio, gives
\begin{equation}
  \mathrm{CoT} \equiv \frac{P_{\mathrm{total}}}{W\,v}
  = \frac{P_{\mathrm{fixed}}}{W\,v} + \frac{\revred{k}\,v^{2}}{W}.
  \label{eq:gcot_def}
\end{equation}
The two terms vary in opposite directions with speed, the overhead term falling as $v^{-1}$ and the drag term rising as $v^{2}$. In our pump-driven boat the useful jet power is about $0.5$~mW against an electrical input of about $5$~W, so the dimensionless propulsive power $\revred{\Pi} = 2P_{\mathrm{prop}}/P_{\mathrm{fixed}}$ is of order $10^{-4}$ \revred{(the jet kinetic power bounds $P_{\mathrm{prop}}$ from above, so this estimate of $\Pi$ is conservative)} and the overhead term dominates. Superpropulsion raises the jet impulse and hence the cruising speed.  $P_{\mathrm{fixed}}$ is set by the pump and drive and changes only modestly, so the dominant term falls as $v^{-1}$ and the cost of transport drops accordingly (Fig.~\ref{f_application}T). The gain therefore comes from covering more distance per unit of overhead rather than from any reduction in energy consumed. The full derivation, including the universal $\mathrm{CoT}/\mathrm{CoT}_{\min}$ relation, is given in the SI (see \revred{\ref{si:actuation_energetics}}).} 

Overarm throwing provides a useful analogue \rev{of this overhead-dominated regime. Elastic} storage and recoil in the shoulder\revred{--}arm system can boost release speed for a light projectile (e.g., a baseball) without a commensurate increase in energetic cost, because
the added projectile kinetic energy is small compared with the  metabolic expenditure of the throwing motion~\cite{Naito2021,roach2013elastic}. The same timing advantage diminishes for heavier loads  once the motion becomes load-limited (e.g., the shot put). Whether superpropulsion yields energetic gains under load-limited operation, where motor torque or pump-pressure ceilings interact with the jet load, remains to be established.

Biologically, the collagen-rich sheath and repeatable phase lag point to a  passive elastohydrodynamic contribution on the pulse timescale (see Figs.~\ref{sif_histology}--\ref{sif_paralyze})~\cite{anderson2000mechanics,bartol2009hydrodynamics,staaf2014aperture,gosline1983patterns}. Disentangling passive recoil from muscle-driven modulation during an escape jet is experimentally difficult \rev{\textit{in vivo}}. Rather than a strict active\revred{--}passive dichotomy, a plausible division of \revred{labor} is that muscle sets the operating point (orientation, geometry and aperture), while passive recoil shapes within-pulse timing. How that operating point shifts across \revred{behaviors}, ontogeny, and species remains an open question. \rev{A further limitation is that we treat the funnel as the sole outlet. Flow escaping at the mantle--head margin, which we did not measure, would lower the funnel flux, the jet pressure and the effective stroke ratio~\cite{staaf2014aperture, gharib1998universal}.}

Compliant outlets coupled to pulsatile through-flows occur well beyond cephalopods, from salp siphons~\cite{sutherland2010comparative} and self-oscillating velvet-worm jets~\cite{concha2015oscillation} to elastic cardiovascular conduits~\cite{vandevosse2011pulse,hughes1973onedimensional,li2004dynamics}. \rev{Jet superpropulsion is one instance of the elastic storage and return known from such conduits. The nozzle acts as an elastic capacitor that charges and discharges within a single pulse, as the arterial wall does over a heartbeat in the Windkessel effect~\cite{westerhof2009arterial}.} Timescale-matching optima also underpin efficient actuation in other systems, from resonant jellyfish swimming~\cite{demont1988resonance,hoover2015jellyfish} and stiffness-tuned fish swimming~\cite{tytell2014role,zhong2021tunable} to resonant bioinspired platforms~\cite{bujard2021resonant,haldane2016robotic}. These parallels suggest that response-time matching may be a reusable evolutionary motif, and an effective design strategy, for shaping pulsatile flows when compliant structures mediate energy exchange between an internal driver and the surrounding fluid.

From an engineering standpoint, we focus on the dominant (first-harmonic) response, but repeated-pulsing experiments reveal secondary peaks (Fig.~\ref{sif_aerialjet}), suggesting a multi-peak response  akin to impedance pumping and Starling-resistor dynamics~\cite{forouhar2006suction,hickerson2006resonance,avrahami2008computational,aghilinejad2025power,jensen1989existence,jensen1990instabilities,luo1996numerical,grotberg2004biofluid,wang2021energetics}. Although these higher modes contribute less to axial impulse amplification over the range tested here, they may couple more strongly to transverse tip motion and symmetry breaking, providing an intrinsic route to jet spreading and mixing similar to other self-oscillating elastohydrodynamic jets~\cite{concha2015oscillation}. The $\tau/T$-tuned framework (Eq.~\ref{eq:lumped_io_main}) should hold across scales in inertia-dominated regimes, whereas strong viscous losses at low Reynolds/Womersley number, or cavitation and compressibility at extreme operating conditions, can attenuate recoil and shift the timing of peak amplification.

\rev{Pulsed jets are used where a fluid is delivered in short bursts, for example in thrusters for underwater vehicles and soft robots, in pulsed sprays for fire suppression and cleaning, and in mixing and cooling. In such systems a nozzle whose response time is matched to the pulse could raise the jet impulse without added actuation, as in the boat, the aerial jet and the dye plume tested here. The gain is limited to pulsed operation, because a flexible nozzle returns stored elastic energy only within a pulse, so a nozzle that discharges steadily, such as that of a fire hose, would benefit at start-up and not in continuous discharge.} \rev{The three demonstrations test the same timing rule, each in a different medium, load and metric. They illustrate the fluid-mechanical mechanism, whereas the validation of a complete propulsion system is left for future work.} 

In ongoing work, we are integrating superpropulsion-tuned compliant nozzles into rotary propulsors to test whether blade-rate forcing can generate a controlled pulse train~\revred{\cite{wallace2025multi,singh2026multi}}. If wake timing and spectra can be tuned this way, soft nozzles may provide a passive route to  improved efficiency and maneuverability for marine vehicles while reducing radiated noise (acoustic signature management). Beyond passive cylinders, compliance patterning and stimuli\revred{-}responsive materials could enable actively tunable nozzles that switch between thrust, vectoring, and mixing, augmenting control authority~\cite{bertoldi2017flexible,kim2022magnetic}.

\section*{\revtc{Materials and Methods}}
\revtc{Live squid (\textit{Doryteuthis opalescens} and \textit{Sepioteuthis lessoniana}) were collected and kept in seawater tanks (collection sites and holding tanks in Fig.~\ref{sif_squid} of the Supplementary Materials), and funnel and mantle tissue of \textit{D.~opalescens} was examined by histology (histological staining in Fig.~\ref{sif_histology}). During escape jets the squid were filmed with a high-speed camera, the mantle-cavity pressure of restrained \textit{D.~opalescens} was recorded at the same time, and the funnel and mantle widths were measured by tracking chromatophores (imaging setup, pressure measurement and tracking method in Fig.~\ref{sif_chromatophore}, validation against 2D image correlation in Fig.~\ref{sif_membrane_dic}, correction of track jumps in Fig.~\ref{sif_tracking_jumps}, validation against 3D image correlation in Fig.~\ref{sif_balloon_dic}). Restrained \textit{D.~opalescens} were also measured after funnel denervation (denervation method, reflex-test setup and validation in Fig.~\ref{sif_paralyze}). Thin-walled silicone nozzles (fabrication and installation in Fig.~\ref{sif_nozzlefabrication}, wall thickness in Fig.~\ref{sif_nozzle_thickness}, silicone stiffness in Fig.~\ref{sif_silicone_stiffness}) were tested in a piston-driven single-pulse jet facility with particle image velocimetry and nozzle-wall tracking (jet facility, PIV and flow analysis in Fig.~\ref{sif_piv}, jet timescales in Fig.~\ref{sif_timescales}, impulse at a common time in Fig.~\ref{sif_common_window}). Some of these recordings were reprocessed from an earlier study~\cite{choi2024mechanism}, and Table~\ref{tab:S1} (experimental and simulation conditions) lists every data set of this work and marks those reported before and those acquired for this study. The impulse was verified independently by direct force measurement with a load cell (load-cell rig and rigid-nozzle impulse in Fig.~\ref{sif_force_rig}, force and impulse histories in Fig.~\ref{sif_force_histories}). Two-way coupled fluid--structure interaction simulations used OpenFOAM, CalculiX and preCICE (simulation setup and impulse in Fig.~\ref{sif_cfd_method}, nozzle deformation and vortex structure in Fig.~\ref{sif_cfd_3dflow}), and a reduced-order model was derived (1D tube model in Fig.~\ref{sif_mathmodel_setting}, analogies in Fig.~\ref{sif_mathmodel_lumped}, mass, impulse and energy of the jet in Fig.~\ref{sif_slug_budget}, collapse model and its validation in Fig.~\ref{sif_mathmodel_collapsemodel}, predictions against experiments and simulations in Fig.~\ref{sif_mathmodel_prediction}). The demonstrations were driven mostly by commercial diaphragm pumps (aerial jets in Fig.~\ref{sif_aerialjet}, pump setup and repeated pulses in Fig.~\ref{sif_repeated_pulse}, downward jets in Fig.~\ref{sif_downwardjet}, boat setup and results in Fig.~\ref{sif_boat}, small-boat setup and results in Fig.~\ref{sif_smallboat}, mixing setup and image analysis in Fig.~\ref{sif_mixing}). Sample sizes and error measures are given in the figure legends and in Table~\ref{tab:S1}, and the Supplementary Materials describe every method in full. All experiments and procedures involving animals were conducted independently by the Stanford University team under their approved experimental protocol, Stanford University Administrative Panel on Laboratory Animal Care (APLAC) Protocol 10643, and in accordance with the Universities Federation for Animal Welfare guidelines and Stanford University Institutional Animal Care guidelines. These animal experiments were not part of the DARPA-funded scope of work.}

\bibliography{references}

@article{anderson2000mechanics,
  author={Anderson, Erik J and DeMont, M Edwin},
  title={The mechanics of locomotion in the squid Loligo pealei: locomotory function and unsteady hydrodynamics of the jet and intramantle pressure},
  journal={J. Exp.\ Biol.},
  volume={203},
  pages={2851--2863},
  year={2000},
  doi={10.1242/jeb.203.18.2851}
}

@article{pereira2022sleap,
  title={{SLEAP}: A deep learning system for multi-animal pose tracking},
  author={Pereira, Talmo D and Tabris, Nathaniel and Matsliah, Arie and Turner, David M and Li, Junyu and Ravindranath, Shruthi and Papadoyannis, Eleni S and Normand, Edna and Deutsch, David S and Wang, Z Yan and McKenzie-Smith, Grace C and Mitelut, Catalin C and Castro, Marielisa Diez and D'Uva, John and Kislin, Mikhail and Sanes, Dan H and Kocher, Sarah D and Wang, Samuel S-H and Falkner, Annegret L and Shaevitz, Joshua W and Murthy, Mala},
  journal={Nat.\ Methods},
  volume={19},
  number={4},
  pages={486--495},
  year={2022},
  publisher={Nature Publishing Group},
  doi={10.1038/s41592-022-01426-1}
}

@article{gilly2006vertical,
  title={Vertical and horizontal migrations by the jumbo squid \textit{Dosidicus gigas} revealed by electronic tagging},
  author={Gilly, William F and Markaida, Unai and Baxter, Colin H and Block, Barbara A and Boustany, Andre and Zeidberg, Louis and Reisenbichler, Kim and Robison, Bruce and Bazzino, Gabriela and Salinas, Carmen},
  journal={Mar.\ Ecol.\ Prog.\ Ser.},
  volume={324},
  pages={1--17},
  year={2006},
  publisher={Inter-Research Science Center},
  doi={10.3354/meps324001}
}

@article{choi2022atomization,
  title={Analysis of liquid column atomization by annular dual-nozzle gas jet flow},
  author={Choi, Daehyun and Byun, Jinho and Park, Hyungmin},
  journal={J. Fluid Mech.},
  volume={943},
  pages={A25},
  year={2022},
  doi={10.1017/jfm.2022.435}
}

@article{dabiri2009optimal,
  title={Optimal vortex formation as a unifying principle in biological propulsion},
  author={Dabiri, John O},
  journal={Annu.\ Rev.\ Fluid Mech.},
  volume={41},
  pages={17--33},
  year={2009},
  publisher={Annual Reviews},
  doi={10.1146/annurev.fluid.010908.165232}
}

@article{dabiri2005starting,
  title={Starting flow through nozzles with temporally variable exit diameter},
  author={Dabiri, John O and Gharib, Morteza},
  journal={J. Fluid Mech.},
  volume={538},
  pages={111--136},
  year={2005},
  publisher={Cambridge University Press},
  doi={10.1017/S002211200500515X}
}

@article{gharib1998universal,
  title={A universal time scale for vortex ring formation},
  author={Gharib, Morteza and Rambod, Edmond and Shariff, Karim},
  journal={J. Fluid Mech.},
  volume={360},
  pages={121--140},
  year={1998},
  publisher={Cambridge University Press},
  doi={10.1017/S0022112097008410}
}

@article{hove2003intracardiac,
  title={Intracardiac fluid forces are an essential epigenetic factor for embryonic cardiogenesis},
  author={Hove, Jay R and K{\"o}ster, Rolf W and Forouhar, Arian S and Acevedo-Bolton, Gabriel and Fraser, Scott E and Gharib, Morteza},
  journal={Nature},
  volume={421},
  number={6919},
  pages={172--177},
  year={2003},
  publisher={Nature Publishing Group},
  doi={10.1038/nature01282}
}

@article{aghilinejad2025power,
  title={Power-frequency relationship of wave dynamics in fluid-filled compliant tubes},
  author={Aghilinejad, Amirhossein and Amlani, Faezeh and Gharib, Morteza},
  journal={Phys.\ Rev.\ Fluids},
  volume={10},
  number={3},
  pages={033102},
  year={2025},
  publisher={American Physical Society},
  doi={10.1103/PhysRevFluids.10.033102}
}

@article{avrahami2008computational,
  title        = {Computational studies of resonance wave pumping in compliant tubes},
  author       = {Avrahami, Idit and Gharib, Morteza},
  journal      = {J. Fluid Mech.},
  volume       = {608},
  pages        = {139--160},
  year         = {2008},
  publisher    = {Cambridge University Press},
  doi          = {10.1017/S0022112008002012},
}

@article{hickerson2006resonance,
  title        = {On the resonance of a pliant tube as a mechanism for valveless pumping},
  author       = {Hickerson, Anna I. and Gharib, Morteza},
  journal      = {J. Fluid Mech.},
  volume       = {555},
  pages        = {141--148},
  year         = {2006},
  publisher    = {Cambridge University Press},
  doi          = {10.1017/S0022112006009220},
  url          = {https://www.cambridge.org/core/journals/journal-of-fluid-mechanics/article/on-the-resonance-of-a-pliant-tube-as-a-mechanism-for-valveless-pumping/7A4EAE5EA67D0D1B920B9F933B7867EB},
}

@article{krueger2003significance,
  title={The significance of vortex ring formation to the impulse and thrust of a starting jet},
  author={Krueger, Paul S. and Gharib, Morteza},
  journal={Phys.\ Fluids},
  volume={15},
  number={5},
  pages={1271--1281},
  year={2003},
  doi={10.1063/1.1564600},
  url={https://aip.scitation.org/doi/10.1063/1.1564600}
}

@article{krueger2006formation,
  title={The formation number of vortex rings formed in uniform background co-flow},
  author={Krueger, Paul S. and Dabiri, John O. and Gharib, Morteza},
  journal={J. Fluid Mech.},
  volume={556},
  pages={147--166},
  year={2006},
  publisher={Cambridge University Press},
  doi={10.1017/S0022112006009347}
}

@article{bartol2009hydrodynamics,
  title        = {Hydrodynamics of pulsed jetting in juvenile and adult brief squid \textit{Lolliguncula brevis}: evidence of multiple jet 'modes' and their implications for propulsive efficiency},
  author       = {Bartol, Ian K. and Krueger, Paul S. and Stewart, William J. and Thompson, Joseph T.},
  journal      = {J. Exp.\ Biol.},
  volume       = {212},
  number       = {12},
  pages        = {1889--1903},
  year         = {2009},
  publisher    = {The Company of Biologists Ltd},
  doi          = {10.1242/jeb.027771},
  url          = {https://journals.biologists.com/jeb/article/212/12/1889/19136/Hydrodynamics-of-pulsed-jetting-in-juvenile-and},
}

@article{bartol2008swimming,
  title={Swimming dynamics and propulsive efficiency of squids throughout ontogeny},
  author={Bartol, Ian K and Krueger, Paul S and Thompson, Joseph T and Stewart, W Jon},
  journal={Integr.\ Comp.\ Biol.},
  volume={48},
  number={6},
  pages={720--733},
  year={2008},
  publisher={Oxford University Press},
  doi={10.1093/icb/icn043}
}

@article{odor2013rocket,
  title={Squid rocket science: How squid launch into air},
  author={O'Dor, Ron and Stewart, Julia and Gilly, William and Payne, John and Borges, Teresa Cerveira and Thys, Tierney},
  journal={Deep Sea Res.\ Part II Top.\ Stud.\ Oceanogr.},
  volume={95},
  pages={113--118},
  year={2013},
  publisher={Elsevier},
  doi={10.1016/j.dsr2.2012.07.002}
}

@article{muramatsu2013oceanic,
  title={Oceanic squid do fly},
  author={Muramatsu, Kazuo and Yamamoto, Jun and Abe, Takuji and Sekiguchi, Katsufumi and Hoshi, Norihisa and Sakurai, Yasunori},
  journal={Mar.\ Biol.},
  volume={160},
  number={5},
  pages={1171--1175},
  year={2013},
  publisher={Springer},
  doi={10.1007/s00227-013-2169-9}
}

@article{weymouth2013ultrafast,
  title={Ultra-fast escape of a deformable jet-propelled body},
  author={Weymouth, G D and Triantafyllou, M S},
  journal={J. Fluid Mech.},
  volume={721},
  pages={367--385},
  year={2013},
  publisher={Cambridge University Press},
  doi={10.1017/jfm.2013.65}
}

@article{moslemi2010propulsive,
  title={Propulsive efficiency of a biomorphic pulsed-jet underwater vehicle},
  author={Moslemi, A A and Krueger, P S},
  journal={Bioinspir.\ Biomim.},
  volume={5},
  number={3},
  pages={036003},
  year={2010},
  publisher={IOP Publishing},
  doi={10.1088/1748-3182/5/3/036003}
}

@article{whittlesey2013optimal,
  title={Optimal vortex formation in a self-propelled vehicle},
  author={Whittlesey, Robert W and Dabiri, John O},
  journal={J. Fluid Mech.},
  volume={737},
  pages={78--104},
  year={2013},
  publisher={Cambridge University Press},
  doi={10.1017/jfm.2013.560}
}

@article{gao2020jfm,
  title        = {Development of the impulse and thrust for laminar starting jets with finite discharged volume},
  author       = {Gao, Lei and Wang, Xin and Yu, Simon C. M. and Chyu, Minking K.},
  journal      = {J. Fluid Mech.},
  volume       = {902},
  pages        = {A27},
  year         = {2020},
  publisher    = {Cambridge University Press},
  doi          = {10.1017/jfm.2020.570},
}

@article{christianson2020cephalopod,
  title        = {Cephalopod-inspired robot capable of cyclic jet propulsion through shape change},
  author       = {Christianson, Caleb and Cui, Yi and Ishida, Michael and Bi, Xiaobo and Zhu, Qiang and Pawlak, Geno and Tolley, Michael T.},
  journal      = {Bioinspir.\ Biomim.},
  volume       = {16},
  number       = {1},
  pages        = {016014},
  year         = {2020},
  publisher    = {IOP Publishing},
  doi          = {10.1088/1748-3190/abbc72}
}

@article{weymouth2015octopus,
  title={Ultra-fast escape maneuver of an octopus-inspired robot},
  author={Weymouth, G. D. and Subramaniam, V. and Triantafyllou, M. S.},
  journal={Bioinspir.\ Biomim.},
  volume={10},
  number={1},
  pages={016016},
  year={2015},
  publisher={IOP Publishing},
  doi={10.1088/1748-3190/10/1/016016}
}

@article{zhang2021siphon,
  title={A Cephalopod-Inspired Soft-Robotic Siphon for Thrust Vectoring and Flow Rate Regulation},
  author={Zhang, Runzhi and Shen, Zhong and Zhong, Hua and Tan, Jiyong and Hu, Yong and Wang, Zheng},
  journal={Soft Robot.},
  volume={8},
  number={4},
  pages={416--431},
  year={2021},
  publisher={Mary Ann Liebert, Inc.},
  doi={10.1089/soro.2019.0152}
}

@article{flores2025robonautilus,
  title={RoboNautilus: a cephalopod-inspired soft robotic siphon for underwater propulsion},
  author={Flores, Dominic and Sandhu, Sahib and White, Alexander and Yin, Alexander and Li, Ang Leo and Kang, Soohyeon and Wang, Yuechao and Chamorro, Leonardo P. and Duduta, Mihai},
  journal={npj Robot.},
  volume={3},
  pages={17},
  year={2025},
  publisher={Nature Publishing Group},
  doi={10.1038/s44182-025-00035-2}
}

@article{gosline1983patterns,
  title={Patterns of Circular And Radial Mantle Muscle Activity in Respiration and Jetting of the Squid \textit{Loligo opalescens}},
  author={Gosline, John M. and Steeves, John D. and Harman, Anthony D. and DeMont, M. Edwin},
  journal={J. Exp.\ Biol.},
  volume={104},
  number={1},
  pages={97--109},
  year={1983},
  doi={10.1242/jeb.104.1.97},
  url={https://journals.biologists.com/jeb/article/104/1/97/40965/Patterns-of-Circular-And-Radial-Mantle-Muscle}
}

@article{young1938functioning,
  title={The functioning of the giant nerve fibres of the squid},
  author={Young, J Z},
  journal={J. Exp.\ Biol.},
  volume={15},
  number={2},
  pages={170--185},
  year={1938},
  doi={10.1242/jeb.15.2.170}
}

@article{hodgkin1952quantitative,
  title={A quantitative description of membrane current and its application to conduction and excitation in nerve},
  author={Hodgkin, Alan L and Huxley, Andrew F},
  journal={J. Physiol.},
  volume={117},
  number={4},
  pages={500--544},
  year={1952},
  doi={10.1113/jphysiol.1952.sp004764}
}

@article{otis1990concerted,
  title={Jet-propelled escape in the squid \textit{Loligo opalescens}: concerted control by giant and non-giant motor axon pathways},
  author={Otis, Thomas S and Gilly, William F},
  journal={Proc.\ Natl.\ Acad.\ Sci.\ U.S.A.},
  volume={87},
  number={8},
  pages={2911--2915},
  year={1990},
  doi={10.1073/pnas.87.8.2911}
}

@article{neumeister2000temperature,
  title={Effects of temperature on escape jetting in the squid \textit{Loligo opalescens}},
  author={Neumeister, Harald and Ripley, Brenda and Preuss, Thomas and Gilly, William F},
  journal={J. Exp.\ Biol.},
  volume={203},
  number={3},
  pages={547--557},
  year={2000},
  doi={10.1242/jeb.203.3.547}
}

@article{li2019hypoxia,
  title={Hypoxia tolerance of giant axon-mediated escape jetting in {California} market squid (\textit{Doryteuthis opalescens})},
  author={Li, Diana H and Gilly, William F},
  journal={J. Exp.\ Biol.},
  volume={222},
  number={8},
  pages={jeb198812},
  year={2019},
  doi={10.1242/jeb.198812}
}

@article{li2023hydrodynamic,
  title={Hydrodynamic diversity of jets mediated by giant and non-giant axon systems in brief squid},
  author={Li, Diana H and Bartol, Ian K and Gilly, William F},
  journal={Integr.\ Comp.\ Biol.},
  volume={63},
  number={6},
  pages={1266--1276},
  year={2023},
  doi={10.1093/icb/icad086}
}

@article{gilly1984threshold,
  title={Threshold channels---a novel type of sodium channel in squid giant axon},
  author={Gilly, William F and Armstrong, Clay M},
  journal={Nature},
  volume={309},
  pages={448--450},
  year={1984},
  doi={10.1038/309448a0}
}

@article{choi2022flow,
  title={Flow--structure interaction of a starting jet through a flexible circular nozzle},
  author={Choi, Daehyun and Park, Hyungmin},
  journal={J. Fluid Mech.},
  volume={949},
  pages={A39},
  year={2022},
  publisher={Cambridge University Press},
  doi={10.1017/jfm.2022.781}
}

@article{choi2024mechanism,
  title={Mechanism of enhanced impulse and entrainment of a pulsed jet through a flexible nozzle},
  author={Choi, Daehyun and Park, Hyungmin},
  journal={J. Fluid Mech.},
  volume={996},
  pages={A6},
  year={2024},
  publisher={Cambridge University Press},
  doi={10.1017/jfm.2024.720},
}

@article{morris2025optimal,
  title={Formation of multiple vortex rings from passively flexible nozzles},
  author={Mitchell, Brysen and Morris, Sarah},
  journal={J. Fluid Mech.},
  volume={1011},
  pages={A3},
  year={2025},
  publisher={Cambridge University Press},
  doi={10.1017/jfm.2025.378}
}

@article{haldane2016robotic,
  title={Robotic vertical jumping agility via series-elastic power modulation},
  author={Haldane, Duncan W and Plecnik, Mark M and Yim, Justin K and Fearing, Ronald S},
  journal={Sci.\ Robot.},
  volume={1},
  number={1},
  pages={eaag2048},
  year={2016},
  publisher={American Association for the Advancement of Science},
  doi={10.1126/scirobotics.aag2048}
}

@article{kroger2011cephalopod,
  title={Cephalopod origin and evolution: A congruent picture emerging from fossils, development and molecules},
  author={Kr{\"o}ger, Bj{\"o}rn and Vinther, Jakob and Fuchs, Dirk},
  journal={BioEssays},
  volume={33},
  number={8},
  pages={602--613},
  year={2011},
  publisher={Wiley Online Library},
  doi={10.1002/bies.201100001}
}

@book{staaf2017squid,
  title={Squid Empire: The Rise and Fall of the Cephalopods},
  author={Staaf, Danna},
  year={2017},
  publisher={ForeEdge},
  address={Lebanon, NH},
  isbn={978-1611689235}
}

@article{rosa2017biology,
  title={Biology and ecology of the world's largest invertebrate, the colossal squid (\textit{Mesonychoteuthis hamiltoni}): a short review},
  author={Rosa, Rui and Lopes, Vanessa M and Guerreiro, Miguel and Bolstad, Kathrin and Xavier, Jos{\'e} C},
  journal={Polar Biol.},
  volume={40},
  number={9},
  pages={1871--1883},
  year={2017},
  publisher={Springer},
  doi={10.1007/s00300-017-2104-5}
}

@article{odor1988forces,
  title={The forces acting on swimming squid},
  author={O'Dor, R. K.},
  journal={J. Exp.\ Biol.},
  volume={137},
  number={1},
  pages={421--442},
  year={1988},
  publisher={The Company of Biologists Ltd},
  doi={10.1242/jeb.137.1.421}
}

@article{bartol2001swimming,
  title={Swimming mechanics and behavior of the shallow-water brief squid \textit{Lolliguncula brevis}},
  author={Bartol, I. K. and Patterson, M. R. and Mann, R.},
  journal={J. Exp.\ Biol.},
  volume={204},
  number={21},
  pages={3655--3682},
  year={2001},
  publisher={The Company of Biologists Ltd},
  doi={10.1242/jeb.204.21.3655}
}

@article{york2020squids,
  title={Squids use multiple escape jet patterns throughout ontogeny},
  author={York, Carly A. and Bartol, Ian K. and Krueger, Paul S. and Thompson, Joseph T.},
  journal={Biol.\ Open},
  volume={9},
  number={11},
  pages={bio054585},
  year={2020},
  publisher={The Company of Biologists Ltd},
  doi={10.1242/bio.054585}
}

@article{weller1998tensorial,
  title={A tensorial approach to computational continuum mechanics using object-oriented techniques},
  author={Weller, H. G. and Tabor, G. and Jasak, H. and Fureby, C.},
  journal={Comput.\ Phys.},
  volume={12},
  number={6},
  pages={620--631},
  year={1998},
  publisher={American Institute of Physics},
  doi={10.1063/1.168744}
}

@manual{dhondt2023calculix,
  title        = {CalculiX: A Free Software Three-Dimensional Structural Finite Element Program},
  author       = {Dhondt, Guido and Wittig, Klaus},
  organization = {Open Source Project},
  year         = {2023},
}

@article{chourdakis2022preCICE,
  title={{preCICE} v2: A sustainable and user-friendly coupling library},
  author={Chourdakis, Gerasimos and Davis, Kyle and Rodenberg, Benjamin and Schulte, Miriam and Simonis, Frédéric and Uekermann, Benjamin and Abrams, Georg and Bungartz, Hans-Joachim and {Cheung Yau}, Lucia and Desai, Ishaan and Eder, Konrad and Hertrich, Richard and Lindner, Florian and Rusch, Alexander and Sashko, Dmytro and Schneider, David and Totounferoush, Amin and Volland, Dominik and Vollmer, Peter and Koseomur, Oguz Ziya},
  journal={Open Res.\ Eur.},
  volume={2},
  pages={51},
  year={2022},
  publisher={F1000 Research Limited},
  doi={10.12688/openreseurope.14445.2}
}

@article{celik2008procedure,
  title={Procedure for estimation and reporting of uncertainty due to discretization in {CFD} applications},
  author={Celik, Ismail B. and Ghia, U. and Roache, Patrick J. and Freitas, Christopher J.},
  journal={J. Fluids Eng.},
  volume={130},
  number={7},
  pages={078001},
  year={2008},
  publisher={American Society of Mechanical Engineers},
  doi={10.1115/1.2960953}
}

@article{pumphrey1938rates,
  title={The Rates of Conduction of Nerve Fibres of Various Diameters in Cephalopods},
  author={Pumphrey, R. J. and Young, J. Z.},
  journal={J. Exp.\ Biol.},
  volume={15},
  number={4},
  pages={453--466},
  year={1938},
  publisher={The Company of Biologists Ltd},
  doi={10.1242/jeb.15.4.453}
}

@article{bullock1948properties,
  title={Properties of a single synapse in the stellate ganglion of squid},
  author={Bullock, Theodore Holmes},
  journal={J. Neurophysiol.},
  volume={11},
  number={4},
  pages={343--363},
  year={1948},
  publisher={American Physiological Society},
  doi={10.1152/jn.1948.11.4.343},
  url={https://journals.physiology.org/doi/10.1152/jn.1948.11.4.343}
}

@book{bullock1965structure,
  title={Structure and Function in the Nervous Systems of Invertebrates},
  author={Bullock, Theodore Holmes and Horridge, G. Adrian},
  volume={2},
  year={1965},
  publisher={W.H. Freeman and Company},
  address={San Francisco}
}

@incollection{hoving2014deep,
  title={The study of deep-sea cephalopods},
  author={Hoving, Henk-Jan T. and P{\'e}rez, Jos{\'e} {\'A}ngel A. and Bolstad, Kathrin S.R. and Braid, Heather E. and Evans, Aaron B. and Fuchs, Dirk and Judkins, Heather L. and Kelly, Jesse T. and Marian, Jos{\'e} E.A.R. and Nakajima, Ryuta and Piatkowski, Uwe and Reid, Amanda and Vecchione, Michael and Xavier, Jos{\'e} C.C.},
  booktitle={Advances in Cephalopod Science: Biology, Ecology, Cultivation and Fisheries},
  editor={Vidal, Erica A.G.},
  series={Advances in Marine Biology},
  volume={67},
  pages={235--359},
  year={2014},
  publisher={Academic Press},
  doi={10.1016/B978-0-12-800287-2.00003-2},
  url={https://www.sciencedirect.com/science/chapter/bookseries/pii/B9780128002872000032}
}

@book{roper2010cephalopods,
  title={Cephalopods of the world. An annotated and illustrated catalogue of cephalopod species known to date. Volume 2. Myopsid and Oegopsid Squids},
  editor={Jereb, P. and Roper, C.F.E.},
  series={FAO Species Catalogue for Fishery Purposes No.~4},
  volume={2},
  pages={605},
  year={2010},
  publisher={FAO},
  address={Rome},
  isbn={978-92-5-106720-8},
  url={https://www.fao.org/4/i1920e/i1920e00.htm}
}

@article{martinez-calvo2020satellite,
  title        = {Natural break-up and satellite formation regimes of surfactant-laden liquid threads},
  author       = {Mart\'{\i}nez-Calvo, A. and Rivero-Rodr\'{\i}guez, J. and Scheid, B. and Sevilla, A.},
  journal      = {J. Fluid Mech.},
  volume       = {883},
  pages        = {A35},
  year         = {2020},
  publisher    = {Cambridge University Press},
  doi          = {10.1017/jfm.2019.874},
}

@article{driessen2014jet,
  title={Control of jet breakup by a superposition of two {Rayleigh--Plateau}-unstable modes},
  author={Driessen, Theo and Sleutel, Pascal and Dijksman, Frits and Jeurissen, Roger and Lohse, Detlef},
  journal={J. Fluid Mech.},
  volume={749},
  pages={275--296},
  year={2014},
  publisher={Cambridge University Press},
  doi={10.1017/jfm.2014.178}
}

@article{challita2023superpropulsion,
  title        = {Droplet superpropulsion in an energetically constrained insect},
  author       = {Challita, Elio J. and Sehgal, Prateek and Krugner, Rodrigo and Bhamla, M. Saad},
  journal      = {Nat.\ Commun.},
  volume       = {14},
  pages        = {860},
  year         = {2023},
  doi          = {10.1038/s41467-023-36376-5},
}

@article{prl119108001,
  title={Superpropulsion of Droplets and Soft Elastic Solids},
  author={Raufaste, Christophe and Ramos Chagas, Gabriela and Darmanin, Thierry and Claudet, Cyrille and Guittard, Fr{\'e}d{\'e}ric and Celestini, Franck},
  journal={Phys.\ Rev.\ Lett.},
  volume={119},
  number={10},
  pages={108001},
  year={2017},
  publisher={American Physical Society},
  doi={10.1103/PhysRevLett.119.108001}
}

@article{celestini2020contactlayer,
  title        = {Contact Layer as a Propelling Advantage in Throwing},
  author       = {Celestini, Franck and Mathiesen, Joachim and Argentina, M{\'e}d{\'e}ric and Raufaste, Christophe},
  journal      = {Phys.\ Rev.\ Appl.},
  volume       = {14},
  number       = {4},
  pages        = {044026},
  year         = {2020},
  month        = {October},
  doi          = {10.1103/PhysRevApplied.14.044026}
}

@article{giombini2022use,
  title        = {Use of compliant actuators for throwing rigid projectiles},
  author       = {Giombini, Guillaume and Mathiesen, Joachim and D'Angelo, Christophe and Argentina, M{\'e}d{\'e}ric and Raufaste, Christophe and Celestini, Franck},
  journal      = {Phys.\ Rev.\ E},
  volume       = {105},
  number       = {2},
  pages        = {025001},
  year         = {2022},
  doi          = {10.1103/PhysRevE.105.025001}
}

@article{reynolds2003bifurcating,
  title        = {Bifurcating and Blooming Jets},
  author       = {W. C. Reynolds and D. E. Parekh and P. J. D. Juvet and M. J. D. Lee},
  journal      = {Annu.\ Rev.\ Fluid Mech.},
  volume       = {35},
  pages        = {295--315},
  year         = {2003},
  doi          = {10.1146/annurev.fluid.35.101101.161128},
  publisher    = {Annual Reviews},
}

@article{gohil2015simulation,
  title        = {Simulation of the blooming phenomenon in forced circular jets},
  author       = {Gohil, Tushar B. and Saha, Arun K. and Muralidhar, K.},
  journal      = {J. Fluid Mech.},
  volume       = {783},
  pages        = {567--604},
  year         = {2015},
  publisher    = {Cambridge University Press},
  doi          = {10.1017/jfm.2015.571},
}

@article{jensen1989existence,
  title        = {The existence of steady flow in a collapsed tube},
  author       = {Jensen, O. E. and Pedley, T. J.},
  journal      = {J. Fluid Mech.},
  volume       = {206},
  pages        = {339--374},
  year         = {1989},
  publisher    = {Cambridge University Press},
  doi          = {10.1017/S0022112089002326},
  url          = {https://www.cambridge.org/core/journals/journal-of-fluid-mechanics/article/existence-of-steady-flow-in-a-collapsed-tube/65B2A122727B6147F5EAA2FC0377827A},
}

@article{jensen1990instabilities,
  title        = {Instabilities of flow in a collapsed tube},
  author       = {Jensen, O. E.},
  journal      = {J. Fluid Mech.},
  volume       = {220},
  pages        = {623--659},
  year         = {1990},
  publisher    = {Cambridge University Press},
  doi          = {10.1017/S0022112090003408},
  url          = {https://doi.org/10.1017/S0022112090003408},
}

@article{luo1996numerical,
  title={A numerical simulation of unsteady flow in a two-dimensional collapsible channel},
  author={Luo, X Y and Pedley, T J},
  journal={J. Fluid Mech.},
  volume={314},
  pages={191--225},
  year={1996},
  publisher={Cambridge University Press},
  doi={10.1017/S0022112096000286}
}

@article{grotberg2004biofluid,
  title={Biofluid mechanics in flexible tubes},
  author={Grotberg, James B and Jensen, Oliver E},
  journal={Annu.\ Rev.\ Fluid Mech.},
  volume={36},
  pages={121--147},
  year={2004},
  publisher={Annual Reviews},
  doi={10.1146/annurev.fluid.36.050802.121918}
}

@article{wang2021energetics,
  title={Energetics of collapsible channel flow with a nonlinear fluid-beam model},
  author={Wang, D Y and Luo, X Y and Stewart, P S},
  journal={J. Fluid Mech.},
  volume={926},
  pages={A2},
  year={2021},
  publisher={Cambridge University Press},
  doi={10.1017/jfm.2021.642}
}

@article{concha2015oscillation,
  title={Oscillation of the velvet worm slime jet by passive hydrodynamic instability},
  author={Concha, Andr{\'e}s and Mellado, Paula and Morera-Brenes, Bernal and Sampaio Costa, Cristiano and Mahadevan, L and Monge-N{\'a}jera, Juli{\'a}n},
  journal={Nat.\ Commun.},
  volume={6},
  pages={6292},
  year={2015},
  publisher={Nature Publishing Group},
  doi={10.1038/ncomms7292}
}

@article{bujard2021resonant,
  title={A resonant squid-inspired robot unlocks biological propulsive efficiency},
  author={Bujard, Thierry and Giorgio-Serchi, Francesco and Weymouth, Gabriel D},
  journal={Sci.\ Robot.},
  volume={6},
  number={50},
  pages={eabd2971},
  year={2021},
  publisher={AAAS},
  doi={10.1126/scirobotics.abd2971}
}

@article{forouhar2006suction,
  title={The embryonic vertebrate heart tube is a dynamic suction pump},
  author={Forouhar, Arian S. and Liebling, Michael and Hickerson, Anna and Nasiraei-Moghaddam, Abbas and Tsai, Huai-Jen and Hove, Jay R. and Fraser, Scott E. and Dickinson, Mary E. and Gharib, Morteza},
  journal={Science},
  volume={312},
  number={5774},
  pages={751--753},
  year={2006},
  publisher={American Association for the Advancement of Science},
  doi={10.1126/science.1123775},
  url={https://www.science.org/doi/10.1126/science.1123775},
}

@article{sutherland2010comparative,
  title={Comparative jet wake structure and swimming performance of salps},
  author={Sutherland, Kelly R and Madin, Laurence P},
  journal={J. Exp.\ Biol.},
  volume={213},
  number={17},
  pages={2967--2975},
  year={2010},
  publisher={Company of Biologists},
  doi={10.1242/jeb.041962}
}

@article{staaf2014aperture,
  title={Aperture effects in squid jet propulsion},
  author={Staaf, Danna J and Gilly, William F and Denny, Mark W},
  journal={J. Exp.\ Biol.},
  volume={217},
  number={9},
  pages={1588--1600},
  year={2014},
  publisher={The Company of Biologists Ltd},
  doi={10.1242/jeb.082271}
}

@article{bertoldi2017flexible,
  title={Flexible mechanical metamaterials},
  author={Bertoldi, Katia and Vitelli, Vincenzo and Christensen, Johan and Van Hecke, Martin},
  journal={Nat.\ Rev.\ Mater.},
  volume={2},
  number={11},
  pages={17066},
  year={2017},
  publisher={Nature Publishing Group},
  doi={10.1038/natrevmats.2017.66}
}

@article{roach2013elastic,
  title={Elastic energy storage in the shoulder and the evolution of high-speed throwing in {Homo}},
  author={Roach, Neil T. and Venkadesan, Madhusudhan and Rainbow, Michael J. and Lieberman, Daniel E.},
  journal={Nature},
  volume={498},
  number={7455},
  pages={483--486},
  year={2013},
  doi={10.1038/nature12267}
}

@inproceedings{wallace2025multi,
  title={Multi-fidelity approach for thrust enhancement in bio-inspired nozzles for underwater vehicles using LES and experiments},
  author={Wallace, Halley J and Choi, Daehyun and Singh, Paras and Samal, Gourav and Bose, Chandan and Bhamla, Saad},
  booktitle={Division of Fluid Dynamics Annual Meeting 2025},
  year={2025},
  organization={APS}
}

@misc{choi2026computational,
  title={Computational and reduced-order modelling of elastic wave-driven impulse enhancement in pulsed jets through passively flexible nozzles},
  author={Choi, Daehyun and Singh, Paras and Bhamla, Saad and Bose, Chandan},
  year={2026},
  eprint={2605.17319},
  archivePrefix={arXiv},
  primaryClass={physics.flu-dyn}
}

@misc{singh2026multi,
  title={Multi-objective {B}ayesian optimization of rigid and flexible nozzles for energy-efficient pulsed jet propulsion},
  author={Singh, Paras and Karki, Yukesh and Hernandez, Victor and Choi, Daehyun and Bhamla, Saad and Bose, Chandan},
  year={2026},
  eprint={2606.19252},
  archivePrefix={arXiv},
  primaryClass={physics.flu-dyn}
}

@article{kim2022magnetic,
  title={Magnetic soft materials and robots},
  author={Kim, Yoonho and Zhao, Xuanhe},
  journal={Chem.\ Rev.},
  volume={122},
  number={5},
  pages={5317--5364},
  year={2022},
  publisher={ACS Publications},
  doi={10.1021/acs.chemrev.1c00481}
}

@book{li2004dynamics,
  title        = {Dynamics of the Vascular System},
  author       = {Li, John K-J},
  publisher    = {World Scientific},
  year         = {2004},
  address      = {Singapore},
  isbn         = {978-981-238-815-5},
  doi          = {10.1142/4923},
}

@article{vandevosse2011pulse,
  title={Pulse Wave Propagation in the Arterial Tree},
  author={van de Vosse, Frans N. and Stergiopulos, Nikos},
  journal={Annu.\ Rev.\ Fluid Mech.},
  year={2011},
  volume={43},
  pages={467--499},
  doi={10.1146/annurev-fluid-122109-160730},
}

@article{hughes1973onedimensional,
  title={On the one-dimensional theory of blood flow in the larger vessels},
  author={Hughes, Thomas J. R. and Lubliner, Jacob},
  journal={Math.\ Biosci.},
  volume={18},
  number={1--2},
  pages={161--170},
  year={1973},
  publisher={Elsevier},
  doi={10.1016/0025-5564(73)90027-8},
  url={https://www.sciencedirect.com/science/article/pii/0025556473900278},
}

@article{Naito2021,
  author  = {Naito, Kozo},
  title   = {Time-varying motor control strategy for proximal-to-distal sequential energy distribution: insights from baseball pitching},
  journal = {J. Exp.\ Biol.},
  year    = {2021},
  volume  = {224},
  number  = {20},
  pages   = {jeb227207},
  doi     = {10.1242/jeb.227207},
}

@article{Ward1972,
  author  = {Ward, D. V. and Wainwright, S. A.},
  title   = {Locomotory aspects of squid mantle structure},
  journal = {J. Zool.},
  volume  = {167},
  number  = {4},
  pages   = {437--449},
  year    = {1972},
  doi     = {10.1111/j.1469-7998.1972.tb01735.x}
}

@article{Kier1985,
  author  = {Kier, William M. and Smith, Kathleen K.},
  title   = {Tongues, tentacles and trunks: the biomechanics of movement in muscular-hydrostats},
  journal = {Zool.\ J. Linn.\ Soc.},
  volume  = {83},
  number  = {4},
  pages   = {307--324},
  year    = {1985},
  doi     = {10.1111/j.1096-3642.1985.tb01178.x}
}

@incollection{Kier1992,
  author    = {Kier, William M.},
  title     = {Hydrostatic skeletons and muscular hydrostats},
  booktitle = {Biomechanics (Structures and Systems): A Practical Approach},
  editor    = {Biewener, Andrew A.},
  publisher = {IRL Press at Oxford University Press},
  address   = {New York},
  year      = {1992},
  pages     = {205--231},
  doi={10.1093/oso/9780199632688.003.0009}
}

@article{demont1988resonance,
  author  = {DeMont, M. Edwin and Gosline, John M.},
  title   = {Mechanics of jet propulsion in the hydromedusan jellyfish, \textit{{P}olyorchis penicillatus}: {III}. {A} natural resonating bell; the presence and importance of a resonant phenomenon in the locomotor structure},
  journal = {J. Exp.\ Biol.},
  volume  = {134},
  number  = {1},
  pages   = {347--361},
  year    = {1988},
  doi     = {10.1242/jeb.134.1.347}
}

@article{hoover2015jellyfish,
  author  = {Hoover, Alexander and Miller, Laura},
  title   = {A numerical study of the benefits of driving jellyfish bells at their natural frequency},
  journal = {J. Theor.\ Biol.},
  volume  = {374},
  pages   = {13--25},
  year    = {2015},
  doi     = {10.1016/j.jtbi.2015.03.016}
}

@article{tytell2014role,
  author  = {Tytell, Eric D. and Hsu, Chia-Yu and Fauci, Lisa J.},
  title   = {The role of mechanical resonance in the neural control of swimming in fishes},
  journal = {Zoology},
  volume  = {117},
  number  = {1},
  pages   = {48--56},
  year    = {2014},
  doi     = {10.1016/j.zool.2013.10.011}
}

@article{zhong2021tunable,
  author  = {Zhong, Qiang and Zhu, Jizhong and Fish, Frank E. and Kerr, Steven J. and Downs, Ashlee M. and Bart-Smith, Hilary and Quinn, Daniel B.},
  title   = {Tunable stiffness enables fast and efficient swimming in fish-like robots},
  journal = {Sci.\ Robot.},
  volume  = {6},
  number  = {57},
  pages   = {eabe4088},
  year    = {2021},
  doi     = {10.1126/scirobotics.abe4088}
}

@article{chourdakis2023openfoam,
  title={Open{FOAM}-pre{CICE}: Coupling Open{FOAM} with external solvers for multi-physics simulations},
  author={Chourdakis, Gerasimos and Schneider, David and Uekermann, Benjamin},
  journal={OpenFOAM J.},
  volume={3},
  pages={1--25},
  year={2023},
  doi={10.51560/ofj.v3.88}
}

@article{marshall2012squid,
  title={Squid can fly to save energy},
  author={Marshall, Jessica},
  journal={Nature},
  year={2012},
  note={News feature, 20 February 2012},
  doi={10.1038/nature.2012.10060}
}

@article{blaber2015ncorr,
  title={Ncorr: open-source {2D} digital image correlation {Matlab} software},
  author={Blaber, Justin and Adair, Benjamin and Antoniou, Antonia},
  journal={Exp.\ Mech.},
  volume={55},
  number={6},
  pages={1105--1122},
  year={2015},
  doi={10.1007/s11340-015-0009-1}
}

@article{solav2018multidic,
  title={{MultiDIC}: an open-source toolbox for multi-view {3D} digital image correlation},
  author={Solav, Dana and Moerman, Kevin M and Jaeger, Aaron M and Genovese, Katia and Herr, Hugh M},
  journal={IEEE Access},
  volume={6},
  pages={30520--30535},
  year={2018},
  doi={10.1109/ACCESS.2018.2843725}
}

@article{messenger2001cephalopod,
  title={Cephalopod chromatophores: neurobiology and natural history},
  author={Messenger, John B},
  journal={Biol.\ Rev.},
  volume={76},
  number={4},
  pages={473--528},
  year={2001},
  doi={10.1017/S1464793101005772}
}

@article{hadjisolomou2021quantifying,
  title={Quantifying the speed of chromatophore activity at the single-organ level in response to a visual startle stimulus in living, intact squid},
  author={Hadjisolomou, Stavros P and El-Haddad, Rita W and Kloskowski, Kamil and Chavarga, Alla and Abramov, Israel},
  journal={Front.\ Physiol.},
  volume={12},
  pages={675252},
  year={2021},
  doi={10.3389/fphys.2021.675252}
}

@article{thompson2001ontogenetic,
  title={Ontogenetic changes in mantle kinematics during escape-jet locomotion in the oval squid, \textit{Sepioteuthis lessoniana} Lesson, 1830},
  author={Thompson, Joseph T and Kier, William M},
  journal={Biol.\ Bull.},
  volume={201},
  number={2},
  pages={154--166},
  year={2001},
  doi={10.2307/1543330}
}

@article{thompson2002ontogeny,
  title={Ontogeny of squid mantle function: changes in the mechanics of escape-jet locomotion in the oval squid, \textit{Sepioteuthis lessoniana} Lesson, 1830},
  author={Thompson, Joseph T and Kier, William M},
  journal={Biol.\ Bull.},
  volume={203},
  number={1},
  pages={14--26},
  year={2002},
  doi={10.2307/1543454}
}

@article{westerhof2009arterial,
  title={The arterial {Windkessel}},
  author={Westerhof, Nico and Lankhaar, Jan-Willem and Westerhof, Berend E},
  journal={Med.\ Biol.\ Eng.\ Comput.},
  volume={47},
  number={2},
  pages={131--141},
  year={2009},
  doi={10.1007/s11517-008-0359-2}
}

\subsection*{\textbf{Acknowledgements}}
We thank Stephen~L.~Edwards, Christopher Warren, Kevin Sloan, Dean~R.~Culver, and Susan Swithenbank for their advice and Alexander Norton for squid collection.

\subsection*{\revtc{\textbf{Funding}}}
This work was supported by the DARPA Young Faculty Award (DARPA-RA-24-01-18-YFA18-FP-004 \revtc{to S.B.}), the National Research Foundation of Korea (RS-2022-NR070924 \revtc{and} RS-2023-00248034 \revtc{to D.C., and} RS-2024-00343259 \revtc{to H.P.}), and the Technology Innovation Program (20026416 \revtc{to H.P.}) funded by MOTIE of \revred{the} Korean government.
The content of the information does not necessarily reflect the position or the policy of the Government, and no official endorsement should be inferred. Approved for public release; distribution is unlimited.
The present FSI simulations \revred{were} carried out using the computational resources provided by the UK national supercomputing facility ARCHER2 through the EPSRC Access To HPC Pioneer Grant \revtc{(to C.B.)}.

\subsection*{\revtc{\textbf{Data, Code, and Materials Availability}}}
\revtc{All data and code needed to evaluate and reproduce the results in the paper are present in the paper and/or the Supplementary Materials. The plotted data of the figures and the custom analysis scripts are archived at Zenodo (\href{https://doi.org/10.5281/zenodo.23677948}{\nolinkurl{https://doi.org/10.5281/zenodo.23677948}}) and are also available on GitHub (\url{https://github.com/bhamla-lab/squid-superpropulsion-2026}). The fluid--structure interaction simulations used the open-source codes OpenFOAM (finite volume flow solver), preCICE (coupling library) and CalculiX (finite element structural solver), set up as described in section~\ref{si:fsi} of the Supplementary Materials. This study did not generate new materials. The compliant nozzles were cast from commercially available silicone elastomers (Smooth-On) as described in section~\ref{si:fabrication}, and the squid were collected from the wild as described in section~\ref{si:squid}.}

\subsection*{\textbf{Author contributions}}
D.C. and S.B. conceived the study.
D.C. designed and performed the biological and engineering experiments, developed the theoretical framework, \revred{analyzed} the data, and wrote the manuscript.
P.S. developed and performed the fluid--structure interaction simulations, \revred{analyzed} the simulation data, and co-wrote the manuscript.
I.B. designed and performed the boat and mixing experiments, \revred{analyzed} the data, and co-wrote the manuscript.
M.K., J.P., H.J.W., and K.Z. assisted with engineering experiments.
S.Y.H., A.T.A., and T.A.U. performed histological preparation and imaging, and contributed to  writing and editing.
W.F.G. supervised and coordinated the live-squid fieldwork, contributed to biological experiments and reviewed the manuscript.
H.P. contributed to engineering experiments, data analysis, theoretical basis, and reviewed the paper.
D.K. contributed to engineering experiments, supervised portions of the work, and reviewed the paper.
C.B. developed the fluid-structure interaction simulation framework, supervised the simulations and analysis of the computational results, and reviewed the manuscript.
S.B. supervised the project, contributed to biological and engineering experiments, data analysis, and theoretical development, and wrote and reviewed the manuscript.

\subsection*{\textbf{Corresponding authors}}
Correspondence to Saad Bhamla.

\subsection*{\textbf{Competing interests}}
D.C. and S.B. are co-inventors on a provisional patent application (US 63/772,372) submitted by Georgia Tech Research Corporation that covers flexible nozzle designs for jet propulsion. The remaining authors declare no competing interests.

\subsection*{\textbf{Additional Information}}
Supplementary Movie is available for this paper. Correspondence and requests for materials should be addressed to Saad Bhamla.

\clearpage
\onecolumngrid

\setcounter{section}{0}
\setcounter{figure}{0}
\setcounter{table}{0}
\setcounter{equation}{0}
\setcounter{secnumdepth}{2}
\setcounter{tocdepth}{2}
\renewcommand{\thesection}{S\arabic{section}}
\renewcommand{\thesubsection}{S\arabic{section}.\arabic{subsection}}
\renewcommand{\thesubsubsection}{S\arabic{section}.\arabic{subsection}.\arabic{subsubsection}}
\renewcommand{\thetable}{S\arabic{table}}
\renewcommand{\thefigure}{S\arabic{figure}}
\renewcommand{\theequation}{S\arabic{equation}}
\renewcommand{\theHfigure}{S\arabic{figure}}
\renewcommand{\theHtable}{S\arabic{table}}
\renewcommand{\theHequation}{S\arabic{equation}}
\renewcommand{\theHsection}{S\arabic{section}}
\renewcommand{\theHsubsection}{S\arabic{section}.\arabic{subsection}}
\makeatletter
\let\addcontentsline\oldaddcontentsline
\makeatother

{\centering\Large\bfseries Supplementary Materials:\\[0.6em]
\revred{Squid funnels time their elastic recoil to amplify pulsed jets}\par}

\vspace{1.2em}

{\centering\normalsize
Daehyun Choi, Paras Singh, Ian Bergerson, Minho Kim, Jieun Park,
Halley J.\ Wallace, Kenny Zhang, Sandy Y.\ Hsieh, Aqua T.\ Asberry,
Theodore A.\ Uyeno, William F.\ Gilly, Hyungmin Park, Daeshik Kang,
Chandan Bose, Saad Bhamla\textsuperscript{*}\\[0.3em]
{\footnotesize \textsuperscript{*}Correspondence: saad.bhamla@colorado.edu}\par}

\vspace{1.5em}

\tableofcontents

\newpage
\FloatBarrier
\section{Squid acquisition}\label{si:squid}

\begin{figure*}[!htbp]
	\centering
	\includegraphics[width=\textwidth]{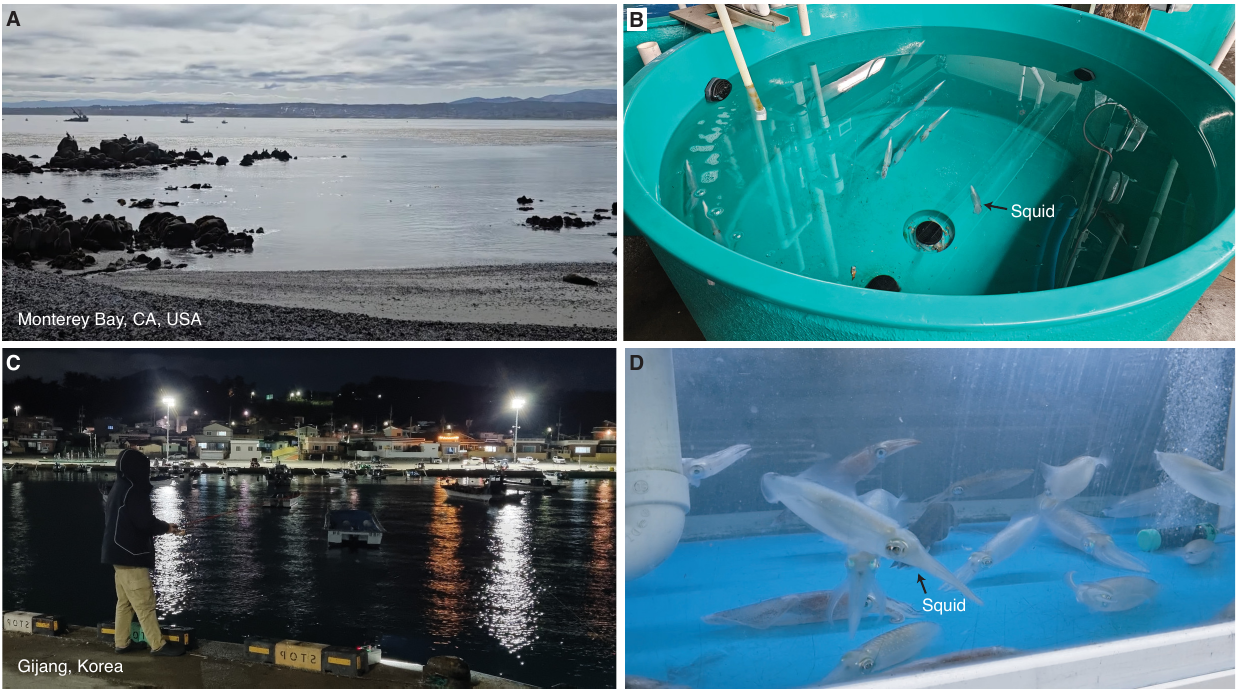}
	\caption{\textbf{Squid acquisition}.
    (\textbf{A})~Monterey Bay, California, USA, site of \textit{Doryteuthis opalescens} collection (coordinates: 36$^\circ$37$'$15$''$N, 121$^\circ$54$'$1$''$W).
    (\textbf{B})~Circulating seawater tank at Hopkins Marine Station, Pacific Grove, California, USA\rev{, holding \textit{Doryteuthis opalescens}}.
    (\textbf{C})~Gijang, Busan, Korea, site of \textit{Sepioteuthis lessoniana} collection (coordinates: 35$^\circ$16$'$15$''$N, 129$^\circ$14$'$47$''$E).
    (\textbf{D})~Circulating seawater tank at Gijang Bay, Busan, Korea\rev{, holding \textit{Sepioteuthis lessoniana}}.
    }
    \label{sif_squid}
\end{figure*}

California market squid (\textit{Doryteuthis opalescens}) and bigfin reef squid (\textit{Sepioteuthis lessoniana}) were collected at two different locations to investigate funnel and mantle deformation during escape behavior. \rev{Both species were chosen because they are readily available, \textit{D. opalescens} in Monterey Bay, California, and \textit{S. lessoniana} in Gijang, Busan, and because both produce rapid escape jets~\cite{otis1990concerted, neumeister2000temperature, li2019hypoxia, thompson2001ontogenetic, thompson2002ontogeny}. \textit{S. lessoniana} also differs in funnel geometry and could be recorded while swimming freely, which tests whether the phase lag is species-specific. Both species show the phase-lagged expansion and recoil of the funnel. Funnel shape and the ability to expel water from the mantle differ between squid species, and the present measurements cover only these two, so the dependence of the recoil on funnel shape is a limitation of this study and requires follow-up work on more species.} 
Forty California market squid were freshly collected from Monterey Bay, California, USA, in May 2025 (Fig.~\ref{sif_squid}, A), with an average mantle length of $141.3 \pm 1.48$~mm ($n = 6$ for the length measurement), and were housed at Hopkins Marine Station (Pacific Grove, CA) in a circulating seawater tank maintained at 13$^\circ$C and 94.2\% dissolved oxygen (Fig.~\ref{sif_squid}, B). All experimental procedures were completed within 72 hours of capture. Quantification of funnel and mantle deformation was performed for both intact and \rev{denervated} squid under restrained conditions in the water tank. Additionally, measurements of funnel and mantle deformation in freely swimming squid were conducted in Gijang, Korea (Fig.~\ref{sif_squid}, C), using bigfin reef squid (\textit{Sepioteuthis lessoniana}). Over 20 individuals were kept in a circulating seawater tank (Fig.~\ref{sif_squid}, D). Of these, five squid (average mantle length $112.3 \pm 1.4$~mm, $n = 3$ for the length measurement) were selected for deformation measurements within 72 hours of capture, conducted in a 270~mm $\times$ 210~mm $\times$ 170~mm seawater tank. Escape responses were triggered by gently touching an arm with a thin metal rod (diameter approximately 3~mm). To promote linear swimming along the focal plane while minimizing body contact, a transparent slit channel was placed inside the tank. 

\FloatBarrier
\section{Histological analysis}\label{si:histology}

\begin{figure*}[!htbp]
	\centering
	\includegraphics[width=\textwidth]{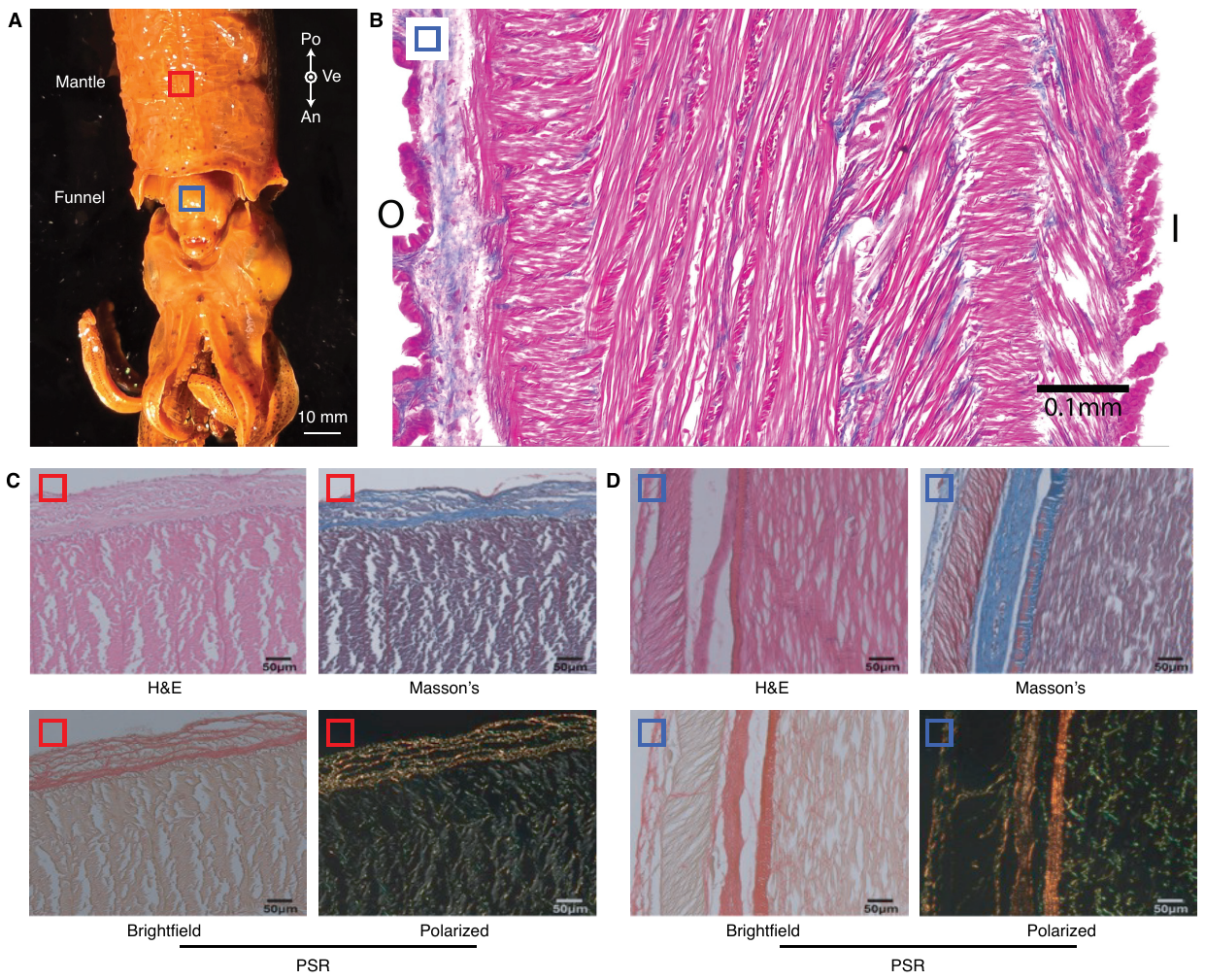}
	\caption{\textbf{Histological structure and collagen staining of squid tissues}.
    (\textbf{A})~Squid (\textit{Doryteuthis opalescens}) fixed in 4\% formaldehyde, showing the intact mantle, funnel, and arms. \rev{Axis labels follow Fig.~\ref{f_squid}, with An anterior, Po posterior and Ve ventral pointing out of the image plane.} 
     (\textbf{B})~Cross section of the funnel wall in \rev{\textit{Doryteuthis opalescens}}. Section stained using Milligan’s Trichrome \cite{Kier1992}, blue = collagen fibers, magenta = muscle fibers. (O = Outside or external surface, I = Internal or luminal surface of the funnel). Note that collagen appears as a thick layer near the external surface and as a network within the muscular layer. The muscular layer is likely arranged as a muscular hydrostat as it is composed of different orientations of muscle fibers organized in both discrete layers and interdigitating transverse, longitudinal, oblique and circular fibers.  
    (\textbf{C--D})~Representative sections of \rev{\textit{Doryteuthis opalescens}}, (\textbf{C}) \revred{mantle} and (\textbf{D}) \revred{funnel}, stained with Hematoxylin \& Eosin (H\&E), Masson's Trichrome, and Picrosirius Red (PSR) to examine tissue morphology and collagen distribution. For each tissue: H\&E (top left), Masson's Trichrome (top right), PSR under brightfield (bottom left), and PSR under crossed polarized illumination (bottom right). In Masson's trichrome, collagen-rich regions stain blue/green. In PSR, collagen appears red/orange in brightfield and is birefringent under polarized illumination; thick collagen fibers typically appear yellow/orange/red, whereas thinner fibers often appear green. Scale bars: 10~mm in (\textbf{A}); 100~$\mu$m in (\textbf{B}); and 50~$\mu$m in (\textbf{C--D}).
    }
	\label{sif_histology}
\end{figure*}

To characterize the tissue architecture and collagen distribution in the funnel and mantle of squid (\emph{Doryteuthis opalescens}), sections from the funnel and mantle were stained with Hematoxylin \& Eosin (H\&E), Masson's Trichrome (Masson's), and Picrosirius Red (PSR). Slides were imaged on a Zeiss Axio Observer 7 equipped with an AxioCam 305 color camera under either brightfield or polarized illumination. For brightfield imaging, proper Köhler illumination was performed to ensure smooth and even illumination across the field of view. For polarized imaging, two polarizers (one above the condenser and one under the nosepiece) were aligned in a crossed configuration to produce a dark background. The exposure time was determined by acquiring a series of test images over a range of exposures  under constant illumination, and selecting \revred{the} longest exposure that avoided pixel saturation. Exposure  was readjusted whenever the objective lens or slide was changed. H\&E staining was used to evaluate general tissue morphology and organization. Masson's trichrome and PSR staining were used to visualize collagen, which was evident in both  funnel and mantle tissues.

In Masson's trichrome, the outer sheath surrounding the musculature stained blue/green, consistent with a collagen-rich layer. In PSR, this sheath appeared red/orange in brightfield and exhibited strong birefringence under polarized illumination, with predominantly yellow/orange/red signal and occasional green regions, consistent with a mixture of thick, densely packed and thinner collagen fibers.

The collagen-rich sheath thickness was 24.3~$\mu$m $\pm$ 5.6~$\mu$m in the mantle and 72.1~$\mu$m $\pm$ 12.4~$\mu$m in the funnel (mean $\pm$ s.d.),  suggesting greater  structural reinforcement  in the funnel to withstand elevated internal pressures during jetting.

\FloatBarrier
\section{Chromatophore-tracking algorithm}\label{si:chromatophore}
\begin{figure*}[!htbp]
	\centering
	\includegraphics[width=\textwidth]{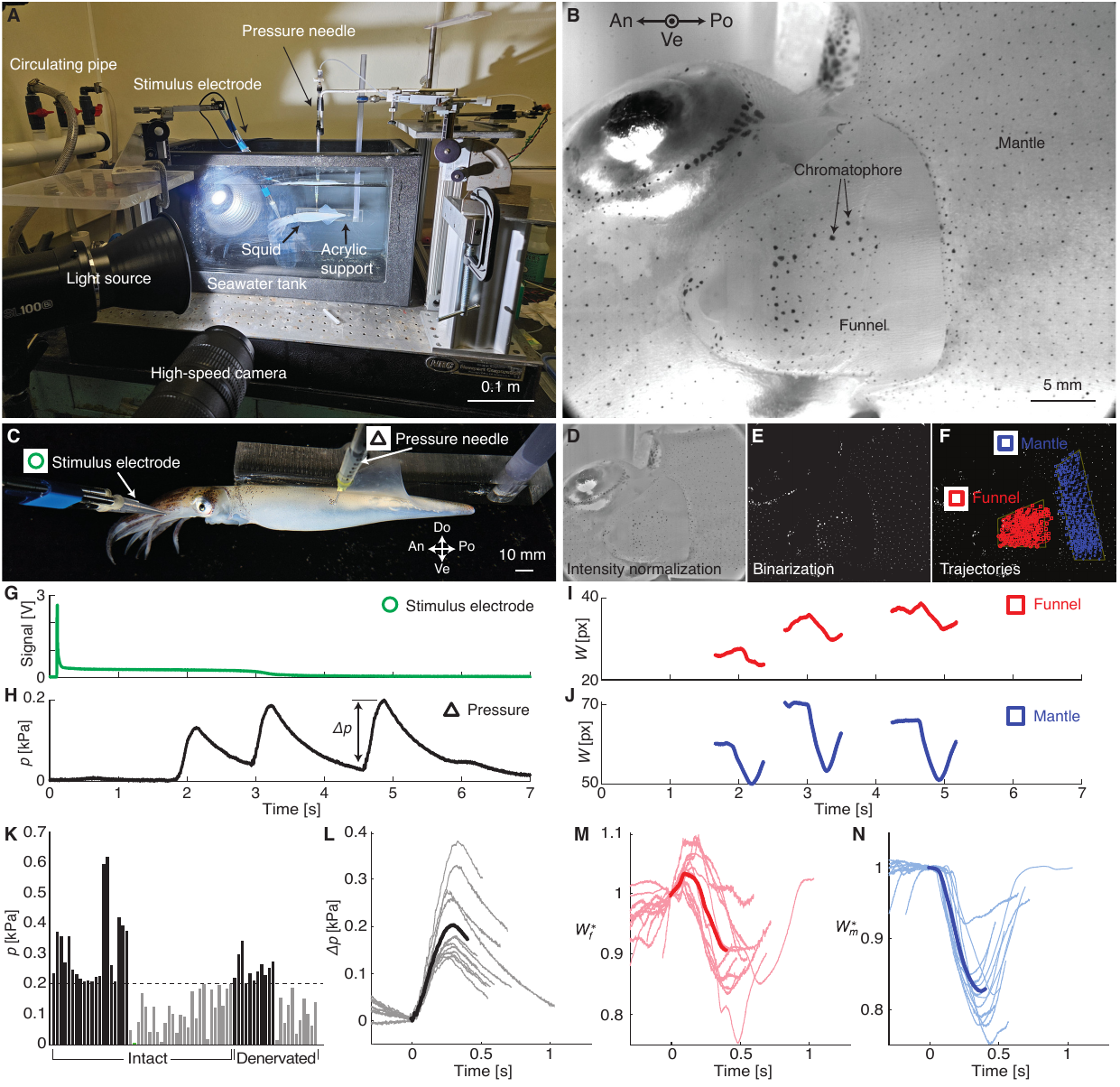}
	\caption{\textbf{Chromatophore-tracking algorithm for quantifying squid funnel and mantle deformation}.
    \rev{All panels show \textit{Doryteuthis opalescens}.}
    (\textbf{A})~High-speed imaging setup for squid measurements. The squid was fixed to an acrylic rod with adhesive inside a circulating seawater tank. Electrical stimulation was delivered via a custom electrode, while internal mantle pressure was monitored with a pressure sensor. Squid deformation was captured using a high-speed camera and a light source.
    (\textbf{B})~Raw high-speed image showing the distribution of chromatophores across the funnel and mantle surfaces. \rev{Axis labels give the orientation, where An is anterior, Po is posterior, Do is dorsal and Ve is ventral. Ve points out of the image plane in (\textbf{B}), whereas both axes lie in the plane in (\textbf{C}).}
    (\textbf{C})~Close-up view showing the position of the triggering electrode (located at the arm) and the pressure sensor needle penetrating the mantle wall.
    (\textbf{D--F})~Steps of the chromatophore-tracking image processing pipeline: (\textbf{D}) Intensity normalization; (\textbf{E}) Binarization to detect chromatophores; (\textbf{F}) Particle tracking to quantify width variation, with region-of-interest polygons (yellow) designating the central mantle (blue) and funnel (red) areas to minimize three-dimensional motion artifacts.
    (\textbf{G--J})~Time series data: (\textbf{G})~Electrical stimulus timing from the electrode; (\textbf{H})~Synchronized pressure sensor output, highlighting three jetting events; (\textbf{I})~Funnel width variation; (\textbf{J})~Mantle width variation.
    (\textbf{K})~Selected strong jets (N=30, n=8; N = number of jets, n = number of individuals) based on a maximum pressure criterion $>0.2$~kPa. In total, 30 out of 70 jets were selected for statistical analysis. Here two conditions were considered: intact and \rev{denervated} squid.
    (\textbf{L--N})~Averaged profiles of pressure difference, $\Delta p = p_{max} - p_{min}$, and normalized funnel and mantle deformations $w_f/w_{f,0}$ and $w_m/w_{m,0}$, where $w_{f,0}$ and $w_{m,0}$ are initial widths measured at pressure minima.
	}
	\label{sif_chromatophore}
\end{figure*}

For measurement of squid deformation, each squid was immobilized underwater on an acrylic rod using adhesive (Fig.~\ref{sif_chromatophore}, C) in a circulating seawater tank (500~mm $\times$ 300~mm $\times$ 300~mm). Escape behavior was triggered by applying an electrical stimulus of approximately 3~V to electrodes located at the arm (Fig.~\ref{sif_chromatophore}, C). To measure the internal pressure of the mantle, a needle was inserted through the mantle wall into the mantle cavity (Fig.~\ref{sif_chromatophore}, C). An oil-filled syringe was connected to a pressure sensor (40PC006G, Honeywell), whose signal was transmitted to a microprocessor and recorded at 2~kHz. 
The pressure sensor was calibrated by measuring output voltage against known hydrostatic pressures generated by a water column in a vertical glass tube. Care was taken to exclude air bubbles from the pressure measurement system, as these could introduce time delays and result in pressure underestimation. 
A high-speed camera (Photron MiniUX100, 1000~fps, 1024$\times$1024~pixels) recorded deformation of the entire funnel (viewed from the ventral side) and anterior mantle (Fig.~\ref{sif_chromatophore}, B), with illumination provided by 100~W LED lights (SL100, Godox). TTL triggering for squid stimulation, pressure sensor readout, and camera operation were synchronized by signal generators (SD9, Grass Instruments) with a timing error of less than 0.1~ms (see Fig.~\ref{sif_chromatophore}, G--J).
The measured relative pressure ($\Delta P \sim 0.2$--$1$~kPa) is consistent with mantle pressures reported for giant axon-mediated escape jets ($\sim$0.5--1~kPa)~\cite{otis1990concerted,li2023hydrodynamic}. Higher pressures ($\sim$5--50~kPa) have been reported in studies conducted at low temperatures or under hypoxic conditions, where enhanced escape performance elevates peak mantle pressures~\cite{neumeister2000temperature,li2019hypoxia}. 
To quantify funnel deformation during the jetting phase, an in-house chromatophore-tracking algorithm was developed, which uses the natural pigment cells (chromatophores) distributed across the funnel and mantle surfaces (Fig.~\ref{sif_chromatophore}, B). From the raw high-speed images, light intensity was normalized by subtracting a Gaussian-blurred background ($\sigma = 30$ pixels, Fig.~\ref{sif_chromatophore}, D). Additionally, stationary noise was removed by identifying intensity features present in more than 80\% of frames. Next, intensity was binarized using the 2nd percentile threshold to isolate chromatophores. Large dots (area $>$100 pixels), corresponding to background or squid eye, were filtered out. 
Particle tracking was performed via nearest-neighbor matching with a maximum displacement threshold of 5 pixels per frame; region-of-interest selection was focused on central areas of the mantle and funnel to minimize three-dimensional motion artifacts (Fig.~\ref{sif_chromatophore}, F). 
Tracks were subsequently filtered to retain only spatially continuous trajectories (maximum displacement $<$5 pixels/frame, minimum length 300 frames, minimum total displacement $>$5 pixels). To quantify the width of the mantle and funnel, principal component analysis (PCA) was performed on chromatophore positions to extract the orientation and deformation of the funnel and mantle. Using the major axis (PC1), the laboratory coordinate frame was aligned with the camera axes at frame 1 and rotated by the cumulative angular change of the tissue's PC1 axis, thereby tracking tissue rotation while maintaining a fixed relationship to the initial tissue orientation. Widths of the funnel and mantle were quantified as changes in the averaged pairwise distances between tracked chromatophores projected on the vertical axes of the rotated laboratory frame.

After the squid was stimulated by the electrode (at $t \simeq 0$~s in Fig.~\ref{sif_chromatophore}, G), the internal pressure peaked (Fig.~\ref{sif_chromatophore}, H), and deformation of the funnel and mantle began (Fig.~\ref{sif_chromatophore}, I and J, respectively). The initiation of deformation was detected by local minima (sharp increase) in the pressure sensor voltage, for example, at $t = 1.8$, 2.9, and 4.6~s. Only strong jetting instances were selected using a criterion of maximum pressure, $p_{max} > 0.2$~kPa, yielding a total of 19 and 11 jets for intact and \rev{denervated} squid, collected from 5 and 3 individuals, respectively (Fig.~\ref{sif_chromatophore}, K). The initial widths of the funnel and mantle at pressure minima were used for normalization (Fig.~\ref{sif_chromatophore}, M and N). It should also be noted that deformation measurements of freely \revred{swimming} squid were conducted on 3 individuals and 4 jets during which pressure measurement was omitted in these cases for unconstrained swimming.

\begin{figure*}[!htbp]
    \centering
    \includegraphics[width=\textwidth]{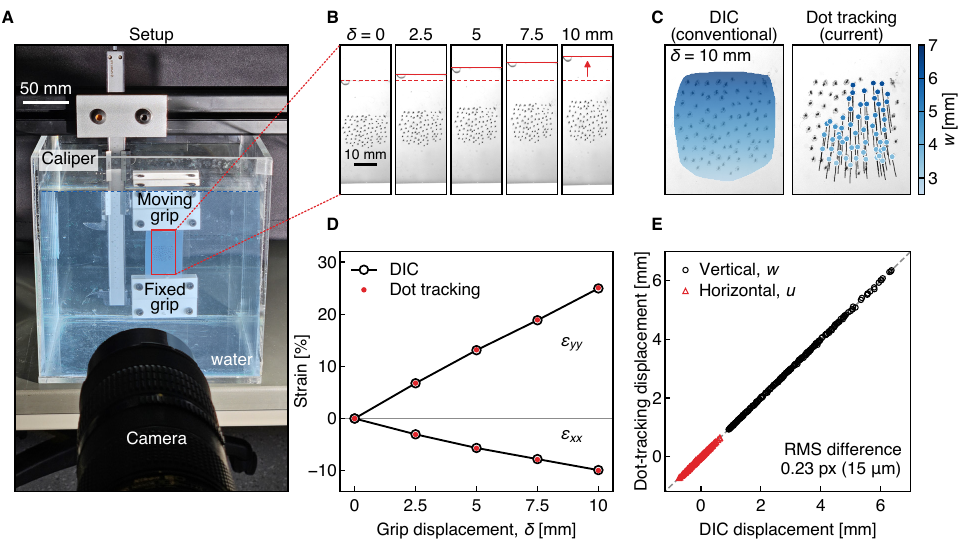}
    \caption{\rev{\textbf{Validation of dot tracking against two-dimensional DIC in a tensile test of an elastic membrane.}}
    \rev{(\textbf{A})~The setup, with the camera in the foreground and the water tinted blue, in which the upper grip rides on the slider of a vernier caliper that sets the grip displacement $\delta$ while the lower grip is fixed. The scale bar applies at the distance of the black aluminum frame behind the caliper, whose front face is 40~mm high. The red box marks the membrane between the grips that the frames of (\textbf{B}) show, joined to the first frame by dotted lines.}
    \rev{(\textbf{B})~Frames at $\delta = 0$, 2.5, 5, 7.5 and 10~mm. The red lines mark the edge of the moving grip at $\delta = 0$ (dashed) and \revred{in each later frame} (solid), 10.4~mm apart in the image \revred{at $\delta = 10$~mm}.}
    \rev{(\textbf{C})~The two measurements on the $\delta = 10$~mm frame. Left, the upward displacement $w$ from DIC (Ncorr, subset radius 30~pixels), drawn in color at the displaced positions. Right, the 58 tracked dots, each drawn at its $\delta = 10$~mm position in the same color scale, with its path through the \revred{four} steps as a black line.}
    \rev{(\textbf{D})~Vertical and horizontal strain from DIC and from dot tracking, each obtained from an affine fit over the same 58 dots.}
    \rev{(\textbf{E})~Displacement of the 58 dots at every step, dot tracking against DIC.}
    }
    \label{sif_membrane_dic}
\end{figure*}

\rev{The tracking was validated against a known deformation in a separate experiment (Fig.~\ref{sif_membrane_dic}). An elastic membrane was held between two grips on a vernier caliper, which stretches it along exactly one direction (Fig.~\ref{sif_membrane_dic}, A). The membrane was extended in four steps to a grip displacement of 10~mm (Fig.~\ref{sif_membrane_dic}, B), and it carries 85 dots marked manually with an oil-based pen. Each dot lies a median of 22~pixels from its nearest neighbor, which is comparable to the 14~pixels (funnel) and 19~pixels (mantle) between the tracked chromatophores of the squid. The dot tracking was compared with Ncorr, a widely used open-source two-dimensional digital image correlation (DIC) code~\cite{blaber2015ncorr}, at the same 58 dots (Fig.~\ref{sif_membrane_dic}, C). The two follow the same trend at every step (Fig.~\ref{sif_membrane_dic}, D), and at the largest stretch the vertical strain is 25.05\% against 24.93\% from DIC while the horizontal strain is $-9.95$\% against $-9.89$\%. At every step both strains are within 1.2\% of their DIC values, and over all four steps the displacements of the dots differ from DIC by a root-mean-square value of 0.23~pixels (Fig.~\ref{sif_membrane_dic}, E).}

\rev{Chromatophores belong to an active coloration system, so two effects of their activity on the tracking were examined. (i)~To our knowledge a chromatophore does not move across the body, since it is held in the skin by its radial muscles~\cite{messenger2001cephalopod}. (ii)~Blinking, the appearance or disappearance of a chromatophore within one or a few frames, was not observed in the recordings. One expansion or retraction of the pigment sac of a chromatophore lasts 217 to 384~ms in the related squid \textit{Doryteuthis pealeii}~\cite{hadjisolomou2021quantifying}, which is more than 200 frames at 1000~fps. With this high-speed imaging a chromatophore is therefore quasi-steady from frame to frame, which minimizes the effect of its activity on the dot tracking.}

\begin{figure*}[!htbp]
    \centering
    \includegraphics[width=\textwidth]{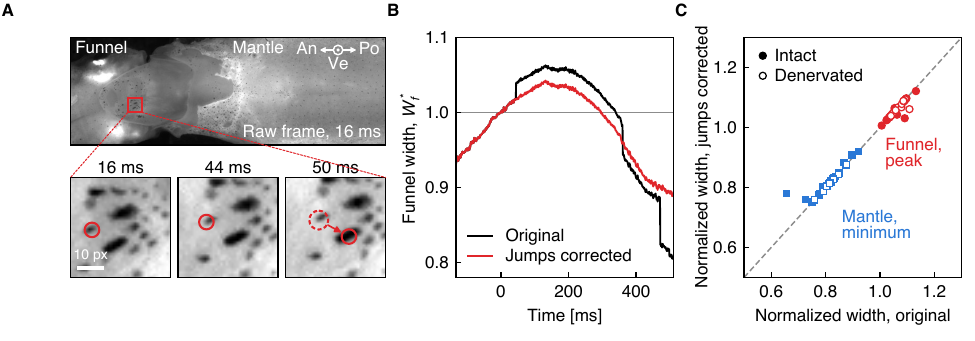}
    \caption{\rev{\textbf{Jumps of chromatophore tracks and their correction.}}
    \revred{All panels show \textit{Doryteuthis opalescens}.}
    \rev{(\textbf{A})~Top, the raw frame 16~ms after the pressure minimum, with the squid's head to the left and the red box marking the region of the frames below. An is anterior, Po is posterior and Ve is ventral, and the circled dot for Ve points out of the image toward the viewer. Below, a funnel track (red circle) 16, 44 and 50~ms after the pressure minimum. At 44~ms the chromatophore is still visible, but the thresholding barely detects it, and by 50~ms the track has moved to a neighboring chromatophore. The dashed circle marks the chromatophore the track started on.}
    \rev{(\textbf{B})~Funnel width of the same jet from the original tracks and with the jumps corrected. The original width makes the jump of (\textbf{A}) within one 1~ms frame.}
    \rev{(\textbf{C})~Peak funnel width (red circles) and minimum mantle width (blue squares) of each jet, original against jumps corrected, for intact (filled) and denervated (open) squid. Widths are normalized by their value at the pressure minimum, as in Fig.~\ref{f_squid}, I.}
    }
    \label{sif_tracking_jumps}
\end{figure*}

\rev{This check, however, led to an unexpected finding. Although there is no blinking, the thresholding can lose a faint chromatophore when its brightness contrast drops, a detection fault that affects 152 of the 1,842 tracks (8\%) in 20 of the 30 jets of intact and denervated squid (Fig.~\ref{f_squid}, I and J). A lost spot normally ends its track, which is then discarded, because only chromatophores that are tracked without a gap are kept. The exception is a lost spot that has another detected spot inside the search radius of the tracker. Its track is then linked to that neighbor instead of ending, which appears as a jump of several pixels within 1~ms that the rest of the region does not make (Fig.~\ref{sif_tracking_jumps}, A).}

\rev{A jump was identified as a frame-to-frame step that departs by more than 3~pixels from the median step of all tracks of its region, and every jump was replaced by that median step (Fig.~\ref{sif_tracking_jumps}, B). The values in Fig.~\ref{f_squid}, I and J are from the corrected tracks. The correction lowers the median peak funnel widening from 5.4\% to 4.4\% in intact squid and from 8.0\% to 6.7\% in denervated squid, and it leaves the median minimum mantle width unchanged (Fig.~\ref{sif_tracking_jumps}, C). After the correction the funnel widens in every jet of intact and denervated squid.}

\begin{figure*}[!htbp]
    \centering
    \includegraphics[width=\textwidth]{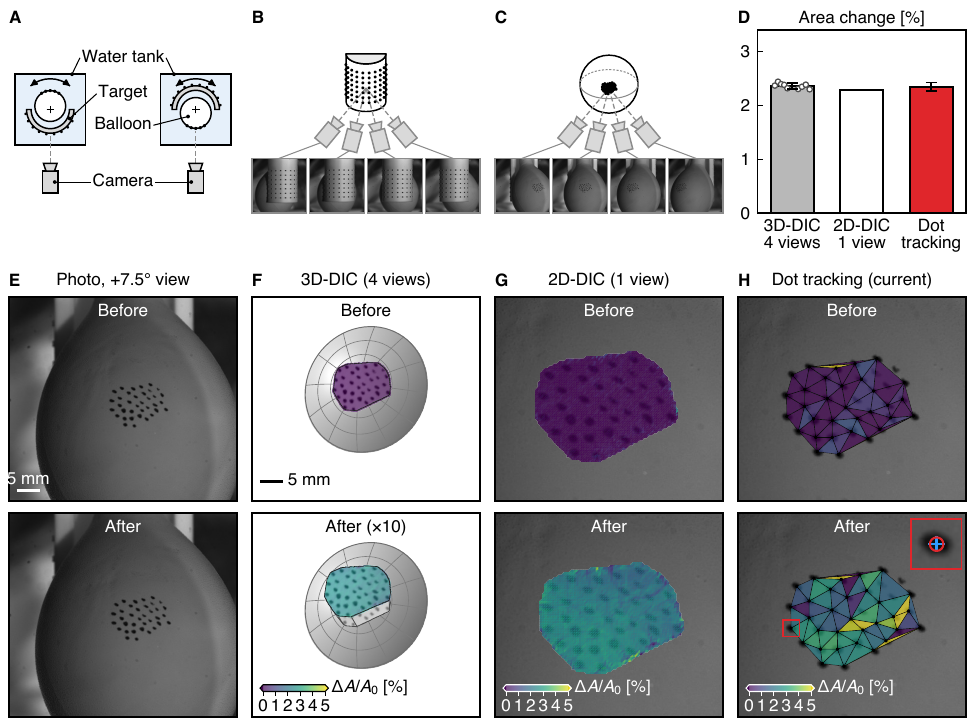}
    \caption{\rev{\textbf{Single-view error measured on an inflated balloon filmed through the wall of a cubic water tank, with multi-view 3D-DIC as the reference.}}
    \rev{(\textbf{A})~Top views of the setup, for the calibration (left) and for the balloon measurement (right). The calibration target is a hollow half cylinder with the balloon inside it, and both stand on a rotation stage in the cubic water tank. One fixed camera images the marked face of the target (left) or the dotted side of the balloon (right) through the wall, and rotating the stage is the same as placing virtual cameras around a fixed specimen.}
    \rev{(\textbf{B})~The four virtual cameras around the calibration target, a 3D-printed half cylinder with $7 \times 10$ embossed marks, rows 5~mm apart on a radius of 18~mm, and below them its four views from $+22.5^\circ$ to $-22.5^\circ$. One pinhole camera and one rotation axis, with the angle of each view free, were fitted to these marks together with the balloon dots.}
    \rev{(\textbf{C})~The same four virtual cameras around the balloon and its four views.}
    \rev{(\textbf{D})~Area change of the dotted patch from 3D-DIC with four views (gray, one SD, the nine camera pairs as circles), from 2D-DIC of the frontal view alone (white) and from the chromatophore-tracking code (red, SD over three expanded frames). Two of the nine pairs and the calibration also use a fifth view at $-37.5^\circ$.}
    \rev{(\textbf{E})~The $+7.5^\circ$ view, the most frontal of the four and the one on which the single-view methods run, before and after inflation, with the 36 dots on the balloon.}
    \rev{(\textbf{F})~Measured patch from 3D-DIC before and after inflation, seen from close to the direction of (\textbf{E}) and colored by the area change, with the displacement magnified ten times.}
    \rev{(\textbf{G})~2D-DIC of the frontal view alone, the area change of each triangle of the correlation grid over the dotted patch, before inflation (a second still frame against the first, which gives the noise floor) and after it.}
    \rev{(\textbf{H})~The 32 dots tracked by the chromatophore-tracking code in the same view, meshed and colored by the area change in the image, before inflation (again two still frames) and after it. The inset enlarges one dot and shows its tracked position (red circle) and its 2D-DIC position (blue cross).}
    }
    \label{sif_balloon_dic}
\end{figure*}

\rev{The widths are measured in the two-dimensional projection of the camera image and are assumed to represent the three-dimensional deformation of the body. This assumption was validated in a further experiment with three-dimensional digital image correlation (3D-DIC) in the MultiDIC toolbox~\cite{solav2018multidic}, using stereo imaging of an inflated balloon (Fig.~\ref{sif_balloon_dic}). The setup is a cubic water tank with sides of 30~cm, in which a rotation stage with fine angle adjustment carries the calibration target, a hollow half cylinder, with the balloon inside it (Fig.~\ref{sif_balloon_dic}, A). One fixed camera outside the tank films them through the tank wall. The camera, the lens and the distances from the object to the tank wall and from the wall to the lens were those of the squid recordings (Fig.~\ref{f_squid}, B and Fig.~\ref{sif_chromatophore}, A). The balloon carries 36 ink dots in place of chromatophores, and a syringe inflates it by injecting about 5~mL of water. Rotating the stage in three steps of $15^\circ$ gives the camera four views, at $\pm 7.5^\circ$ and $\pm 22.5^\circ$ from the center of the range, of the target (Fig.~\ref{sif_balloon_dic}, B) and of the balloon (Fig.~\ref{sif_balloon_dic}, C). The balloon was recorded in these views before and after the inflation (Fig.~\ref{sif_balloon_dic}, E).}

\rev{For the calibration, one pinhole camera and one rotation axis, with the angle of each view free, were fitted to the marks of the target (Fig.~\ref{sif_balloon_dic}, B) and to the balloon dots in every view. The fitted camera has an effective focal length of $5950 \pm 130$~pixels through the wall and the water, its principal point fixed at the image center and no distortion term, which gives 18.0~pixels/mm at the patch, 330~mm from the camera, and 16.0~pixels/mm at the rotation axis. It reproduces the marks to about 1~pixel (0.06~mm) and the balloon dots to 0.05 to 0.4~pixels. The fitted angle of each view differs from its nominal stage angle by $0.25^\circ$ on average and $0.57^\circ$ at most, with a standard uncertainty of $0.2^\circ$, and the reconstruction uses the fitted angles.}

\rev{The single-view methods run on the most frontal of the four views (Fig.~\ref{sif_balloon_dic}, E), whereas 3D-DIC uses all four and gives $2.36 \pm 0.06$\% for the area change of the dotted patch (Fig.~\ref{sif_balloon_dic}, F). On that view 2D-DIC gives 2.28\% (Fig.~\ref{sif_balloon_dic}, G), and the chromatophore-tracking code, with only its dot-size, search-distance and track-length limits adjusted for six still frames of larger dots, gives $2.34 \pm 0.08$\% (Fig.~\ref{sif_balloon_dic}, H). The small differences come from the rigid motion of the patch, which moved 0.74~mm toward the camera and tilted by $0.64^\circ$, and not from the tracking. The three methods agree on the area change within 0.08 percentage points (Fig.~\ref{sif_balloon_dic}, D). For the squid, effects of this size are small compared with the funnel widening of up to 12.1\% (Fig.~\ref{f_squid}, I), and the tilt effect is smallest where the surface faces the camera, as it does in the central funnel and mantle regions that are tracked (Fig.~\ref{sif_chromatophore}, F). An error of this size is unlikely to produce the funnel expansion--recoil or to move its peak appreciably, although it can shift a width by a few tenths of a percentage point in either direction, which is a limitation of the single-view measurement.}

\FloatBarrier
\section{Funnel \rev{denervation}}\label{si:denervation}

\begin{figure*}[!htbp]
    \centering
    \includegraphics[width=0.8\textwidth]{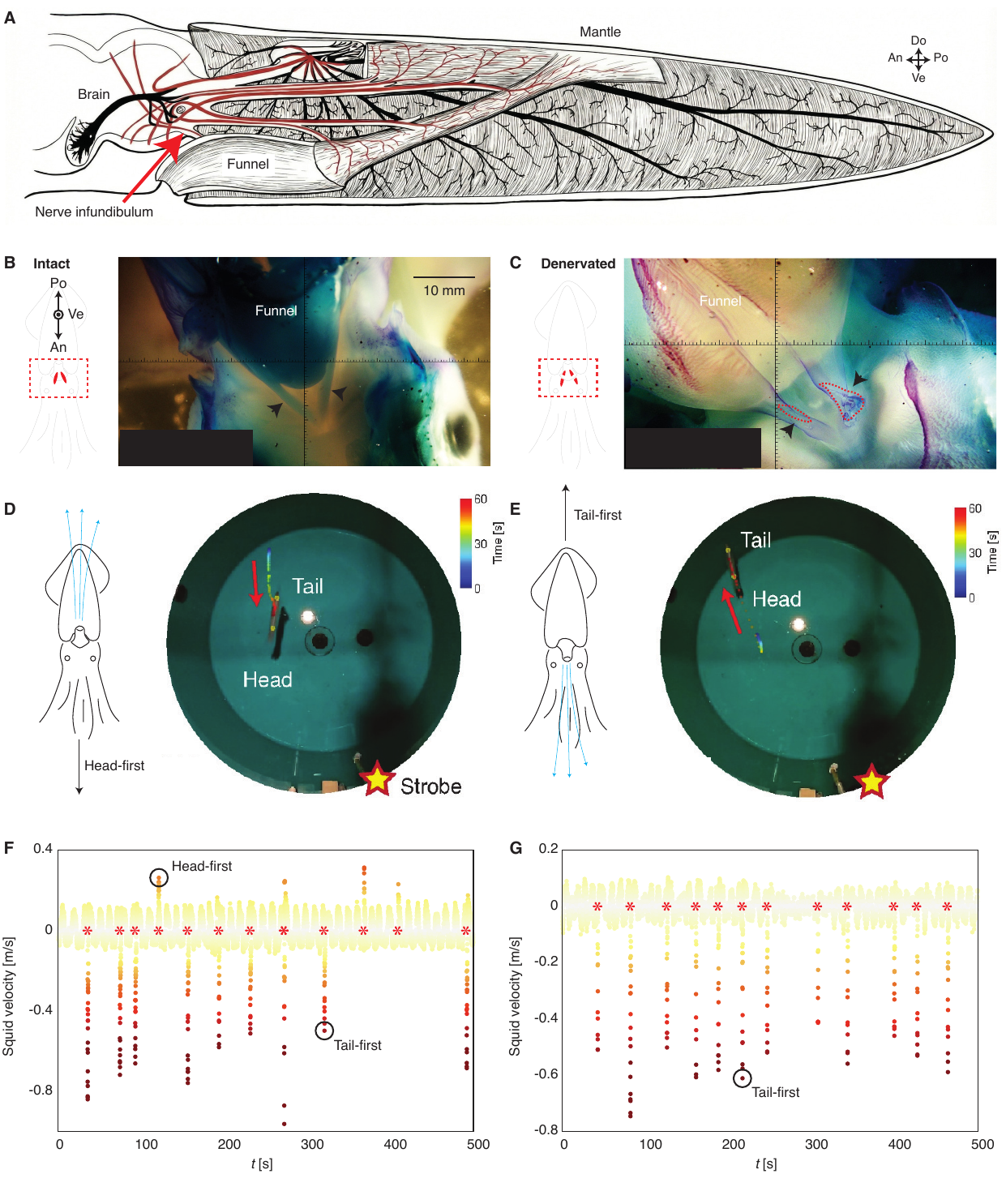}
    \caption{\textbf{Funnel \rev{denervation} by nerve incision.}
    \rev{All panels show \textit{Doryteuthis opalescens}.}
    (\textbf{A})~Schematic of squid nervous system showing the location of the nerve infundibulum innervating the funnel (redrawn from~\cite{bullock1965structure}). \rev{Axis labels give the orientation, where An is anterior, Po is posterior, Do is dorsal and Ve is ventral, all lying in the image plane.}
    (\textbf{B--C})~Schematic of the nerve infundibulum location (left) and microscope images showing the connection between head and funnel (right): (\textbf{B})~\revred{intact} condition; (\textbf{C})~funnel-\rev{denervated} condition due to nerve incision.
    (\textbf{D--E})~Snapshots of squid reflexes in response to strobe light stimulus in a water pool, with movement tracked using SLEAP~\cite{pereira2022sleap}: (\textbf{D})~head-first reflex; (\textbf{E})~tail-first reflex.
    (\textbf{F--G})~Directional velocity plots for (\textbf{F}) intact and (\textbf{G}) funnel-\rev{denervated} squid, showing that funnel-\rev{denervated} squid always move tail-first due to their inability to vector the funnel, whereas intact squid can move either head-first or tail-first. The star symbols indicate the strobe incidents. Total number of reflexes is 45 for intact squid and 34 for funnel-\rev{denervated} squid.
    }
    \label{sif_paralyze}
\end{figure*}

To investigate the relationship between funnel deformation and muscle activity, we developed a novel surgical method to selectively \rev{denervate} the funnel of the squid (\emph{Doryteuthis opalescens}).
\rev{Denervation} was achieved by severing the nerve infundibulum\revred{,} the neural pathway innervating and regulating the funnel (Fig.~\ref{sif_paralyze}, A)\revred{,} via a small incision on the ventral (aboral) surface, accessible by gently bending back the funnel structure (Fig.~\ref{sif_paralyze}, B and C).
The incision was carefully controlled to approximately one third of the membrane thickness: shallower cuts produced incomplete \rev{denervation}, whereas overly deep incisions severely compromised animal health, often resulting in mortality within a few hours. When performed at the optimal depth, the procedure consistently produced effective funnel \rev{denervation} while allowing the squid to swim normally and survive for at least 24 hours.

Funnel \rev{denervation} was verified using a light reflex assay. Freshly captured squid (see Fig.~\ref{sif_squid}, B) were acclimated in a circular, temperature-controlled (13\textdegree{}C), filtered seawater pool (1.5~m diameter, 1~m depth, 94.2\% dissolved oxygen). A strobe light positioned at one end of the pool (Fig.~\ref{sif_paralyze}, D) was flashed every 5 minutes for a total of 10 trials (50 minutes) to evoke escape responses. Following this initial measurement, squid were transferred to the shallow tank, and the funnel nerve was surgically incised in less than one minute before being returned to the same pool. The strobe stimulus regime was then repeated to assess changes in escape direction (see Fig.~\ref{sif_paralyze}, D and E). Prior to \rev{denervation}, squid could escape either head-first (11.1\%) or tail-first (88.9\%, where total number of escapes is 45 for three individuals) (Fig.~\ref{sif_paralyze}, D and F), whereas after funnel \rev{denervation}, all escapes were tail-first (100\%, where total number of escapes is 34 for three individuals) (Fig.~\ref{sif_paralyze}, \rev{E and G}), reflecting the inability to actively direct the funnel.
Fisher's exact test yields \revred{$p = 0.067$ (two-sided)}, which is marginally above the conventional significance threshold ($p < 0.05$). \rev{This marginal significance arises largely because} intact squid already exhibit a strong preference for tail-first escape (88.9\%), leaving only a small fraction of head-first events (5 out of 45) available to detect a difference. Nonetheless, the complete absence of head-first escapes after \rev{denervation} (0 out of 34) is biologically meaningful. For example, if the pre-\rev{denervation} rate had been maintained, approximately 3--4 head-first escapes would have been expected. The observed outcome is consistent with the mechanistic expectation that funnel \rev{denervation} eliminates the ability to vector thrust for head-first locomotion. 

\FloatBarrier
\section{Fabrication of flexible nozzle}\label{si:fabrication}
Three different methods of manufacturing and installation have been developed depending on nozzle thickness; all ensure the precise geometry of the flexible nozzle, robust attachment, and minimize pre-straining during installation. 

\begin{figure*}[!htbp]
    \centering
    \includegraphics[width=\textwidth]{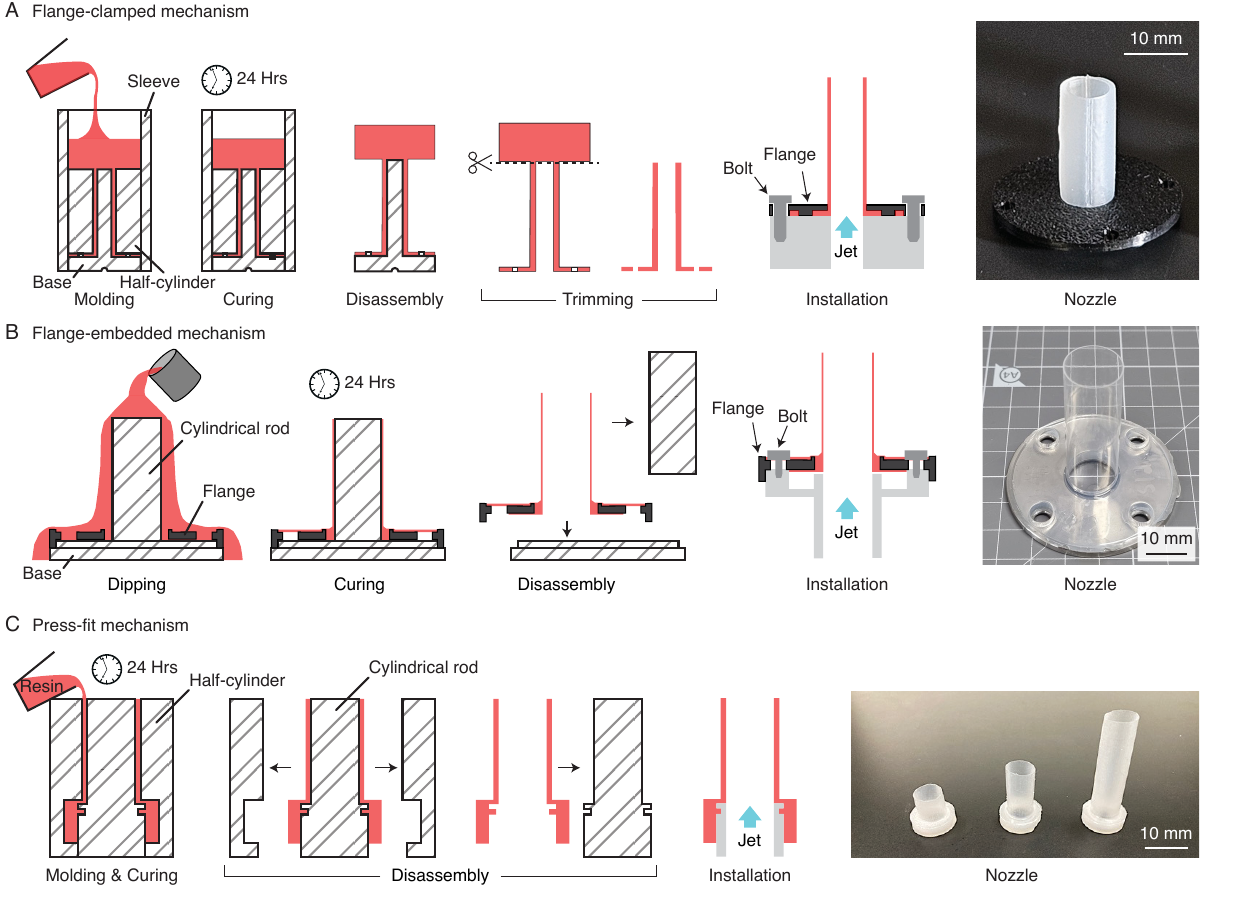}
    \caption{\textbf{Fabrication and installation methods for flexible silicone nozzles.}
    (\textbf{A})~Flange-clamped configuration for thick nozzles ($h > 0.5$~mm). Silicone is cast using internal and external molds to create the nozzle geometry, and then secured to the structure using a clamping plate and flange.
    (\textbf{B})~Flange-embedded configuration for thin nozzles ($h < 0.5$~mm). Silicone is poured onto an internal mold to achieve thin wall thickness, after which an embedded flange at the nozzle base is used to anchor the nozzle to the structure.
    (\textbf{C})~Press-fit configuration. Silicone is molded with internal and external components to form both the nozzle and an integrated base with slit shape, which is then inserted and retained within a slot on a rigid support.
    }
    \label{sif_nozzlefabrication}
\end{figure*}

First, relatively thick nozzles (e.g., wall thickness of 0.7~mm) were fabricated using a flange-clamped mechanism (see Fig.~\ref{sif_nozzlefabrication}, A). The molds consisted of three main parts: a base, two half-cylinders, and a sleeve. The base shaped the lower part of the nozzle and provided support for the inner diameter. The inner side of two half-cylinders established the outer diameter of the flexible nozzle while the outer side of them supported the sleeves which held the mold tightly and prevented leakage of liquid-state silicone during curing. Each mold was designed with an excess resin pool at the top so that empty spots would fill in after $\sim$\revred{10}~min of degassing in a vacuum chamber. To obtain the desired nozzle height, flexible nozzles were cut by repositioning them onto the base mold with soapy water to prevent stretching, aligning a blade at the desired height using customized mold sleeves, and rotating the blade circumferentially to remove excess material. Using this method, six flexible nozzles were fabricated with an inner diameter of 7~mm and wall thickness of 0.7~mm. Each nozzle was trimmed to a specific length (5, 10, 15, 20, 25, and 30~mm) to achieve different compliance values (\revred{$\tau/T = 0.08$--$0.46$}). A rigid nozzle of approximately 25~mm length ($\tau/T = 0$) was 3D-printed from PLA. Each nozzle was fabricated with alignment holes for mounting. Nozzles were secured to the structure using a clamping plate with screws positioned 90$^\circ$ apart, by sandwiching between the plate and support structure (see the picture of the flexible nozzle in Fig.~\ref{sif_nozzlefabrication}, A), with care taken to avoid over-tightening that could deform the silicone. This method has been used to fabricate nozzles for the squid-boat test (Fig.~\ref{sif_boat}), aerial jet with repeated jet generator (Fig.~\ref{sif_aerialjet}), and the mixing experiment (Fig.~\ref{sif_mixing}).

Second, the relatively thin nozzle (\revred{65--260~$\mu$m} wall thickness) was fabricated using a flange-embedded mechanism (see Fig.~\ref{sif_nozzlefabrication}, B). The mold consisted of three main parts: a base, an internal cylindrical rod, and a flange. The base was responsible for fixing and centering the flange and the internal cylindrical rod on which the thin wall of silicone was formed. Liquid-state silicone (SortaClear 40A, \rev{Smooth-On}) was poured into the mold and cured at room temperature for 24 hours. Due to the thinning effect of gravity, a thin nozzle wall formed, with thickness variation along its height within 20\% (see detailed validation in \cite{choi2022flow}). After curing, the internal cylindrical rod was carefully removed to avoid any permanent deformation of the nozzle, while keeping the flange embedded at the base of the flexible nozzle. To obtain the desired nozzle height, the flexible nozzles were cut by repositioning them onto the mold base with soapy water (to prevent stretching), aligning a blade at the desired height using custom mold sleeves, and rotating the blade circumferentially to remove excess material. 
The nozzle with the embedded flange was then installed onto the structure using four bolts. Four nozzles with different structural stiffness levels ($E h$ = 7.0, 14.4 and 43.2~N/m for flexible nozzles, and $\infty$ for the rigid nozzle) were fabricated, each with an inner diameter of 15~mm and a height of 30~mm.
The wall thickness and \rev{stiffness} of the nozzle varied during the curing process due to differences in the viscosity of each silicone (SortaClear 40A, Smooth-On), depending on the mass ratio (0--\revred{41.2} wt\%) of silicone thinner (\rev{Silicone} Thinner, Smooth-On) added (see \cite{choi2022flow} for details). This process resulted in a wall thickness ranging from \revred{65 to 260~$\mu$m}. The rigid nozzle was manufactured from aluminum by milling. This method has been used to generate nozzles for flow field measurement and energy analysis, which require precise geometry and optical transparency to visualize the internal flow field (Fig.~\ref{sif_piv}).

Third, the press-fit mechanism was employed by integrating the nozzle with a thickened base designed to fit securely into a slit on the rigid support (Fig.~\ref{sif_nozzlefabrication}, C). The nozzle was molded (Ecoflex 0030, Smooth-On) such that the wall and thick base with an internal rim were formed as a single piece, allowing the rim to slot easily into the support slit. The \revred{thick} base does not affect the dependence of performance on the time ratio, $\tau/T$, due to its higher structural stiffness (see Fig.~\ref{sif_aerialjet}). Here the nozzles with different diameter (10 and 20~mm) and heights (\revred{15}--60~mm) were fabricated with the wall thickness of 0.7~mm. This method has been used to generate nozzles for the aerial jet with single pulsed-jet generator (Fig.~\ref{sif_aerialjet}, A).

\begin{figure*}[!htbp]
    \centering
    \includegraphics[width=\textwidth]{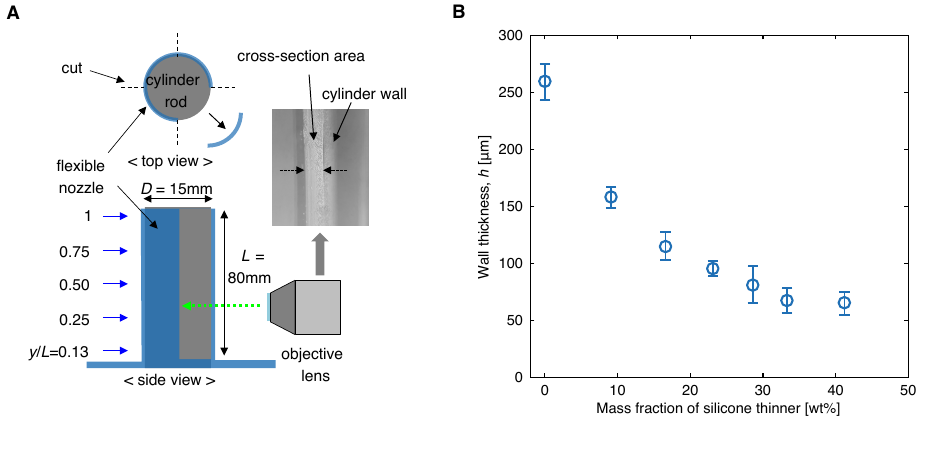}
    \caption{\rev{\textbf{Wall thickness of the thin-walled nozzles.}}
    \rev{(\textbf{A})~A cured nozzle (blue) on its acrylic rod (gray) was sectioned at four azimuthal positions (top view), and the cross-section of its wall was imaged through an objective lens at five axial locations from $y/L = 0.13$ to 1, with $y$ measured from the support (side view).}
    \rev{(\textbf{B})~Mean wall thickness $h$ against the mass fraction of silicone thinner. Error bars give the spread across the measurement locations.}
    }
    \label{sif_nozzle_thickness}
\end{figure*}

\begin{figure*}[!htbp]
    \centering
    \includegraphics[width=\textwidth]{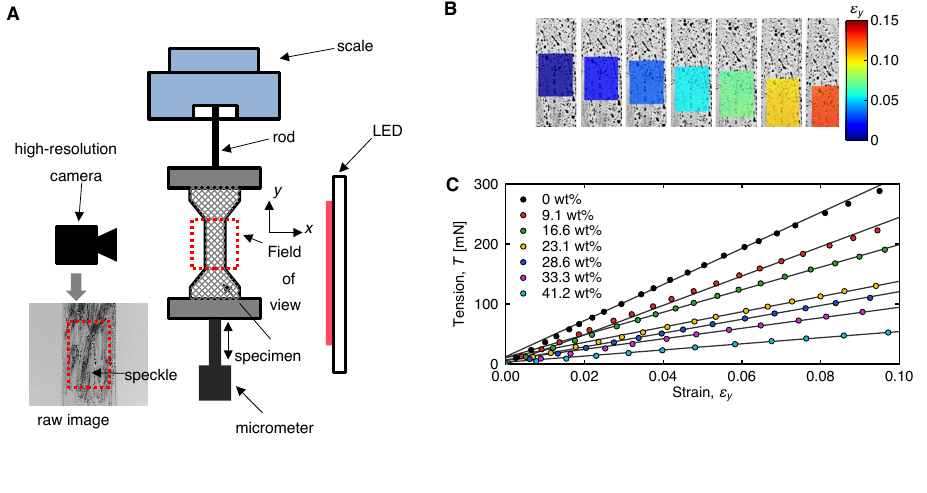}
    \caption{\rev{\textbf{Stiffness of the cured silicone from tensile tests with digital image correlation.}}
    \rev{(\textbf{A})~Tensile stage. A micrometer extends the speckled specimen (cross-hatched) and a scale reads the tension, while a high-resolution camera views the gauge section.}
    \rev{(\textbf{B})~Raw speckle images and the longitudinal strain field $\varepsilon_y$ obtained from them by two-dimensional digital image correlation.}
    \rev{(\textbf{C})~Tension $T$ against strain for each mass fraction of silicone thinner. Young's modulus is taken from the initial slope at $\varepsilon_y = 0$, with the tension divided by the cross-section of the undeformed specimen.}
    }
    \label{sif_silicone_stiffness}
\end{figure*}

\rev{Because the thin wall of these nozzles forms under gravity (Fig.~\ref{sif_nozzlefabrication}, B), its uniformity and the stiffness of the cured silicone were characterized before the nozzles were used~\cite{choi2022flow}. The inner diameter is set by the rigid internal rod (acrylic, 15~mm in diameter) and is fixed during fabrication, so the quantity that can vary is the wall thickness. Cured nozzles, still on the rod, were sectioned along their axis at four azimuthal positions, and the wall thickness was measured under a microscope at five axial locations from $y/L = 0.13$ to 1, with $y$ the distance from the support and $L$ the length of the nozzle (Fig.~\ref{sif_nozzle_thickness}, A). The variation around the circumference was negligible. Along the axis the wall thickened gradually toward the support, at a rate of 12.5\% per nozzle diameter, without a localized accumulation at the lower end. For a 45-mm nozzle, for example, the thickness was 51.1~$\mu$m at the tip and 70~$\mu$m at the support, and for the 30-mm nozzles used here the same rate gives a tip about 20\% thinner than the support. No localized thickening associated with pooling of the silicone was observed.}

\rev{Because the axial variation is gradual, each nozzle is characterized by its mean wall thickness $h$ over the five axial locations and the four azimuthal positions. Across six independently fabricated nozzles this mean was repeatable within 6.6\%. It decreased from approximately 260 to 65~$\mu$m as the mass fraction of the thinner increased from 0 to 41.2~wt\% (Fig.~\ref{sif_nozzle_thickness}, B). The reduced-order model uses this mean thickness and reproduces the measured velocity, impulse, and energy histories (Fig.~\ref{f_mechanism}K--M).}

\rev{Air bubbles were controlled during fabrication. After the silicone was poured, the mold assembly was placed under vacuum for 10~min to degas the material. Any rare bubble that remained visible after degassing was removed before curing, and nozzles with persistent bubbles or visible defects were discarded.}

\rev{Because the response of the nozzle depends on both the wall thickness and the stiffness of the material, the Young's modulus $E$ of the cured silicone was measured independently for each thinner fraction (Fig.~\ref{sif_silicone_stiffness}). A custom tensile stage extended a dumbbell-shaped specimen in steps set by a micrometer with a resolution of 1~$\mu$m, and a scale with a resolution of 10~mg read the tension (Fig.~\ref{sif_silicone_stiffness}, A). A speckle pattern printed on the specimen with oil-based ink was recorded by a high-resolution camera, and the strain was obtained from it by two-dimensional digital image correlation with an in-house code (Fig.~\ref{sif_silicone_stiffness}, B). The tension was converted to stress with the cross-section of the undeformed specimen, and $E$ was taken as the initial slope of the stress--strain curve, obtained by a polynomial fit (Fig.~\ref{sif_silicone_stiffness}, C). The modulus decreased from approximately 340 to 60~kPa over the same range of 0 to 41.2~wt\%. The values of $E$ and $h$ used in the analysis are therefore the measured ones and not nominal fabrication values.}

\FloatBarrier
\section{Single pulsed jet generator}\label{si:jet}
\begin{figure*}[!htbp]
	\centering
	\includegraphics[width=\textwidth]{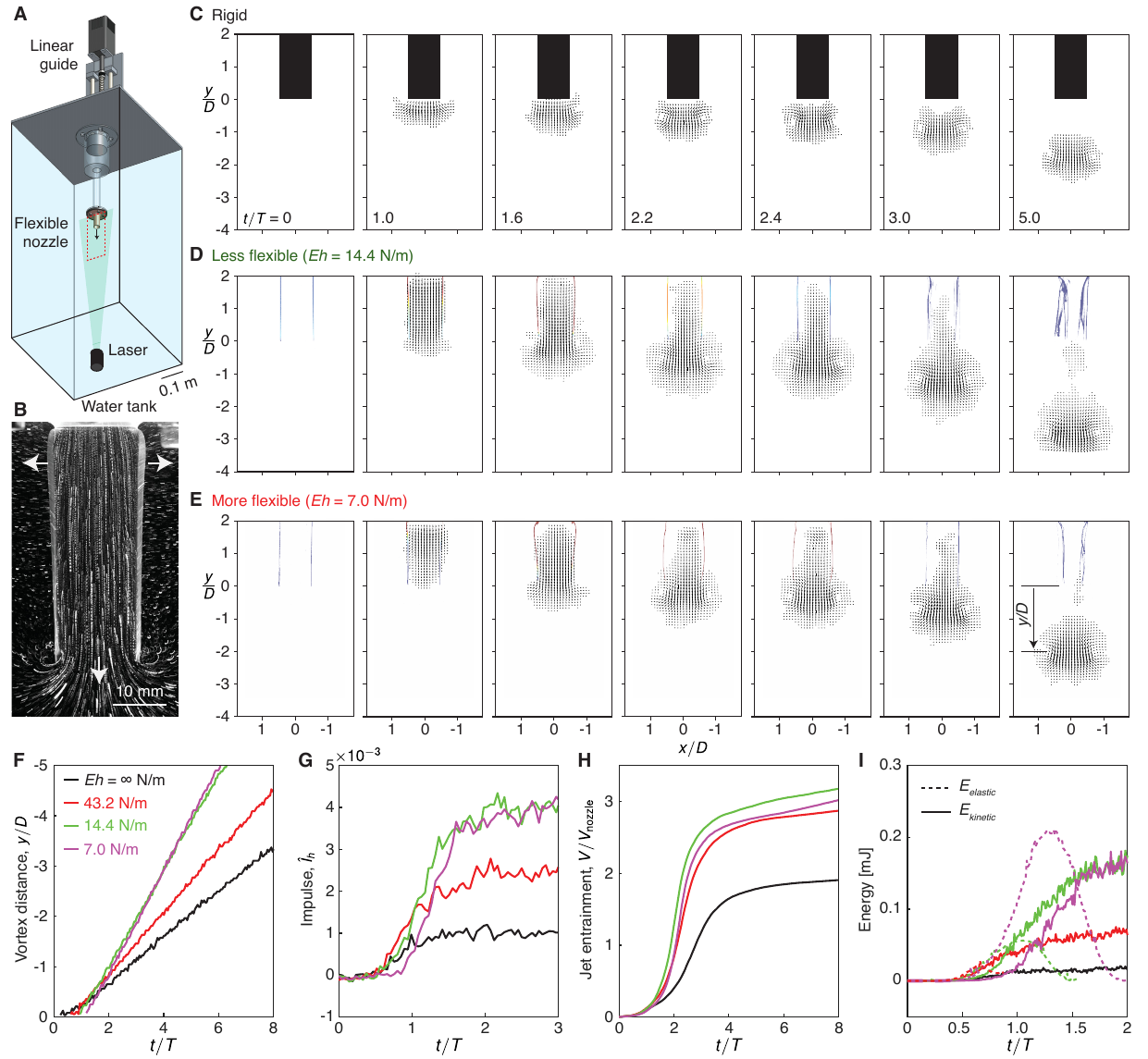}
	\caption{\textbf{Particle image velocimetry}.
    (\textbf{A})~Single pulsed jet generator using a syringe and linear guide in a water tank, where the flow was visualized by a laser sheet and a high-speed camera for particle image velocimetry (PIV).
    (\textbf{B})~Visualization of jet development and nozzle deformation by overlaying raw images from high-speed imaging.
    (\textbf{C--E})~Vortex ring development of the jet (acceleration time $T = 0.05$~s) from nozzles with different stiffnesses ($Eh$, where $E$ is the Young's modulus of the nozzle and $h$ is the nozzle thickness): (\textbf{C}) rigid ($Eh = \infty$), (\textbf{D}) less flexible ($Eh = 14.4$~N/m), and (\textbf{E}) more flexible ($Eh = 7.0$~N/m). Arrows indicate velocity vectors with lengths proportional to the velocity magnitudes. The cross-section of the nozzle is color-coded by its horizontal deformation $w/D_0$, with the color scale ranging from 1.0 (blue, undeformed) to 1.1 (red, 10\% expansion), normalized by the initial nozzle diameter $D_0$.
    (\textbf{F--I})~Quantitative analysis of jet performance under a slower jet condition (acceleration time $T = 0.11$~s) for nozzle stiffnesses $Eh = \infty$, 43.2, 14.4, and 7.0~N/m.
    (\textbf{F})~Travel distance of the vortex ring as a function of $t/T$.
    (\textbf{G})~Hydrodynamic impulse $\hat{I}_h$ of the jet as a function of $t/T$. \rev{The impulse rises and then plateaus, and its terminal value, the mean over $3T \le t \le 5T$, is reported.}
    (\textbf{H})~Cumulative jet entrainment $V/V_{\mathrm{nozzle}}$ measured at $y/D = -1$, normalized by the nozzle volume.
    (\textbf{I})~Energy partitioning between elastic strain energy $E_{\mathrm{elastic}}$ (dashed lines) and kinetic energy $E_{\mathrm{kinetic}}$ (solid lines) as a function of $t/T$. Under both jet conditions, the intermediate stiffness nozzle produces the fastest downstream advection of vortex structures, the highest hydrodynamic impulse, and maximum flow kinetic energy.
    }
	\label{sif_piv}
\end{figure*}

The pulsed jet was generated by a piston\revred{--}motor system driving flow through the nozzle into a water tank (300~mm$\times$~300~mm$\times$~700~mm; see Fig.~\ref{sif_piv}, A), which has been validated in \rev{previous studies} \cite{choi2022flow, choi2024mechanism}. A 50~mm diameter piston pressurized the water inside the cylindrical nozzle and was actuated along a linear guide by a 40~W motor (ELDM6020, LeadShine). 
The piston motion was precisely controlled by a motor driver (ELD5-400, LeadShine) to consistently generate the desired jet-exit velocity profile, with an acceleration time of 0.05--0.34~s and a maximum centerline jet velocity of \revred{0.07}--0.33~m/s. This corresponds to a Reynolds number ($Re = v_{max}D/\nu$, where $D$ is the inner nozzle diameter, $v_{max}$ is the maximum centerline jet velocity, and $\nu$ is the kinematic viscosity of water) of 1,100--5,000. The center of the undeformed nozzle exit was designated as the origin.
\rev{For all cases, the piston stroke was held constant, so that every nozzle displaced the same volume of water.} Five pulsed-jet velocity profiles were generated by controlling both the speed and response time of the motor (see Fig.~\ref{sif_piv}, A). 
The nozzle was fabricated using the flange-embedded method (Fig.~\ref{sif_nozzlefabrication}, B) to achieve a sufficiently thin wall thickness (\revred{70--170~$\mu$m~\cite{choi2022flow}}), enabling significant fluid-structure interaction corresponding to a dimensionless time ratio of $\tau/T \approx 0.3$.

\section{2D Particle image velocimetry and nozzle deformation measurement}\label{si:piv}
Glass tracer particles (50~$\mu$m, HSG-10, Dantec Dynamics) were seeded into the flow and illuminated by a green laser sheet (RayPower 5000, Dantec Dynamics) to perform particle image velocimetry (PIV) at the center plane ($z = 0$). Particle motion was recorded using a high-speed camera (NX5, IDT) at a frame rate of 500--1000~Hz, with a sensor size of 375~$\times$~1000 pixels (3$D$~$\times$~8$D$, where $D$ is the internal diameter of the nozzle). To quantify the nozzle deformation, the raw images were processed to highlight the nozzle wall, and the sub-pixel wall location was determined using Gaussian interpolation. To extract particle-only images for velocity analysis, the wall intensity was subtracted from the raw images. Velocity fields were calculated using a two-stage cross-correlation algorithm (window sizes: 64~$\times$~64 and 32~$\times$~32 pixels), which effectively reduces particle loss and influx within interrogation windows and is robust for high-shear flows \cite{choi2022flow, choi2024mechanism}. The resulting vector field contained approximately 1800 velocity vectors (spatial resolution: 56.3~$\mu$m~pixel$^{-1}$), with uncertainties estimated to be within 5\% using error propagation methods \cite{choi2022flow, choi2024mechanism}.

For combined visualization of nozzle deformation and flow field for the flexible nozzles (Fig.~\ref{sif_piv}, D--E and Fig.~\ref{f_mechanism}, B), the nozzle contour was extracted from masking data (obtained previously \cite{choi2022flow, choi2024mechanism}) and overlaid with the flow field data with color-coded nozzle wall with the diameter variation of the nozzle. Here the expanding (or contracting) nozzle diameter was computed from left and right boundary coordinates at each vertical position, and normalized width $w/D_0$ (where $w$ is the local width and $D_0$ is the undeformed inner nozzle diameter), with the color scale ranging from 1.0 (blue, undeformed) to 1.1 (red, 10\% expansion). This filled contour was overlaid on the raw background image with 50\% transparency, and PIV velocity vectors were superimposed to simultaneously visualize nozzle deformation and fluid motion.

\section{Data provenance and reuse}\label{si:provenance}
The piston-driven jet facility and synchronized PIV/nozzle-deformation recordings were previously used to characterize vortex dynamics and impulse/entrainment trends in flexible nozzles \cite{choi2024mechanism}. The PIV datasets for acceleration times $T = 0.05$, $0.11$, and $0.18$~s with nozzle stiffnesses $Eh = 10^{7}$ (rigid), $43.2$, $14.4$, and $7.0$~N\,m$^{-1}$ are reused here at the raw-data level (image sequences and wall kinematics) and reprocessed with a new analysis pipeline to compute time-resolved jet kinetic power/energy and nozzle elastic power/strain energy (the energy-transfer accounting in Fig.~\ref{f_mechanism}H and associated theory comparison), which were not reported in the prior study. Two additional cases ($Eh = 7.0$~N\,m$^{-1}$ at $T = 0.20$ and $0.34$~s), not included in Choi \& Park~\cite{choi2024mechanism}, were acquired for this work to extend the $\tau/T$ sweep. Table~\ref{tab:S1} summarizes all experimental conditions, specifying sample sizes and indicating for each case whether it was previously reported or newly acquired for this study.

\begin{table*}[ht]
    \centering
    \caption{Summary of experimental and simulation conditions.
    $N$ denotes the total number of independent trials or simulation cases; analysis  letter codes are defined in footnote $\ddagger$.}
    \label{tab:S1}
    \footnotesize
    \setlength{\tabcolsep}{3pt}
    \renewcommand{\arraystretch}{1.2}
    \newcommand*{\tabSoneL}[2]{\begin{tabular}[c]{@{}l@{}}#1\\#2\end{tabular}}
    \newcommand*{\tabSoneC}[2]{\begin{tabular}[c]{@{}c@{}}#1\\#2\end{tabular}}
    \begin{tabular}{l cc c c c c c c c cc}
    \hline\hline
    \multirow{2}{*}{Measurement} & \multirow{2}{*}{\shortstack{$D$\\(mm)}} & \multirow{2}{*}{\shortstack{$L$\\(mm)}} & \multirow{2}{*}{\shortstack{$Eh^{*}$\\(N\,m$^{-1}$)}} & \multirow{2}{*}{\shortstack{Jet\\type}} & \multirow{2}{*}{\shortstack{$T$\\(s)}} & \multirow{2}{*}{$\tau/T$} & \multirow{2}{*}{\revred{$Re^{\P}$}} & \multirow{2}{*}{\shortstack{Data from \\ \cite{choi2024mechanism}}} & \multirow{2}{*}{$N$} & \multicolumn{2}{c}{Analysis$^{\ddagger}$} \\
    \cline{11-12}
     & & & & & & & & & & Old~\cite{choi2024mechanism} & New \\
    \hline
    \tabSoneL{PIV}{(Fig.~\ref{sif_piv})} & 15 & 30 & \tabSoneC{$\sim\!10^{7}$ (rigid),}{7.0--43.2} & Single & \revred{0.055--0.176} & \revred{0.00--0.79} & \tabSoneC{\revred{1{,}100--}}{\revred{5{,}000}} & \checkmark & 12 & \textit{v}, \textit{i} & \textit{e}, \textit{t} \\
    \tabSoneL{PIV}{(Fig.~\ref{sif_piv})} & 15 & 30 & 7.0 & Single & 0.20--0.34 & \revred{0.13--0.22} & \tabSoneC{\revred{1{,}100--}}{\revred{5{,}000}} & $\times$ & 2 & --- & \textit{v}, \textit{i}, \textit{e}, \textit{t} \\
    \tabSoneL{CFD}{(Fig.~\ref{sif_cfd_method})} & 15 & 41 & \tabSoneC{$\infty$ (rigid),}{\revtc{75}} & Single & \revred{0.0275--0.183} & \revred{0.00--0.67} & \revred{4{,}400} & $\times$ & \revred{9} & --- & \textit{v}, \textit{i}, \textit{e} \\
    \tabSoneL{Load cell}{(\rev{Fig.~\ref{sif_force_rig}})} & \rev{18.6} & \rev{50} & \rev{rigid, 76} & Single & \rev{0.03--0.11} & \rev{0.00--0.75} & \tabSoneC{\revred{3{,}300--}}{\revred{15{,}000}} & $\times$ & \rev{300} & --- & \textit{f}\rev{, \textit{i}} \\
    \tabSoneL{HSI$^{\S}$}{(Fig.~\ref{sif_aerialjet})} & \revred{10, 20} & \revred{15--60} & \revred{70} & Single & \revred{0.10} & \revred{0--0.32} & \tabSoneC{\revred{7{,}500--}}{\revred{30{,}000}} & $\times$ & 49 & --- & \textit{a} \\
    \tabSoneL{HSI$^{\S}$}{(Figs.~\ref{sif_aerialjet}--\ref{sif_mixing})} & 7 & 5--30 & 70 & Repeated & \revred{0.02} & \revred{0.00--0.46} & \tabSoneC{\revred{700--}}{\revred{2{,}800}} & $\times$ & 53$^{\dagger}$ & --- & \textit{a}, \textit{b}, \textit{m} \\
    \tabSoneL{HSI$^{\S}$}{(Fig.~\ref{sif_smallboat})} & 2 & 10 & \revred{25} & Repeated & 0.005 & \revred{0.57} & \revred{5{,}000} & $\times$ & 6 & --- & \textit{b} \\
    \hline\hline
    \end{tabular}
    \\[4pt]
    \raggedright \footnotesize
    $^{*}$\,Nozzle fabrication (Fig.~\ref{sif_nozzlefabrication}): rows 1--2, flange-embedded (panel\,B); \rev{row 4, clamped by its flange (Fig.~\ref{sif_force_rig}, A); row 5,} press-fit (panel\,C); row 6, flange-clamped (panel\,A); row 7, dip-coated. \\
    $^{\dagger}$\,Boat: $N = 5$ trials per nozzle $\times$ 7 nozzles $= 35$; mixing: $N = 3$ trials per nozzle $\times$ 6 nozzles $= 18$; total $N = 53$. \\
    $^{\ddagger}$\,\textit{v}\,=\,velocity/vortex dynamics, \textit{i}\,=\,impulse, \textit{e}\,=\,energy/power, \textit{t}\,=\,theory, \textit{f}\,=\,force, \textit{a}\,=\,aerial jet, \textit{b}\,=\,boat, \textit{m}\,=\,mixing. \\
    $^{\S}$\,HSI: high-speed imaging. \\
    \revred{$^{\P}$\,$Re = UD/\nu$, with the jet velocity $U$ ($\sqrt{2gH}$ from the jet height $H$ for the single-pulse aerial jets) and the kinematic viscosity $\nu$ of water.}
    \end{table*}

\begin{figure*}[!htbp]
    \centering
    \includegraphics[width=\textwidth]{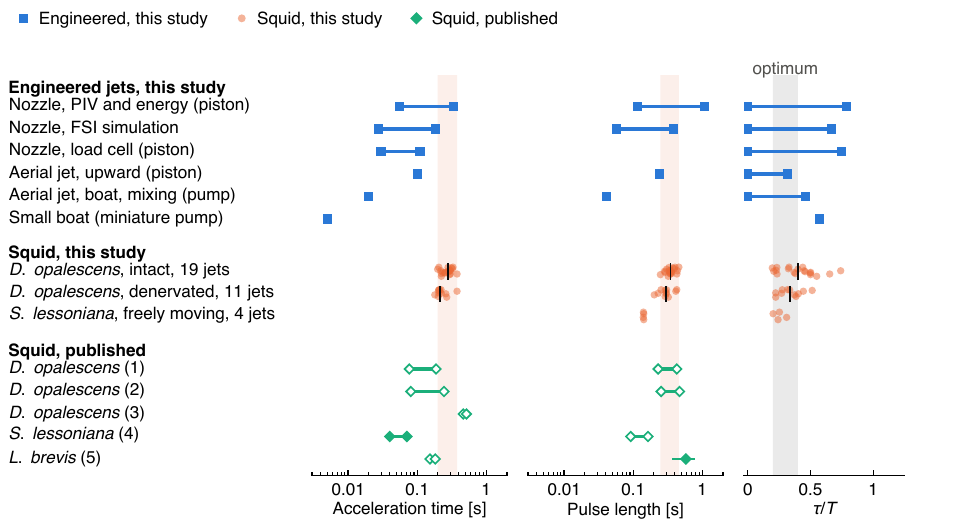}
    \caption{\revred{\textbf{Jet timescales of every experiment in this study and of squid.}}
    \revred{The three plots give the acceleration time, the pulse length and $\tau/T$. Each circle is one jet of our squid, with the median marked in black. Published values stated in the papers are filled, and values read from their traces are open, with the peak taken at the middle of a rounded or flat top. The light bands span our intact squid in the first two plots and the optimum $\tau/T$ in the third. For the pump at 24~Hz and for \textit{L.~brevis} the pulse-length plot shows the period. Published values are from (1)~Otis and Gilly~\cite{otis1990concerted}, (2)~Neumeister \textit{et al.}~\cite{neumeister2000temperature}, (3)~Li and Gilly~\cite{li2019hypoxia}, (4)~Thompson and Kier~\cite{thompson2002ontogeny} and (5)~Li \textit{et al.}~\cite{li2023hydrodynamic}.}
    }
    \label{sif_timescales}
\end{figure*}

\revred{The timescales of every experiment in this study are set beside those of squid in Fig.~\ref{sif_timescales}. In our recordings of restrained \textit{D.~opalescens} the mantle pressure peaks 0.2 to 0.4~s after the contraction begins and the contraction lasts 0.25 to 0.46~s, which agrees with published escape jets of the same species~\cite{otis1990concerted, neumeister2000temperature, li2019hypoxia}. The acceleration time of the engineered jets spans $T = 0.005$ to 0.34~s. The piston generator used for the results of Fig.~\ref{f_mechanism} (0.05 to 0.34~s) covers most of the squid values, and the pump-driven demonstrations (0.005 to 0.02~s), built for engineering use, are faster. We did not match $T$ to the contraction of the squid, because the gain is set by the ratio $\tau/T$ and not by $T$ itself (Fig.~\ref{f_mechanism}E). The response time $\tau = L/c$ grows with the nozzle length $L$, so the ratio also accounts for size, and the experiments spread $\tau/T$ from 0 to 0.79 (Table~\ref{tab:S1}), which covers the \textit{in vivo} range.}

\section{Flow field analysis}\label{si:flow}
As a characterization of the spatial evolution of vortex rings shed from the nozzle, vortex core tracking was performed based on the vorticity fields (Fig.~\ref{sif_piv}, C--E) obtained from particle-image velocimetry (PIV) (Fig.~\ref{sif_piv}, F).
For each time point, the positions of the maximum and minimum vorticity were identified in the region $y < 0$ to identify the vortex cores. The vertical coordinates of these extrema, $y_\mathrm{left}$ and $y_\mathrm{right}$, were averaged to determine the instantaneous vortex core location: $y_\mathrm{center} = (y_\mathrm{left} + y_\mathrm{right})/2$. The motion of this vortex center was then tracked over time and normalized by the nozzle diameter, $D$, to allow for comparison across nozzle configurations. As shown in Fig.~\ref{sif_piv}, F, the nozzle with intermediate stiffness ($Eh = 14.4$~N/m) produces the most rapid downstream propagation of vortex structures, 86\% faster than the rigid nozzle case ($Eh = \infty$). 

Here, the velocity field snapshots in Fig.~\ref{sif_piv}, C--E were obtained using a jet acceleration time of $T = 0.05$~s, while the quantitative analysis in Fig.~\ref{sif_piv}, F--I employed a slower jet condition ($T = 0.11$~s) to explore a broader range of nozzle stiffnesses ($Eh = \infty$, 43.2, 14.4, and 7.0~N/m).

\begin{figure*}[!htbp]
    \centering
    \includegraphics{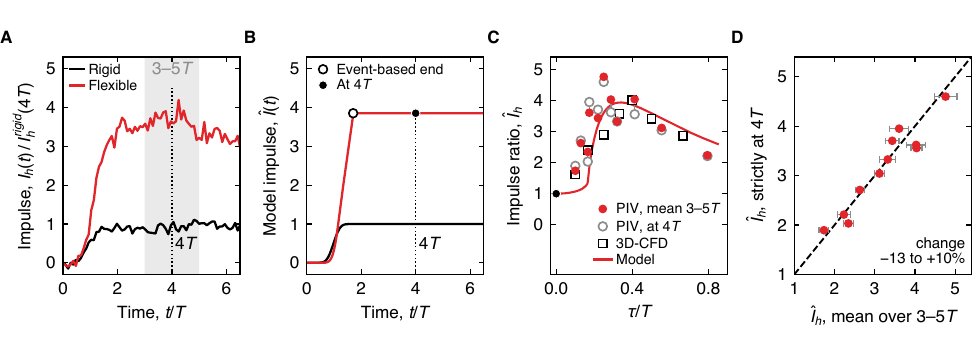}
    \caption{\rev{\textbf{Impulse amplification evaluated at the common time $t = 4T$.}}
    \rev{(\textbf{A})~PIV impulse of the rigid (black) and flexible (red, $Eh = 14.4$~N/m, $T = 0.11$~s) nozzles, with the reported window $3T \le t \le 5T$ in gray.}
    \rev{(\textbf{B})~Reduced-order model for the same case, with the end of its outflow (open) and its value at $4T$ (filled).}
    \rev{(\textbf{C})~Impulse ratio against $\tau/T$ for the PIV values as reported (filled) and strictly at $4T$ (open), 3D-CFD (squares) and the model (line).}
    \rev{(\textbf{D})~Strict against reported value, with the SD of $I_h(t)$ in the window as bars.}
    }
    \label{sif_common_window}
\end{figure*}

To quantify the total momentum generated by the vortex ring, hydrodynamic impulse was calculated from the vorticity field (Fig.~\ref{sif_piv}, G), \revred{assuming axisymmetric flow, as} $\mathbf{I}_h = \rho/2 \cdot \int \mathbf{x} \times \boldsymbol{\omega}\,dV$\revred{.} In 2D PIV, $\boldsymbol{\omega}$ reduces to a single out-of-plane scalar $\omega_z = \partial v/\partial x - \partial u/\partial y$. Under axisymmetry ($dV = 2\pi r\,dA$), the axial component becomes: $I_h = \pi\rho \int r^2\,\omega_z\,dA$ where $r = |x|$ is the radial distance from the centerline and $\omega_z$ is signed, changing sign across $x = 0$ due to the counter-rotating vortex pair in the meridional plane. The impulse was normalized by the value of the rigid nozzle with the same jet condition $\hat{I} = I_f / I_r$, where $I_f$ is the hydrodynamic impulse of the flexible nozzle and $I_r$ is the hydrodynamic impulse of the rigid nozzle. \rev{$I_h$ is evaluated in every frame, and the value reported is its terminal value, evaluated at $t = 4T$ as the mean over $3T \le t \le 5T$, which is the same for all nozzles.} For the intermediate-stiffness nozzle ($Eh = 14.4$~N/m), \rev{the terminal hydrodynamic impulse, evaluated at $t = 4T$ as the mean over $3T \le t \le 5T$, is about 300\% larger than that of the rigid nozzle ($Eh = \infty$) and about 260\% larger strictly at $t = 4T$ (Fig.~\ref{sif_common_window}, C and D).}

Jet entrainment was quantified by computing the volume flux across a horizontal measurement plane at $y/D = -1$ (Fig.~\ref{sif_piv}, H) as $\dot{V} = \int v \cdot 2\pi r \, dr$, where $v$ is the vertical velocity component and $r$ is the radial distance from the jet centerline. The cumulative entrained volume was obtained by time integration and normalized by the nozzle volume $V_{\mathrm{nozzle}} = \pi (D/2)^2 L$. The intermediate-stiffness nozzle ($Eh = 14.4$~N/m) exhibits 66\% greater cumulative entrainment ($V/V_{\mathrm{nozzle}} = 2.95$) compared to the rigid nozzle ($V/V_{\mathrm{nozzle}} = 1.78$), indicating enhanced entrainment capacity due to the stronger vortex ring generated by the flexible nozzle. Here, $I_h$ and $\dot{V}$ are integrated over $r < 2D$, where $D$ is the internal diameter of the nozzle.

Energy partitioning between elastic and kinetic components was analyzed to clarify the transfer of energy from the deforming nozzle to the ejected fluid (Fig.~\ref{sif_piv}, I). The kinetic energy of the ejected fluid was determined by integrating the term $(1/2)\rho (u^2+v^2)$, where $u$ and $v$ are the horizontal and vertical velocity components, respectively, over the downstream control volume ($y < 0$), under the assumption of axisymmetric flow: $E_k = \int_V (1/2)\rho (u^2+v^2) \, dV$, where $dV = 2\pi r \, dx \, dy$ is the axisymmetric volume element. The elastic strain energy stored in the nozzle wall during deformation was estimated using thin-wall cylinder theory by integrating along the nozzle length: $U = \int_0^L (Eh \cdot \pi D \cdot \epsilon^2) / [2(1-\nu)] \, dy$, where $Eh$ is the structural stiffness, $D$ is the local diameter, $\epsilon = (D - D_0)/D_0$ is the local diametric strain relative to the initial diameter $D_0$, and $\nu$ is Poisson's ratio. For the nozzle with intermediate stiffness ($E h = 14.4$~N/m), the elastic strain energy peaks at $t/T = 1.0$ (Fig.~\ref{sif_piv},~I). This temporal alignment between nozzle contraction and jet acceleration maximizes energy transfer and underlies the observed power amplification.

\rev{The impulse reported in this study is the time integral of the force that the nozzle exerts on the fluid, $I=\int_0^{t}F\,dt'$, and each method evaluates it as follows. In the PIV experiments it is the hydrodynamic impulse of the measured vorticity field, $I_h=\tfrac{\rho}{2}\int\mathbf{x}\times\boldsymbol{\omega}\,dV=\pi\rho\int r^{2}\omega_z\,dA$, which equals this integral for a jet started from rest. In the 3D-CFD it is the integral of the momentum flux and of the over-pressure over the exit plane $A$~\cite{krueger2003significance, gao2020jfm}, $I_h=\rho\int_0^{t}\!\int_A u_0|u_0|\,dS\,dt'+\int_0^{t}\!\int_A(p_0-p_\infty)\,dS\,dt'$, where $u_0$ and $p_0$ are the axial velocity and the pressure on $A$. In the load-cell experiments it is the same integral over the exit plane, obtained from the measured force as $I_{\mathrm{jet}}=\int_0^{t_{\mathrm{end}}}F_{\mathrm{loadcell}}\,dt-m_sV_c-\rho A\ell V_c$ (Section~\ref{si:force_measurement}). The reduced-order model retains the momentum flux alone, with a uniform velocity over a fixed exit area, and gives the ratio of the flexible to the rigid nozzle directly, $\hat{I}=\int_{q_{\mathrm{out}}>0}q_{\mathrm{out}}^{2}\,dt\big/\!\int q_{\mathrm{in}}^{2}\,dt$, because its inlet flow is the outflow of the rigid nozzle. This ratio is dimensionless, whereas the impulses of the other three methods are in N\,s and are divided by the value of the rigid nozzle under the same jet condition.}

\rev{The PIV, 3D-CFD and model impulses are evaluated at $t = 4T$ after the onset of the pulse, the PIV value as the mean over $3T \le t \le 5T$, and the load-cell impulse at the end of the push, $t_{\mathrm{end}}$. In the model the outflow has ended before $4T$ for $0.17 \le \tau/T \le 1.5$, so this time gives the same value as the integral over $q_{\mathrm{out}}>0$.}

\rev{The experimental points of Fig.~\ref{f_mechanism}E are the mean of $I_h(t)$ over $3T \le t \le 5T$, a window that is fixed in advance, identical for both nozzles and centered on $4T$ (Fig.~\ref{sif_common_window}, A). Evaluated strictly at $t = 4T$, the eleven ratios change by between $-13\%$ and $+10\%$, with a mean change of $-1\%$ (Fig.~\ref{sif_common_window}, D). The changes are of the size of the fluctuation of $I_h$ about its plateau, because a value taken at one instant retains the fluctuation that the average smooths out. The largest ratio moves from 4.8 to 4.6, and the optimum stays at $\tau/T \approx 0.2$ to 0.4 (Fig.~\ref{sif_common_window}, C). The reduced-order model gives the same value at $4T$ (Fig.~\ref{sif_common_window}, B).}

\FloatBarrier
\section{Force measurement}
\label{si:force_measurement}
\begin{figure*}[!htbp]
    \centering
    \includegraphics[width=\textwidth]{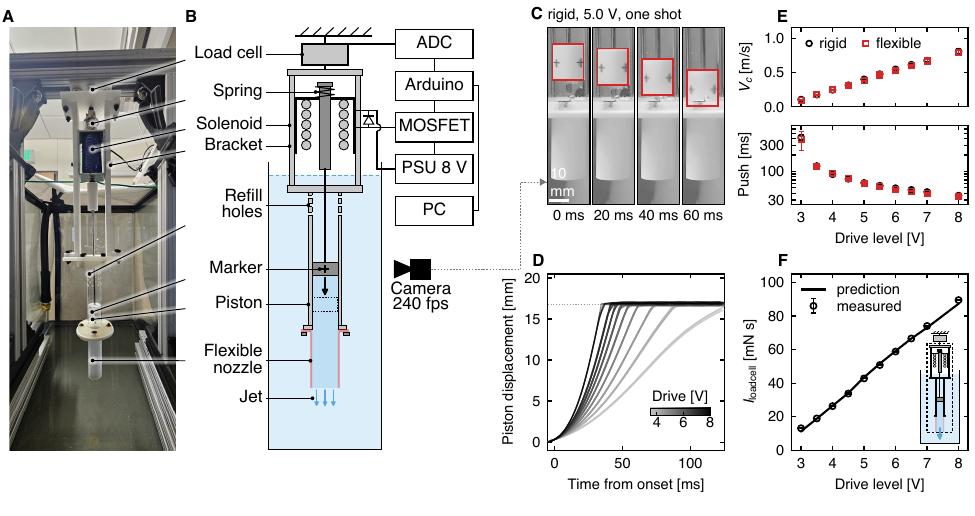}
    \caption{\rev{\textbf{The load-cell rig and the rigid-nozzle impulse.}}
    \rev{(\textbf{A})~The rig. The load cell carries a bracket that holds the solenoid and, below it, the transparent cylinder in the water tank, so that the whole assembly hangs from the load cell. The plunger, pushed back by a return spring, drives the piston ($D = 18.6$~mm, $S = 16.7$~mm) through a thin rod, and the flexible nozzle is clamped by its own flange under the cylinder flange.}
    \rev{(\textbf{B})~Schematic of the rig (flexible nozzle in red) with its drive and recording electronics, a fixed 8~V supply switched by a MOSFET under the PWM of an Arduino, with a flyback diode across the coil, the load cell read by an ADS1256 ADC through the Arduino into the PC, and the 240~fps side camera, which tracks the cross marked on the piston. The piston is drawn at the start of its stroke $S$ and dashed at its end.}
    \rev{(\textbf{C})~One rigid-nozzle shot at 5.0~V, 0 to 60~ms after the onset, red boxes \revred{mark} the tracked template of the piston.}
    \rev{(\textbf{D})~Piston displacement of every filmed rigid shot that reached the end stop, 3.5 to 8~V (85 shots, darker for higher drive).}
    \rev{(\textbf{E})~Impact speed $V_c$ and push duration against drive level, mean $\pm$ SD, rigid (black) and flexible (red).}
    \rev{(\textbf{F})~Rigid nozzle, measured $I_{\mathrm{loadcell}}$ (circles, with the standard error of the mean, SEM, inside the symbols) and its prediction (Eq.~\ref{eq:force_impulse_relation}). Inset, the rig of (\textbf{B}) with the control volume of the momentum balance (dashed box), which encloses everything suspended from the load cell and ends at the nozzle exit.}
    }
    \label{sif_force_rig}
\end{figure*}

\rev{The force and the impulse of single pulsed jets from a rigid and a flexible nozzle were measured with a 5~kg load cell (FBF~333 series, KTOYO) in a separate solenoid-driven rig. The rig and its schematic are shown in Fig.~\ref{sif_force_rig}, A and B. A solenoid-driven plunger and piston (stroke $S=16.7$~mm) pushes water through either a rigid or silicone nozzle of the same length ($L=50$~mm) and inner diameter ($D=18.6$~mm). The load cell supports the entire submerged assembly, so the measured impulse contains the momentum of the moving solid, the water column, and the exiting jet (Eq.~\ref{eq:force_impulse_balance}). The solenoid was driven at 3--8~V, and force was recorded at an effective sampling rate of 2224~Hz. The load cell was calibrated with weights of 1 to 4~g. Three runs were performed in sequence, namely an empty-cylinder run in air, a rigid-nozzle run and a flexible-nozzle run in water, with repeated shots at each drive level.}

\rev{A 240~fps side-view camera tracked the piston displacement relative to the fixed flange (Fig.~\ref{sif_force_rig}, C and D), providing the onset time, push duration (about 126 to 36~ms from 3.5 to 8~V), and impact speed $V_c$ (about 0.18 to 0.81~m/s, Fig.~\ref{sif_force_rig}, E). In terms of the time ratio, this range corresponds to $\tau/T=0.22$ to 0.75, where $\tau=25$~ms is the response time of the silicone nozzle and $T$ is the time in which the piston covers 90\% of its stroke. Each video record was paired with its load-cell record. Force traces were re-zeroed at stroke onset, and the impulse $I_{\mathrm{loadcell}}$ was integrated from the onset to $t=t_{\mathrm{end}}$, the end of the push before the end-stop collision and subsequent ringing. Failed shots and outliers were removed, so that each $I_{\mathrm{loadcell}}$ is the mean of $n = 9$ or 10 shots from 3.5 to 7~V and of $n = 5$ shots at 8~V.}

\rev{The measured impulse $I_{\mathrm{loadcell}}$ was corrected using the momentum balance of a fixed control volume that encloses everything suspended from the load cell (dashed box in the inset of Fig.~\ref{sif_force_rig}, F). For the velocity component $u$ along the jet,}

\begin{equation}
\color{black}
\frac{d}{dt}\int_{\mathrm{CV}}\rho u\,d\mathcal{V}+\int_{\mathrm{CS}}\rho u\,(\mathbf{u}\cdot\mathbf{n})\,dA=\sum F_{\mathrm{ext}}=F_{\mathrm{loadcell}}-\int_A\left(p-p_\infty\right)dA.
\label{eq:force_cv_momentum}
\end{equation}

\noindent \rev{The two terms on the left are the rate of change of the momentum inside the control volume (CV) and the momentum flux through its surface (CS). The external forces along the jet reduce to the load-cell force $F_{\mathrm{loadcell}}$, positive for a thrust, and the over-pressure $p-p_\infty$ on the nozzle exit plane of area $A$~\cite{krueger2003significance}, which acts against the jet, with $p_\infty$ the hydrostatic pressure. Weight and buoyancy are removed by the zeroing, friction on the inner walls acts between two parts of the control volume and cancels, like every internal force, and the refill flow carries no net momentum, since it is perpendicular to the axis and the refill holes are arranged axisymmetrically. Integrating Eq.~(\ref{eq:force_cv_momentum}) over the push, from rest at $t=0$ to $t=t_{\mathrm{end}}$, where the piston velocity $V$ equals \revred{its terminal value} $V_c$, and moving the over-pressure to the left-hand side gives}

\begin{equation}
\color{black}
\left(m_s+\rho A\ell\right)V_c+\int_0^{t_{\mathrm{end}}}\!\!\int_A\left[\rho u^2+\left(p-p_\infty\right)\right]dA\,dt=I_{\mathrm{loadcell}}.
\label{eq:force_impulse_balance}
\end{equation}

\noindent \rev{Here the first term is the momentum inside the control volume, with $m_sV_c$ for the moving solid and $\rho A\ell V_c$ for the incompressible water, $\bigl[\int_{\mathrm{CV}}\rho u\,d\mathcal{V}\bigr]_{t_{\mathrm{end}}}=m_sV_c+\rho A\ell V_c$. The water column has the length $\ell$, measured from the refill holes to the exit of the rigid nozzle. The second term is defined as the impulse of the jet, $I_{\mathrm{jet}}=\int_0^{t_{\mathrm{end}}}\!\int_A\left[\rho u^2+\left(p-p_\infty\right)\right]dA\,dt$, the momentum that has left through the surface together with the impulse of the over-pressure, both on the exit plane~\cite{krueger2003significance, gao2020jfm}. The right-hand side is the measured impulse, $I_{\mathrm{loadcell}}=\int_0^{t_{\mathrm{end}}}F_{\mathrm{loadcell}}\,dt$, and Eq.~(\ref{eq:force_impulse_balance}) becomes}

\begin{equation}
\color{black}
I_{\mathrm{jet}}=\underbrace{I_{\mathrm{loadcell}}}_{\text{measured}}-\underbrace{m_sV_c}_{\text{moving solid}}-\underbrace{\rho A\ell V_c\rlap{.}}_{\text{water column}}
\label{eq:force_impulse_relation}
\end{equation}

\begin{figure*}[!htbp]
    \centering
    \includegraphics[width=\textwidth]{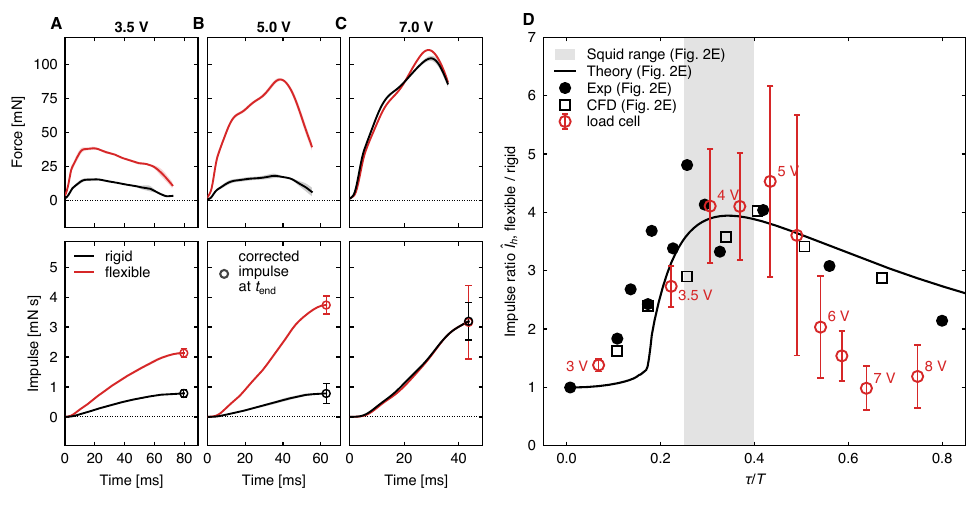}
    \caption{\rev{\textbf{Load-cell force and impulse histories of the rigid and the flexible nozzle.}}
    \rev{(\textbf{A}--\textbf{C})~\revred{Measured force (top) and impulse (bottom) during the push at 3.5, 5.0 and 7.0~V, each scaled by $I_{\mathrm{jet}}/I_{\mathrm{loadcell}}$ at $t_{\mathrm{end}}$}, means of the kept trials ($n \ge 8$) with shading for $\pm1$~SD of those trials\revred{, scaled by the same factor (the spread of the traces only), and open circles for $I_{\mathrm{jet}}$ at $t_{\mathrm{end}}$ with bars for its uncertainty}.}
    \rev{(\textbf{D})~Flexible-to-rigid ratio of $I_{\mathrm{jet}}$ (red open circles) over the data of Fig.~\ref{f_mechanism}, E, redrawn from its source data (filled circles PIV, open squares CFD, line the model, gray band the squid range).}
    }
    \label{sif_force_histories}
\end{figure*}

\rev{The moving-solid term was measured directly in the empty-cylinder run, for which Eq.~(\ref{eq:force_impulse_relation}) reduces to $I_{\mathrm{loadcell}}=m_sV_c$. Its impulse was linear in $V_c$, and the fitted line ($R^2=0.996$), whose slope gives $m_s=82.4$~g, was used for $m_sV_c$. In the rigid-nozzle run in water, $I_{\mathrm{jet}}$ was taken as the momentum flux of a uniform exit velocity equal to the piston velocity $V$, $\rho A\int_0^{t_{\mathrm{end}}}V^2\,dt$, which leaves $\ell$ as the only unknown of Eq.~(\ref{eq:force_impulse_relation}). A fit to the measured impulse (Fig.~\ref{sif_force_rig}, F) gave $\ell=83.9$~mm, which is shorter than the observed water-column length of 91.4~mm because the inflow through the radial refill holes becomes axial only after an entrance length. We then compared the three terms of Eq.~(\ref{eq:force_impulse_relation}), which are the moving solid, the water column and the jet. At 5~V they contributed 75.7\%, 21.0\%, and 3.3\% of their sum, respectively, and at 7~V they contributed 75.9\%, 21.0\%, and 3.1\%, showing that the jet accounts for only a few percent of the raw load-cell impulse. Because the jet term is only about 3\% of the measured impulse, the experiment required the following measures against electrical noise, vibration and shot-to-shot scatter. A MOSFET under pulse-width modulation set the drive level accurately from a fixed supply, at 31~kHz so that the switching noise lies far above the measurement band, and a flyback diode across the coil suppressed its back electromotive force (Fig.~\ref{sif_force_rig}, B). Vibration of the table was minimized, the interval between shots was kept constant, 300 shots were recorded over the three runs, and the two nozzle runs followed each other on the same rig, with only the nozzle carefully exchanged. \revred{The standard error of the mean of the measured impulse over the trials is at most 0.5\% of the impulse, under one sixth of the jet term.}}

\revred{Fig.~\ref{sif_force_histories}, A--C shows the force and impulse histories of the two nozzles at three drive levels. At 3.5 and 5.0~V the flexible nozzle gives a larger force and impulse than the rigid nozzle, so the deformation of the nozzle amplifies the jet impulse. At 5.0~V the force of the flexible nozzle also peaks later in the push, at about 60\% of it against about 25\% at 3.5~V. This delay appears to arise because the jet now accelerates faster relative to the response time of the nozzle ($\tau/T=0.43$ against 0.22), so the wall responds later in the push. At 7.0~V the two nozzles give similar histories. The flexible nozzle appears to store a little energy early in the push and to return a little of it toward the end, but it is too flexible for so short a push ($\tau/T=0.64$), and the impulse is not amplified. It should be noted that, because Eq.~(\ref{eq:force_impulse_relation}) gives $I_{\mathrm{jet}}$ only at $t_{\mathrm{end}}$, these histories are the measured traces scaled by $I_{\mathrm{jet}}/I_{\mathrm{loadcell}}$ at $t_{\mathrm{end}}$, and their shapes are qualitative. Only their end values are quantitative, and the conclusions rest on these values alone (Fig.~\ref{sif_force_histories}, D).}

\rev{Over all drive levels the flexible nozzle gives the larger corrected impulse up to $\tau/T\approx0.6$, and the resulting flexible-to-rigid ratios reproduce the main trend of the PIV, CFD, and reduced-order-model results (Fig.~\ref{sif_force_histories}, D). These ratios are the gray open circles of Fig.~\ref{f_mechanism}, E. The ratio rises to about 4 over \revred{$\tau/T\approx0.3$--0.45, which overlaps the range reported for squid (0.25--0.40),} and decreases at larger $\tau/T$. Because this rig differs from the PIV experiment in geometry, orientation, and measurement principle, this agreement provides an independent check of the impulse gain. For $\tau/T$ above about 0.5 the load-cell ratios fall below the PIV, CFD and model values. \revred{We emphasize the comparison between the two nozzles (Fig.~\ref{sif_force_histories}, A to C) and their ratio (Fig.~\ref{sif_force_histories}, D) rather than the absolute values.}}

\FloatBarrier
\section{Fluid-structure interaction simulations for understanding jet characteristics through flexible nozzles}\label{si:fsi}

\begin{figure*}[!htbp]
	\centering
	\includegraphics[width=0.9\columnwidth]{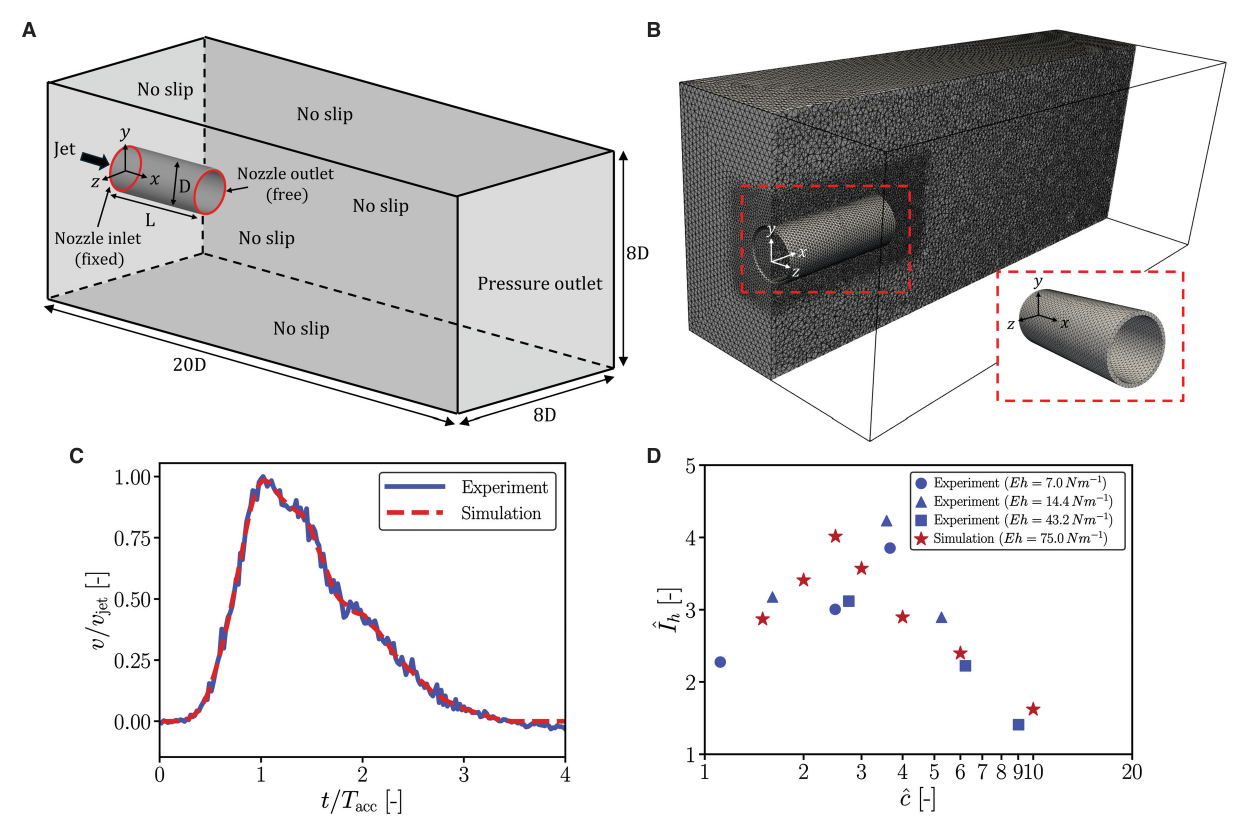}
	\caption{\revred{\textbf{Fluid--structure interaction simulations.}} Schematic representation of (\textbf{A}) the computational domain with boundary conditions; (\textbf{B}) surface and volume mesh for the fluid domain and the inset displaying the mesh considered for the flexible nozzle; (\textbf{C}) time history of jet velocity at the nozzle inlet with jet acceleration time, $T_{acc}$ = 0.05 s and peak jet velocity, $v_{jet}=0.293~\mathrm{m\,s^{-1}}$. (\textbf{D}) Variation in the normalized hydrodynamic impulse ($\hat{I}_h = I_h/I_{h,rigid}$)\rev{, evaluated at the end of the window $0\le t\le4T_{\mathrm{acc}}$}, generated by the flexible nozzle for different dimensionless wave speeds (\revred{$\hat{c} = cT/L = T/\tau$}), computed from experiments and simulations.} 
	\label{sif_cfd_method}
\end{figure*}

The two-way coupled fluid-structure interaction (FSI) simulations were performed using a partitioned strong-coupling approach~\revred{\citep{choi2026computational}}. In the adopted FSI framework, the incompressible fluid flow was simulated using the finite volume method-based {\tt pimpleFoam} solver, available in {\tt OpenFOAM} \citep{weller1998tensorial}, governing structural equations were solved using the finite element method-based open-source code, CalculiX \citep{dhondt2023calculix}, and the fluid-solid coupling was achieved using the preCICE library \citep{chourdakis2022preCICE}. For exchanging information at the fluid-solid interface, the Radial Basis Function - thin plate spline (RBF-TPS) interpolation technique was used, and strong coupling was achieved through the parallel-implicit scheme with the interface quasi-Newton inverse least-squares (IQN-ILS) acceleration method \citep{chourdakis2023openfoam}.

\begin{figure*}[!htbp]
	\centering
	\includegraphics[width=0.6\columnwidth]{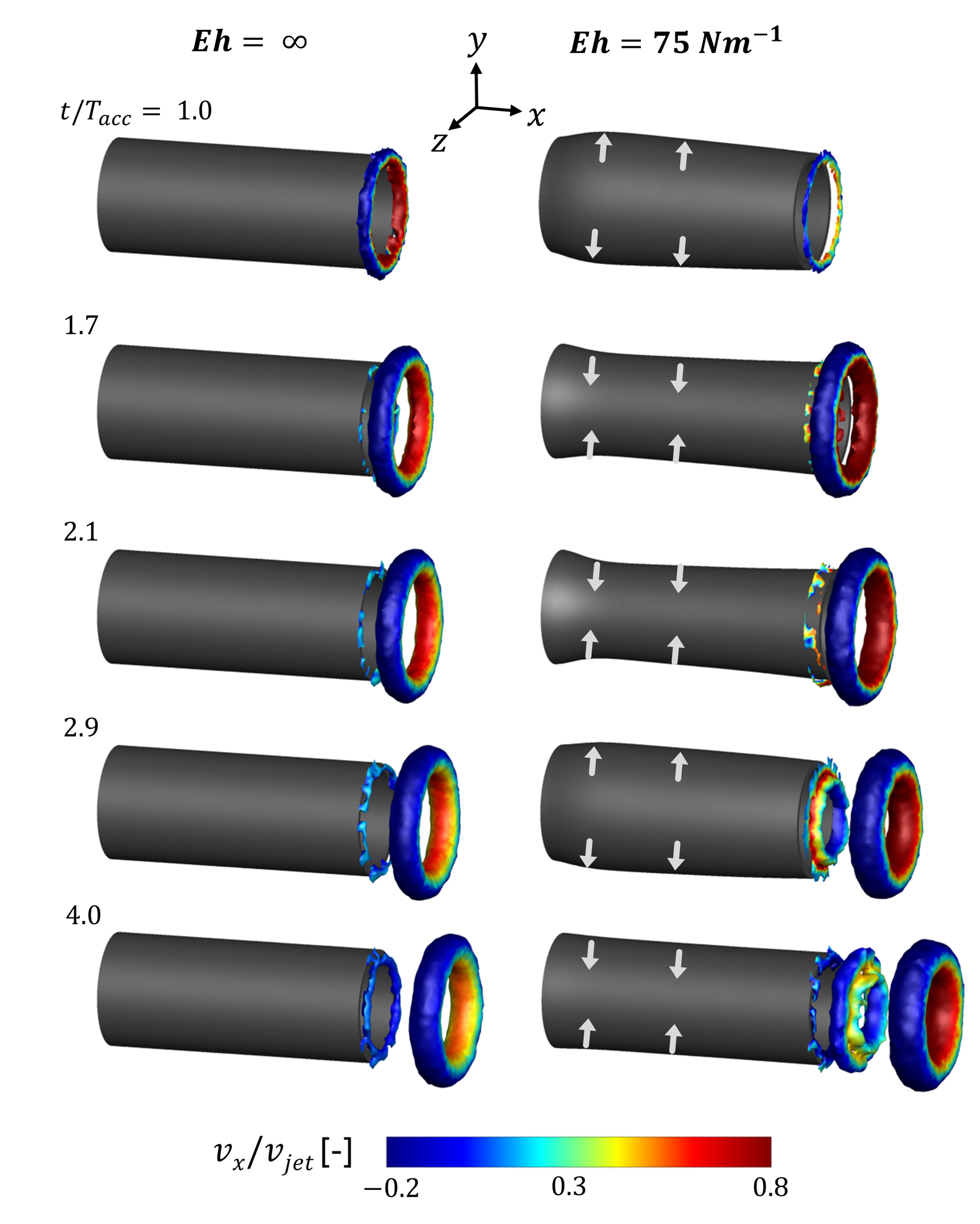}
	\caption{\revred{\textbf{Simulated nozzle deformation and vortex structure.}} Snapshots of 3D nozzle deformation and iso-surface of $Q$ criterion ($Q = 200$) for rigid and flexible nozzle ($Eh=75~\mathrm{N\,m^{-1}}$) colored by the normalized axial component of jet velocity ($\mathbf{v_x}$).} 
	\label{sif_cfd_3dflow}
\end{figure*}

The fluid flow was assumed to be laminar and incompressible, governed by the incompressible Navier-Stokes equations as given by
\begin{equation}
\nabla \cdot \mathbf{v}_f = 0,
\end{equation}
\begin{equation}
\frac{\partial \mathbf{v}_f}{\partial t}
+ \big[(\mathbf{v}_f - \mathbf{v}_m)\cdot\nabla\big]\mathbf{v}_f
= -\frac{1}{\rho}\nabla p + \nu\nabla^2 \mathbf{v}_f,
\end{equation}

\noindent where \revred{$\rho$, $\nu$}, and $\mathbf{v}_f$ are the density, \revred{kinematic} viscosity, and velocity of the fluid, $\mathbf{v}_m$ is the fluid grid point velocity, and $p$ is the fluid pressure. The spatial and temporal discretizations used in the present simulations are second-order accurate. 

The flexible nozzle, considered to be made of a linear-elastic material, exhibits elastic wave propagation when subjected to the pulsed jet. The structural governing equation is given by
\begin{equation}
\rho_s \frac{\partial^2 \mathbf{d}_s}{\partial t^2} - \nabla \cdot \boldsymbol{\sigma}_s = 0,
\end{equation}
\noindent where $\rho_s$ is the density of the solid, $\mathbf{d}_s$ is the displacement of the solid, and $\boldsymbol{\sigma}_s$ is the Cauchy stress tensor. The interface conditions on the fluid-solid interface are
\begin{equation}
\mathbf{v}_f = \mathbf{v}_s,
\end{equation}
\begin{equation}
\boldsymbol{\sigma}_f \cdot \mathbf{n}_f + \boldsymbol{\sigma}_s \cdot \mathbf{n}_s = 0,
\end{equation}
\noindent where $\mathbf{n}_f$ and $\mathbf{n}_s$ are the unit outward normals on the interface between fluid and solid.

To ensure that the numerical solution is independent of the spatio-temporal discretization resolution, the nozzle exit velocity at a probe located at the center for the rigid nozzle was compared using three different grid resolutions (Mesh 1: $0.5 \times 10^6$ cells, Mesh 2: $1.0 \times 10^6$ cells, Mesh 3: $2.0 \times 10^6$ cells), and using three time step resolutions (computed using Mesh 2). Additionally, a grid sensitivity study using the Richardson extrapolation method \cite{celik2008procedure} was carried out. These results indicate satisfactory convergence of the parameters under consideration; hence, Mesh 2 with a time step of $\Delta t = 5 \times 10^{-4}$~s was used for all subsequent simulations.

The evolution of the flow structures as the jet develops and deforms the nozzle is illustrated in Fig.~\ref{sif_cfd_3dflow} \revred{by} iso-surfaces of the $Q$-criterion (the second invariant of the velocity gradient tensor), colored by the normalized axial velocity. Due to the initial expansion phase of the $Eh=75~\mathrm{N\,m^{-1}}$ nozzle, the roll-up of the shear layer and the formation of the primary vortex ring are delayed due to jet entrainment, which can be observed at $t/T_{acc} = 1$. Subsequently, as the nozzle starts to contract, it accelerates the entrained fluid and amplifies the vortex ring strength and the convection speed, as observed at $t/T_{acc} = 4$. Additionally, the unsteady nozzle deformation promotes the formation of secondary vortex rings, reflecting the repeated acceleration\revred{--}deceleration imposed on the jet by the traveling deformation wave.

\FloatBarrier
\section{Theoretical prediction of jet performance using a collapse model}\label{si:model}

\begin{figure*}[!htbp]
    \centering
    \includegraphics[width=0.8\textwidth]{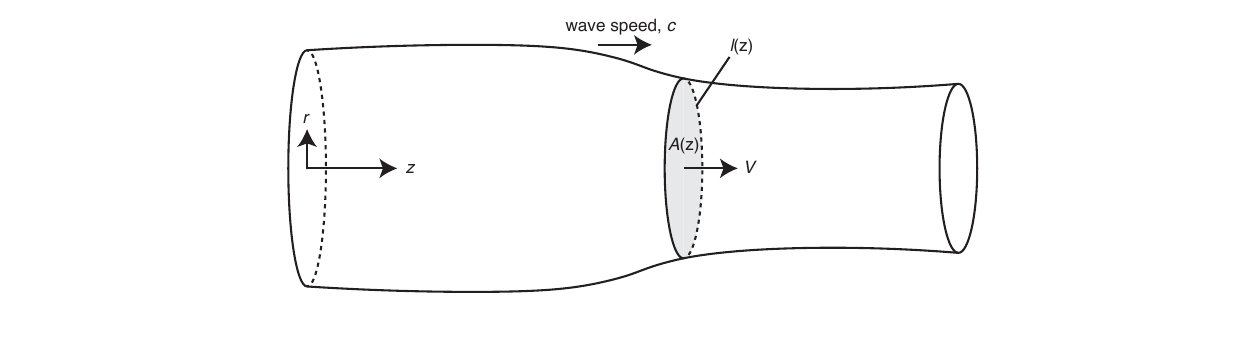}
    \caption{\textbf{Schematic of the 1D tube model.}
    Sketch illustrating tube diameter $D$ (radius $R_0$), axial coordinate $z$, and the wave speed $c$ used in the governing equations and scaling.}
    \label{sif_mathmodel_setting}
\end{figure*}

The internal flow of flexible nozzle can be described by mass conservation and momentum equation for one-dimensional pulsatile flow in a non-permeable compliant tube (Fig.~\ref{sif_mathmodel_setting}) using Reynolds transport theorem~\citep{li2004dynamics,vandevosse2011pulse,hughes1973onedimensional}:

\begin{align}
C\,\frac{\partial p}{\partial t} + \frac{\partial q}{\partial z} &= 0 \label{eq:continuity_1d} \\
\frac{\partial q}{\partial t} + \frac{\partial}{\partial z}\!\left(A\,\overline{v_z^{\,2}}\right) + \frac{A}{\rho}\frac{\partial p}{\partial z} &= \oint_{l}\frac{\tau_w}{\rho}\,dl
\end{align}

\noindent where $p$ is the transmural pressure across the nozzle wall, $q$ is the volume flow rate, $A$ is the cross-sectional area, $z$ is the axial coordinate, $t$ is time, $\rho$ is the fluid density, $v_z$ is the axial velocity, $\tau_w$ is the wall shear stress, $l$ is the perimeter, and $C=\partial A/\partial p$ is the areal compliance of the tube wall.

If we scale following van~de~Vosse and Stergiopulos~(2011) \citep{vandevosse2011pulse} to $v_z = V\,\hat{v}_z$, $p = (\rho V c)\,\hat{p}$, $t = (1/\omega)\,\hat{t}$, $q = (A_0 V)\,\hat{q}$, $A = A_0 \hat{A}$, $z = (\lambda/2\pi)\,\hat{z}$, $\tau_w = (\eta V/R_0)\,\hat{\tau}_w$, and $l = 2\pi R_0\,\hat{l}$, the scaled dimensionless form is

\begin{align}
\frac{\partial \hat{p}}{\partial \hat{t}} + \frac{\partial \hat{q}}{\partial \hat{z}} &= 0\\
\frac{\partial \hat{q}}{\partial \hat{t}} + \frac{V}{c}\frac{\partial}{\partial \hat{z}}\left( \hat{A} \hat{v}_z^{\,2} \right) + \hat{A}\frac{\partial \hat{p}}{\partial \hat{z}} &= \frac{2}{\alpha^{2}} \oint_{\hat{l}} \hat{\tau}_w\, d\hat{l}
\end{align}

\noindent where $V$ is the characteristic velocity, $A_0$ is the reference cross-sectional area, $R_0$ is the reference radius, $\omega$ is the angular frequency, $\eta$ is the dynamic viscosity, $c=\sqrt{E\,h/(\rho\,D)}$ is the wave speed (Moens--Korteweg relation) with Young's modulus $E$, wall thickness $h$, and tube diameter $D$, $\alpha^{2}=\rho\,\omega\,R_0^{2}/\eta$ is the squared Womersley number, and $\lambda=2\pi c/\omega$ is the wavelength. 

Using water ($\rho=10^{3}\ \mathrm{kg/m^{3}}$, $\eta=10^{-3}\ \mathrm{Pa\cdot s}$), $c\approx2\ \mathrm{m/s}$, $V\approx0.2\ \mathrm{m/s}$, $T=0.4\ \mathrm{s}$ ($\omega\approx15.7\ \mathrm{s^{-1}}$), and $D=0.015\ \mathrm{m}$ ($R_0=7.5\times10^{-3}\ \mathrm{m}$), we obtain $V/c\approx0.10$ and $\alpha^{2}=\rho\omega R_0^{2}/\eta\approx8.8\times10^{2}$ so $2/\alpha^{2}\approx2.3\times10^{-3}$. Hence, the convective and wall-shear terms in the non-dimensional momentum equation are $O(10^{-1})$ and $O(10^{-3})$, respectively, and are negligible to leading order. For broader applicability, we temporarily retain the viscous term in the equations and will drop it at the end.

Assuming axisymmetric laminar (Poiseuille) flow in a circular, non-permeable tube, the axial velocity profile is parabolic, $u(r)=2\,\bar{u}\,(1-r^{2}/R^{2})$ with $\bar{u}=q/A$\revred{, which gives} the wall shear $\tau_w=\mu(\partial u/\partial r)|_{r=R}=4\mu\,\bar{u}/R=(4\mu)/(\pi R^{3})\,q$. Hence $(1/\rho)\oint_{l}\tau_w\,dl=(2\pi R/\rho)\,\tau_w=(8\pi\mu)/(\rho A)\,q$, \rev{which is} a linear friction with resistance per unit length $\mathcal{R}=8\pi\mu/A$, so the axial momentum balance becomes $\partial q/\partial t+(A/\rho)\,\partial p/\partial z=-(\mathcal{R}/\rho)\,q$.
Taking $\partial/\partial t$ of the momentum equation and eliminating $p$ using the continuity relation (Eq.~\ref{eq:continuity_1d}) with $C = A/(\rho c^2)$ yields the damped wave equation:

\begin{equation}
\frac{\partial^2 q}{\partial t^2}
 + \frac{8\pi \mu}{\rho A}\,\frac{\partial q}{\partial t}
 = c^2\,\frac{\partial^2 q}{\partial z^2}.
\label{eq:q_wave_laminar}
\end{equation}

\begin{figure*}[!htbp]
    \centering
    \includegraphics[width=\textwidth]{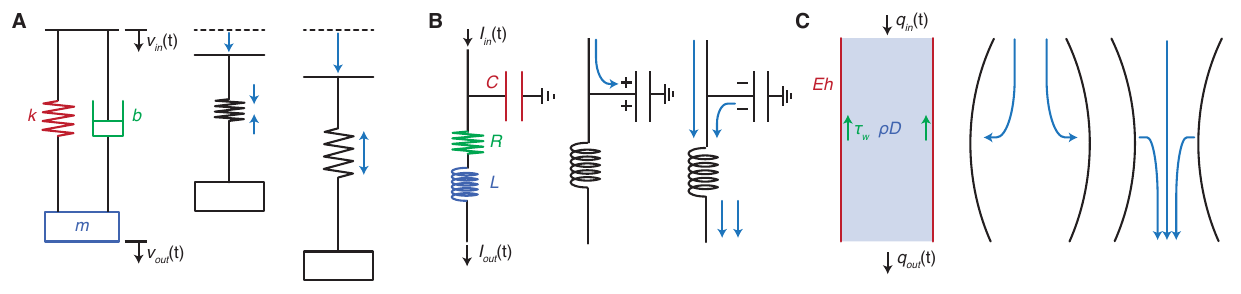}
    \caption{\textbf{Superpropulsion analogies.}
    (\textbf{A})~Spring\revred{--}damper\revred{--}mass resonator. 
    (\textbf{B})~Electrical RLC resonator. 
    (\textbf{C})~Current flexible nozzle system. 
    Right: schematic of superpropulsion as a store\revred{--}and\revred{--}release process, in which elastic energy stored during acceleration is subsequently released to amplify the response. Here, $k$ is the spring constant, $b$ is the damping coefficient, $m$ is the mass, $v$ is velocity, $C$ is capacitance, $R$ is resistance, $L$ is inductance, $Eh$ is structural stiffness, $\tau_w$ is wall shear stress, and $\rho D$ is fluidic mass per unit length inside the flexible nozzle.}
    \label{sif_mathmodel_lumped}
\end{figure*}

\noindent \revred{Eq.~(\ref{eq:q_wave_laminar}) has the form of the telegrapher's equation for the current in an RLC transmission line and of the continuum limit of a spring--damper--mass chain written for the velocity (Fig.~\ref{sif_mathmodel_lumped}). The flow rate $q$ plays the role of current or velocity, $c^{2}$ that of $1/(\mathcal{L}\mathcal{C})$ or $k/m_\ell$, and the viscous coefficient $8\pi\mu/(\rho A)$ that of $\mathcal{R}/\mathcal{L}$ or $c_d/m_\ell$, with inductance $\mathcal{L}$, resistance $\mathcal{R}$, capacitance $\mathcal{C}$, mass $m_\ell$, damping $c_d$ and axial stiffness $k$ per unit length. In the inertia-dominated regime of this study the damping term is negligible, and the three systems reduce to the same undamped oscillator.}

The underlying physics of superpropulsion can be understood through a spring-mass system propelled by a harmonically oscillating plate \citep{celestini2020contactlayer}. The actuating plate undergoes unidirectional motion, accelerating the projectile while compressing the spring, which stores elastic energy (Fig.~\ref{sif_mathmodel_lumped}, A). An optimally tuned elastic projectile delays the force release so that the contact force becomes in-phase with the plate velocity, maximizing the instantaneous mechanical power and resulting in the energy transfer factor up to 300\% when the frequency ratio $f_0/f \approx 1.62$ (where $f_0$ and $f$ mean eigenfrequency and driving frequency, respectively) \citep{celestini2020contactlayer}. \revred{The same mechanism has been identified in elastic projectiles \citep{giombini2022use} and in droplet ejection, including the rapid excretion of the sharpshooter \citep{prl119108001, challita2023superpropulsion}.}

This superpropulsion mechanism is particularly advantageous when the projectile mass $m$ is much smaller than the thrower mass $M$ \rev{(so that} $m/M \ll 1$\rev{)}, a regime where conventional rigid-body throwing becomes inefficient since most kinetic energy remains in the thrower rather than the projectile. 
Squid jet propulsion operates in this regime: a squid ejects \revred{part} of its body mass as a water jet during each mantle contraction, making the jet mass comparable to or smaller than the effective \emph{thrower} mass of the mantle wall.
The flexible nozzle, analogous to the soft contact layer in projectile throwing, provides a compliant interface that can store and release elastic energy \revred{in step with} the driving pressure pulse.

We now restate Eq.~(\ref{eq:q_wave_laminar}) by neglecting viscosity. In the following, we use dot notation for time derivatives (e.g., $\dot{q}=\partial q/\partial t$, $\ddot{q}=\partial^2 q/\partial t^2$) and prime notation for spatial derivatives (e.g., $q'=\partial q/\partial z$, $q''=\partial^2 q/\partial z^2$). Using the continuity relation $(A/\rho c^2)\dot{p}+q'=0$ and the inviscid momentum $\dot{q}+(A/\rho)p'=0$, elimination of $p$ yields the undamped wave equation for flow rate
\begin{equation}
\ddot{q}=c^{2}\,q''.
\label{eq:q_wave_undamped}
\end{equation}

\noindent In nondimensional form, using $\hat{t}=t/T$, $\hat{z}=z/L$, and $\hat{q}=q/q_{\max}$, Eq.~(\ref{eq:q_wave_undamped}) becomes
\begin{equation}
\frac{\partial^{2}\hat{q}}{\partial \hat{t}^{2}}
= \Big(\frac{T}{\tau}\Big)^{2}\,\frac{\partial^{2}\hat{q}}{\partial \hat{z}^{2}},
\label{eq:q_wave_nondim}
\end{equation}
where $T$ is the acceleration time (when the inlet jet velocity reaches its maximum from zero), $L$ is the nozzle length, $q_{\max}$ is the maximum flow rate, and $\tau=L/c$ is the wave transit time through the flexible nozzle. For notational simplicity, we henceforth drop the hat notation ($\hat{\ }$) for nondimensional variables. To derive the lumped input--output model, we consider the boundary conditions $q(0,t)=q_{\mathrm{in}}(t)$ at the inlet (nonhomogeneous Dirichlet) and $q'|_{z=1}=0$ at the outlet (homogeneous Neumann, stress-free). 
The nonhomogeneous inlet is handled by decomposing
\begin{equation}
q(z,t)=q_{\mathrm{in}}(t)+\tilde{q}(z,t),
\label{eq:decomposition}
\end{equation}
where $q_{\mathrm{in}}(t)$ is spatially uniform and $\tilde{q}(z,t)$ captures spatial variations. This ensures homogeneous boundary conditions for $\tilde{q}$: namely $\tilde{q}(0,t)=0$ and $\tilde{q}'(1,t)=0$. Substituting into Eq.~(\ref{eq:q_wave_nondim}) gives
\begin{equation}
\ddot{\tilde{q}} = \Big(\frac{T}{\tau}\Big)^2\tilde{q}'' - \ddot{q}_{\mathrm{in}},
\end{equation}
showing that the inlet acceleration appears as a source term.

The Galerkin method approximates the spatial variation using basis functions satisfying the homogeneous boundary conditions:
\begin{equation}
\tilde{q}(z,t)\approx\sum_n Q_n(t)\phi_n(z).
\label{eq:galerkin}
\end{equation}
For a general basis, the resulting equations are coupled; however, the eigenfunctions of the spatial operator $d^2/dz^2$ decouple these equations due to their orthogonality and the property $\phi_n''=-\kappa_n^2\phi_n$ (where primes denote spatial derivatives with respect to $z$). For our boundary conditions, the eigenfunctions and natural frequencies are
\begin{equation}
\phi_n(z)=\sin\!\Big[\frac{(2n-1)\pi z}{2}\Big], \quad \kappa_n=\frac{(2n-1)\pi}{2}, \quad \omega_n=\frac{(2n-1)\pi}{2}\frac{T}{\tau},
\label{eq:eigenfunctions}
\end{equation}
forming an odd-harmonic series $\omega_1,3\omega_1,5\omega_1,\ldots$ The single-mode approximation is justified by the long-wave condition~\revred{\cite{choi2026computational}}. With the forcing timescale $T$ (jet acceleration time), the frequency ratio for the $n$-th mode is $\omega_n/\omega_{\mathrm{in}}=[(2n-1)/4](T/\tau)$. From our experiments and simulations, $T/\tau\in[1.26,9.99]$, so the second mode frequency exceeds the first by a factor of 3, placing higher modes far from resonance. Away from resonance, amplitude scales as $1/\omega_n^2\propto 1/(2n-1)^2$, so the second mode contributes only $\sim 1/9$ of the first mode amplitude. Furthermore, for smooth inlet signals varying on timescale $T$, Fourier components above $\sim 1/T$ decay rapidly, and since $\omega_2=3\omega_1$, higher modes receive negligible excitation.

Restricting to the first mode, $\tilde{q}(z,t)\approx Q(t)\phi(z)$ with $\phi(z)=\sin(\pi z/2)$, and applying the Galerkin projection (multiplying by $\phi$ and integrating over $z\in[0,1]$), we obtain
\begin{equation}
\ddot{Q} + \omega_0^2 Q = -\frac{4}{\pi}\ddot{q}_{\mathrm{in}}, \qquad \omega_0 = \frac{\pi}{2}\frac{T}{\tau}.
\label{eq:Q_ode}
\end{equation}
The factor $4/\pi$ arises from the normalization integral $\int_0^1 \phi(z)\,dz = \int_0^1 \sin(\pi z/2)\,dz = 2/\pi$, which when squared gives $(2/\pi)^2 = 4/\pi^2$ and combined with the orthogonality relation yields $4/\pi$.
At the outlet ($z=1$), $\phi(1)=1$, therefore, from Eq.~(\ref{eq:decomposition}) we have $q_{\mathrm{out}}=q_{\mathrm{in}}+Q$. Substituting $Q=q_{\mathrm{out}}-q_{\mathrm{in}}$ into Eq.~(\ref{eq:Q_ode}) yields

\begin{equation}
\ddot{q}_{\mathrm{out}} + \omega_0^2 q_{\mathrm{out}} = \omega_0^2 q_{\mathrm{in}} + \Big(1-\frac{4}{\pi}\Big)\ddot{q}_{\mathrm{in}}.
\label{eq:qout_exact}
\end{equation}
The second term on the right-hand side scales as $|\ddot{q}_{\mathrm{in}}|/(\omega_0^2|q_{\mathrm{in}}|)\sim 1/\omega_0^2=(4/\pi^2)(\tau/T)^2$. From our experiments and simulations, $T/\tau\in[1.26,9.99]$, so this ratio lies in the range $[0.004,0.26]$. As an example, for $c\approx 2~\mathrm{m/s}$, $T=0.11~\mathrm{s}$, and $L=40.7~\mathrm{mm}$ (giving $\tau=L/c\approx 0.020~\mathrm{s}$), we obtain $T/\tau\approx 5.4$ and $(4/\pi^2)(\tau/T)^2\approx 0.014$. Even at the lower bound $T/\tau=1.26$, multiplying by the prefactor $(1-4/\pi)\approx -0.27$ gives a net contribution below 7\% of the leading term. Neglecting this term, the lumped input\revred{--}output ODE

\begin{equation}
\ddot{q}_{\mathrm{out}} + \omega_0^2\,q_{\mathrm{out}} = \omega_0^2\,q_{\mathrm{in}}(t),\qquad \omega_0=\frac{\pi}{2}\frac{T}{\tau}.
\label{eq:lumped_io}
\end{equation}

\noindent With this lumped input\revred{--}output ODE, the outlet flow rate $q_{\mathrm{out}}(t)$ can be calculated by solving the ODE with the given inlet profile $q_{\mathrm{in}}(t)$. 

\begin{figure*}[!htbp]
    \centering
    \includegraphics[width=\textwidth]{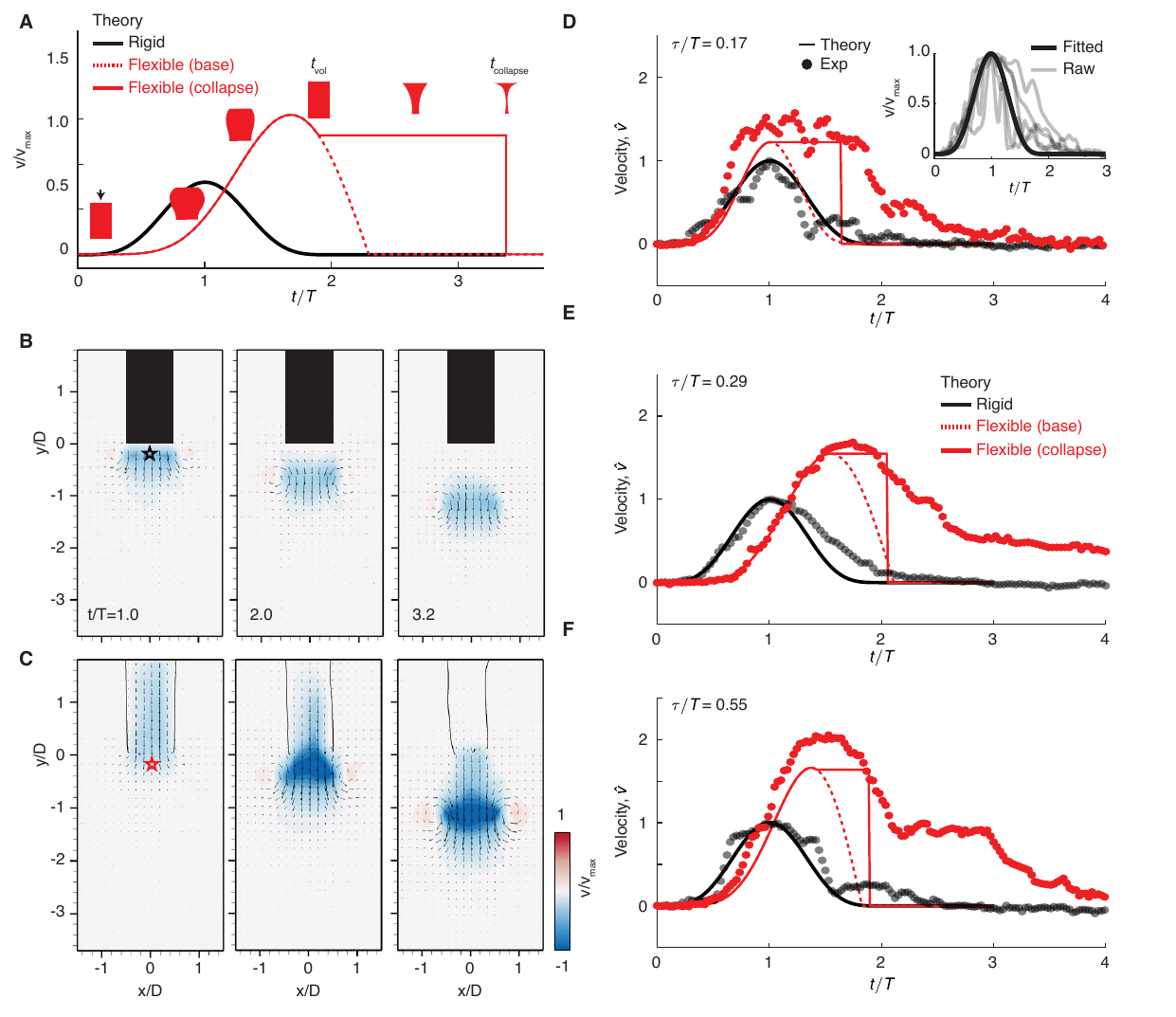}
    \caption{\textbf{Nozzle collapse model and experimental validation.}
    (\textbf{A})~Theoretical exit velocity profiles for rigid (black solid) and flexible (red) nozzles normalized by maximum inlet velocity $V_{\max}$. The flexible nozzle exhibits two phases: a base model (dotted red) representing the pure ODE solution; and a collapse model (solid red) which deviates from the base model for $t_{\mathrm{vol}} < t < t_{\mathrm{collapse}}$. Red schematics illustrate: nozzle expansion and contraction for $t < t_{\mathrm{vol}}$; onset of collapse at $t = t_{\mathrm{vol}}$ when cumulative outflow matches total inflow; and nozzle collapse for $t > t_{\mathrm{vol}}$, ending at $t_{\mathrm{collapse}}$ when the stored nozzle volume $V_{\mathrm{nozzle}}$ is fully exhausted.
    (\textbf{B},~\textbf{C})~Instantaneous velocity fields at $t/T = 1.0$, 2.0, and 3.2 for (\textbf{B}) the rigid and (\textbf{C}) flexible nozzles. Color contours show streamwise velocity $v/v_{\max}$; vectors indicate flow direction. Star markers denote the measurement location for exit velocity. The flexible nozzle produces a more coherent and penetrating vortex ring due to enhanced momentum from wall recoil.
    (\textbf{D}--\textbf{F})~Comparison of theoretical predictions (solid lines) and experimental measurements (dots) of exit velocity histories for three representative response time ratios: $\tau/T = 0.17$ (\textbf{D}), 0.29 (\textbf{E}), and 0.55 (\textbf{F}). Black and red indicate rigid and flexible nozzles, respectively. The collapse model successfully captures the delayed secondary peak in flexible nozzle velocity resulting from nozzle collapse. Here, the black solid line for the rigid nozzle is fitted using Eq.~\ref{eq:inlet_profile}, as shown in the inset of (\textbf{D}), where the raw PIV data (gray) is compared to its fitted curve (black).}
    \label{sif_mathmodel_collapsemodel}
\end{figure*}

The $q_{\mathrm{in}}$ of each flexible nozzle is determined by the $q_{\mathrm{out}}$ of the rigid nozzle, assuming the rigid nozzle has similar values between $q_{\mathrm{in}}$ and $q_{\mathrm{out}}$ (consistent with \citep{choi2024mechanism, choi2022flow}). 
Five different outlet jet velocity profiles from the rigid nozzle, with maximum jet velocities ranging from $v_{\max} = 0.07$ to $0.33$~m/s and jet acceleration times from $T = 0.05$ to \revred{$0.34$}~s, are normalized by $v_{\max}$ and $T$, and fitted with the following parametric family to accurately mimic the piston-generated jet, which exhibits a narrower peak width compared to typical sinusoidal profiles (see the inset of Fig.~\revred{\ref{sif_mathmodel_collapsemodel}D}):

\begin{equation}
q_{\mathrm{in}}(t)=
\begin{cases}
\Big[\dfrac{1-\cos(\pi t)}{2}\Big]^{n}, & 0\le t\le 1,\\
\Big[\dfrac{1-\cos\!\big(\pi (2-t)\big)}{2}\Big]^{n}, & 1< t\le 2,\\
0, & \text{otherwise.}
\end{cases}
\label{eq:inlet_profile}
\end{equation}
The shape exponent $n$ is fitted by least squares on the rising branch $0 < t < 1$, where the major energy transfer occurs, yielding $n = 2.66$ with $R^2 = 0.73$ and RMSE $= 0.15$. The response of the flexible nozzle is sensitive to the shape of the input jet profile, including both the actuation time $T$ and the acceleration history, as characterized by $n$ in this study. For a perfect sinusoid ($n = 1$), the optimal $\tau/T$ is exactly 0.5, since the nozzle resonance precisely matches the jet acceleration period. When $n < 1$, the velocity rises quickly and remains near its maximum longer, effectively decreasing the acceleration time and shifting the optimal $\tau/T$ above 0.5; conversely, when $n > 1$, the velocity stays low for a longer time and then rises sharply, delaying the acceleration time and reducing the optimal $\tau/T$ below 0.5. Velocity profiles in the present experiments and CFD, which use the experimental input profile, consistently exhibit $n > 1$. This explains why the observed optimal $\tau/T = 0.2$--$0.4$ is lower than the idealized sinusoidal prediction of 0.5.

The lumped response solves the nondimensional ODE
\begin{equation}
\ddot{q}_{\mathrm{out}} + \omega_0^{2}\,q_{\mathrm{out}}
= \omega_0^{2}\,q_{\mathrm{in}}(t),\qquad
\omega_0=\frac{\pi}{2}\,\frac{T}{\tau},
\label{eq:lumped_response}
\end{equation}

To calculate the $q_{\mathrm{out}}$ using the lumped response model, the second-order ODE is decomposed into two first-order equations:
\begin{equation}
\begin{cases}
\dot{q} = v\\
\dot{v} = \omega_0^2(q_{\mathrm{in}} - q),
\end{cases}
\end{equation}
At each time step $n$, the numerical scheme updates using a semi-implicit Euler method with the given inlet profile $q_{\mathrm{in}}$ and time step $\Delta t \approx 10^{-5}$, ensuring calculation stability with $\omega_0 \cdot \Delta t \approx 0.0016 \ll 2$:
\begin{equation}
a^n = \omega_0^2(q_{\mathrm{in}}^{n-1} - q^{n-1}),\quad
v^n = v^{n-1} + a^n \Delta t,\quad
q^n = q^{n-1} + v^n \Delta t.
\label{eq:fitted}
\end{equation}

To represent finite-volume jet termination and account for the physical collapse of the flexible nozzle, two models are implemented. The \textit{base} model is the pure ODE solution (Eq.~\ref{eq:lumped_io}) without any collapse adjustment, representing the idealized nozzle response, \rev{shown as the} ideal spring model in Fig.~\ref{sif_mathmodel_lumped}. The \textit{collapse} model builds upon this by identifying the time $t_{\mathrm{vol}}$ when the cumulative outlet volume $\int_0^{t} q_{\mathrm{out}}\,dt'$ first equals and starts to exceed the total inlet volume $\int_0^{t} q_{\mathrm{in}}\,dt'$, marking the onset of nozzle collapse.
From $t_{\mathrm{vol}}$ onward, the stored nozzle volume is expelled at the constant rate $q_0 = q_{\mathrm{out}}(t_{\mathrm{vol}})$ until $t_{\mathrm{collapse}} = t_{\mathrm{vol}} + V_{\mathrm{nozzle}}/q_0$, when the nozzle volume, $V_{\mathrm{nozzle}}$, is fully exhausted. This plateau closure ensures volume conservation while capturing the physical collapse dynamics observed experimentally. The model scans over $\tau/T \in [0.01, 1.5]$. Experimental $\tau/T$ values are computed from nozzle stiffness via $c = \sqrt{Eh/(\rho D)}$ and $\tau = L/c$, using the measured acceleration time $T$ for each piston speed, and compared against CFD simulations ($Eh = 75~\mathrm{N\,m^{-1}}$, $L/D = 2.73$).

\begin{figure*}[!htbp]
    \centering
    \includegraphics[width=\textwidth]{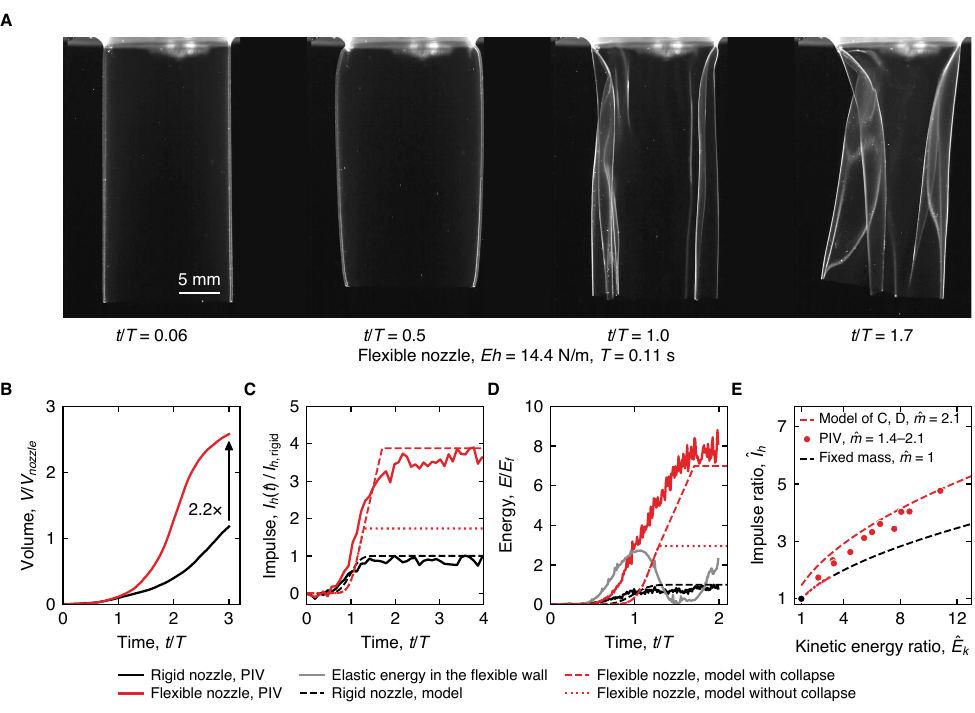}
    \caption{\rev{\textbf{Mass, impulse and kinetic energy of the flexible-nozzle jet relative to the rigid-nozzle jet.}}
    \rev{(\textbf{A})~Frames of the flexible nozzle with $Eh = 14.4$~N/m at $T = 0.11$~s, in which the wall dilates and then collapses~\cite{choi2024mechanism}.}
    \rev{(\textbf{B})~Cumulative volume through a plane one diameter downstream of the exit for the rigid nozzle and the nozzle of (\textbf{A}). The arrow marks the ratio of the two volumes at $t/T = 3$.}
    \rev{(\textbf{C})~Hydrodynamic impulse against time, divided by the rigid value at $t/T = 4$, with the collapse model (dashed) and the model without collapse (dotted), in which the outflow ends once the stroke volume has left the nozzle, at the same $\tau/T = 0.29$.}
    \rev{(\textbf{D})~The same for the jet kinetic energy in units of the kinetic energy of the rigid-nozzle jet, $E_f$, with the elastic energy stored in the flexible wall (gray).}
    \rev{(\textbf{E})~Impulse ratio against kinetic-energy ratio for the eleven flexible nozzles (red circles) and the rigid nozzle (black circle). The lines are $\hat{I}_h = (\hat{m}\hat{E}_k)^{1/2}$ for a fixed mass, $\hat{m} = 1$ (black), and for the collapse model of (\textbf{C}) and (\textbf{D}), $\hat{m} = 2.1$ (red). The model without collapse stays on the fixed-mass line (red dotted).}
    }
    \label{sif_slug_budget}
\end{figure*}

\begin{figure*}[!htbp]
    \centering
    \includegraphics[width=\textwidth]{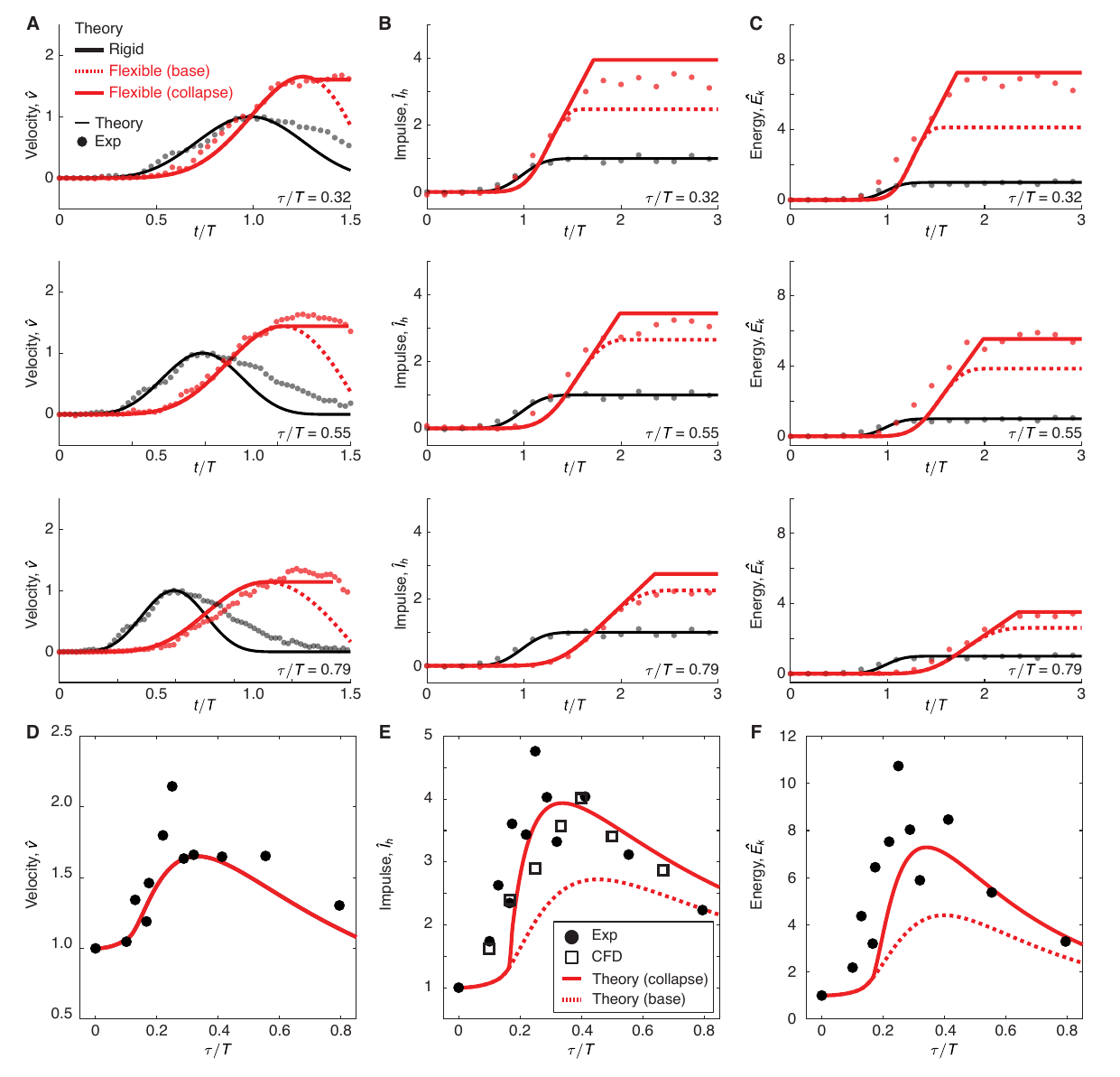}
    \caption{\textbf{Model predictions versus experiments and simulations.}
    (\textbf{A}--\textbf{C})~Time histories of the dimensionless exit velocity $\hat{v}$ (A), hydrodynamic impulse $\hat{I}_h$ (B), and jet kinetic energy $\hat{E}_k$ (C), each normalized by their respective values for the rigid nozzle, are shown for three representative response time ratios: $\tau/T = 0.32$ (top row), 0.55 (middle row), and 0.79 (bottom row). All panels share identical jet conditions (maximum velocity $v_\text{max} = 0.33\,\mathrm{m/s}$ and actuation time $T = 0.05$~s) and nozzle geometry (inner diameter and length), differing only in nozzle modulus ($Eh = 7$--$43.2~\mathrm{N/m}$). Black solid lines show the rigid nozzle prediction fitted using Eq.~\ref{eq:inlet_profile}, as in Fig.~\ref{sif_mathmodel_collapsemodel}; gray scatter points represent rigid nozzle experiments. Red dotted lines denote flexible nozzle predictions without collapse (base model), while red solid lines include the collapse correction. Red scatter points depict flexible nozzle experimental data. The base model captures the phase shift and amplitude modulation, whereas the collapse model further refines the prediction by accounting for the additional flow volume generated by nozzle collapse.
    (\textbf{D}--\textbf{F})~Summary of (\textbf{D}) peak exit velocity $\hat{v}$, (\textbf{E}) jet hydrodynamic impulse $\hat{I}_h$, \rev{and} (\textbf{F}) jet kinetic energy $\hat{E}_k$ as functions of the time ratio $\tau/T$. Filled circles: experiments; open squares: FSI simulations (CFD); \rev{solid} red line: collapse theory; \rev{dotted} red line: base theory. Both impulse and energy exhibit a resonant peak near $\tau/T \approx 0.2$--$0.4$, consistent with the superpropulsion mechanism where optimal frequency matching maximizes energy transfer to the jet.}
    \label{sif_mathmodel_prediction}
\end{figure*}

To quantify the performance of the compliant nozzle, we compute three key metrics that compare the output jet to the input forcing: the ratio of maximum velocity, the hydrodynamic impulse, and the jet kinetic energy.
Observations show that the outlet diameter of the flexible nozzle remains constant throughout the jet cycle before collapse, as ambient pressure is imposed at the outlet. Consequently, the flow rate $q_{\mathrm{out}} = A v$ is approximately proportional to the jet velocity, $q_{\mathrm{out}} \sim v$. Therefore, the velocity ratio $\hat{v} = \max(v_{\mathrm{out}})/\max(v_{\mathrm{in}}) = \max(q_{\mathrm{out}})/\max(q_{\mathrm{in}})$ serves as a proxy for the peak-velocity gain. Similarly, the hydrodynamic impulse $I = \int \rho u^2 A\,dt \sim \int q^2\,dt$ and the kinetic energy $E = \int \rho u^3 A\,dt \sim \int q^3\,dt$ provide proxies for momentum and energy amplification, respectively. Accordingly, the velocity ratio $\hat{v}$, normalized impulse $\hat{I}$, and normalized kinetic energy $\hat{E}_k$ are computed as follows:

\begin{equation}
\hat{v}=\frac{\max(q_{\mathrm{out}})}{\max(q_{\mathrm{in}})},
\end{equation}
\begin{equation}
\hat{I}=\frac{\int_{q_{\mathrm{out}}>0}q_{\mathrm{out}}^{2}\,dt}{\int q_{\mathrm{in}}^{2}\,dt},
\end{equation}
\begin{equation}
\hat{E}_k=\frac{\int_{q_{\mathrm{out}}>0}q_{\mathrm{out}}^{3}\,dt}{\int q_{\mathrm{in}}^{3}\,dt}.
\end{equation}
The integration is restricted to intervals where $q_{\mathrm{out}} > 0$, thereby excluding suction (negative flow) events in the base model and terminating at nozzle collapse ($t = t_{\mathrm{collapse}}$) in the collapse model. \revred{The value reported in the main text is taken at $t = 4T$, which equals this integral for $0.17 \le \tau/T \le 1.5$ (Section~S9).}

The necessity of the collapse model becomes evident when comparing its predictions with those of the base model (Fig.~\ref{sif_mathmodel_collapsemodel}). While the base model successfully captures the initial phase shift and amplitude modulation of the exit velocity, it underestimates the plateauing region at $t > t_{\mathrm{vol}}$ that arises from nozzle wall collapse. The collapse model incorporates this additional flow volume, yielding improved agreement with experimental measurements (Fig.~\ref{sif_mathmodel_collapsemodel}, D--F). The flow field visualizations (Fig.~\ref{sif_mathmodel_collapsemodel}, B and C) confirm that the flexible nozzle produces a more coherent and energetic vortex ring compared to the rigid case, consistent with the enhanced momentum predicted by the collapse model.

The predictive capability of both models is systematically evaluated across a range of response time ratios in Fig.~\ref{sif_mathmodel_prediction}. For the time-resolved quantities (Fig.~\ref{sif_mathmodel_prediction}, A--C), the collapse model (solid red lines) consistently outperforms the base model (dotted lines), particularly in capturing the magnitude and timing of the secondary velocity peak. The improvement is most pronounced at $\tau/T \approx \rev{0.3}$ where resonant amplification is strongest. The summary plots (Fig.~\ref{sif_mathmodel_prediction}, D--F) demonstrate that both models predict the resonant peak in impulse and energy near $\tau/T = 0.2$--$0.4$, but the collapse model provides quantitatively better agreement with experimental data (filled circles) and FSI simulations (open squares). These results validate the collapse model as a necessary refinement for accurate prediction of flexible nozzle performance in the superpropulsion regime.

\rev{The profile of the input jet was not varied systematically in this study, so the dependence of the optimum on the profile shape described above was not tested directly. The experiments nevertheless cover a range of profiles from four actuators and one simulation, namely the motor-driven piston with five velocity profiles (Fig.~\ref{sif_piv}, A), the solenoid-driven piston of the force measurement at drive levels from 3 to 8~V (Fig.~\ref{sif_force_rig}), the diaphragm pump at twelve frequencies between 10 and 80~Hz (Fig.~\ref{sif_repeated_pulse} and Fig.~\ref{sif_aerialjet}, G), the miniature pump of the small boat (Fig.~\ref{sif_smallboat}) and the inlet profile of the FSI simulation (Fig.~\ref{sif_cfd_method}). With the motor-driven piston and with the pump at 24~Hz, the gain of the flexible nozzle peaks at a similar $\tau/T$ of 0.2 to 0.4 (Fig.~\ref{f_mechanism}, E and Fig.~\ref{f_application}, \revred{N--P}), and the squid, which contract freely, lie in the same range (Fig.~\ref{f_squid}, J). How the profile changes the optimum $\tau/T$ and the gain in impulse remains to be quantified in follow-up work.}

\rev{The gain in impulse can be read in terms of the mass and the velocity of the jet. If the rigid nozzle expels a slug of mass $m$ at velocity $U$, its impulse is $mU$ and its kinetic energy is $\tfrac{1}{2}mU^{2}$. If the flexible nozzle expels $\hat{m}$ times this mass at velocity $U_e$, the impulse ratio is $\hat{I}_h = \hat{m} U_e/U$ and the kinetic-energy ratio is $\hat{E}_k = \hat{m} (U_e/U)^{2}$, so that $\hat{I}_h = (\hat{m} \hat{E}_k)^{1/2}$ and $\hat{m} = \hat{I}_h^{2}/\hat{E}_k$. If the mass were fixed ($\hat{m} = 1$) and the wall only stored an elastic energy equal to $\eta$ times the kinetic energy of the rigid-nozzle jet and returned it, then $\hat{E}_k = 1 + \eta$ and $\hat{I}_h = (1+\eta)^{1/2}$, and a fourfold impulse would require a sixteenfold kinetic energy. The flexible nozzle expels more mass, because its wall collapses within the pulse and adds the fluid inside the nozzle to the jet. With about twice the mass at about twice the velocity ($\hat{m} \approx 2$), a fourfold impulse corresponds to an eightfold kinetic energy, close to the measured values (Fig.~\ref{f_mechanism}, L and M).}

\revred{The wall is not the only source of this energy. Its stored elastic energy peaks near $2.7E_f$ (gray line in Fig.~\ref{sif_slug_budget}, D), which is less than the increase in the kinetic energy of the jet. The remainder is additional work done by the piston. The piston motion is prescribed, so the energy input is not fixed, and the flexible nozzle changes the pressure against which the piston works. In the slug argument, $\eta$ is therefore the total additional energy delivered to the jet, of which the elastic return is one part.}

\rev{The measurements support this added mass (Fig.~\ref{sif_slug_budget}). The wall of the flexible nozzle dilates and then collapses within the pulse (Fig.~\ref{sif_slug_budget}, A and Fig.~\ref{f_mechanism}, B), and the volume carried through a plane one diameter downstream of the exit is 2.2 times that of the rigid nozzle at $t/T = 3$ (Fig.~\ref{sif_slug_budget}, B and Fig.~\ref{sif_piv}, H). The comparison is made at $t/T = 3$ because the ambient fluid that the jets themselves entrain later dilutes the ratio. The measured impulse and kinetic energy give $\hat{m} = 1.4$ to 2.1 for the eleven flexible nozzles and 1.9 to 2.1 for the three nozzles whose impulse ratio is 4 or more (Fig.~\ref{sif_slug_budget}, E). For the nozzle of Fig.~\ref{sif_slug_budget}, A, the impulse at $t/T = 4$ is 3.6 times that of the rigid nozzle (4.0 times as the mean over $3 \le t/T \le 5$, Fig.~\ref{f_mechanism}, E) and the jet kinetic energy is 7.7 to 8.8 times. The collapse model comes close to both (Fig.~\ref{sif_slug_budget}, C and D). If the outflow is instead ended once the stroke volume has left the nozzle, so that the mass is fixed, the model underestimates them by about 50 to 65\% and stays on the line $\hat{m} = 1$ (Fig.~\ref{sif_slug_budget}, C to E). The collapse closure thus represents the added mass.}

\FloatBarrier
\section{Repeated pulse jet generation}\label{si:repeated}

Repeated pulse jets were produced using a commercial diaphragm pump (Chapin 12V / 1.0 GPM) connected to the nozzle outlet (Fig.~\ref{sif_repeated_pulse}, A). The pump consists of a flexible diaphragm membrane, inlet and outlet check valves, and a cam-driven actuation mechanism that converts rotary motor motion into a pulsating flow (Fig.~\ref{sif_repeated_pulse}, B). Power was supplied by a 300~W DC power supply (SPS-3010, JESVERTY) set between 2 and 12~V. Jet velocity was characterized using particle image velocimetry (PIV) with high-speed imaging and a laser sheet (see Fig.~\ref{sif_repeated_pulse}, C). The centerline velocity was measured at $x = 0.1D$, $y = 0$, as indicated in Fig.~\ref{sif_repeated_pulse}, C. Pump frequencies ranging from 10 to 80~Hz were set by adjusting the supplied voltage (see Fig.~\ref{sif_repeated_pulse}, D and E), resulting in mean jet velocities from 0.1 to 0.4~m/s (Fig.~\ref{sif_repeated_pulse}, F). With increasing frequency, the normalized velocity fluctuation (defined as \revred{$\sqrt{\tfrac{1}{N}\sum_{i=1}^{N}(u_i - u_\mathrm{mean})^2} / u_\mathrm{mean}$} where $u_i$ is centerline jet velocity at each time step \revred{and $N$ the number of time steps}) decreased, indicating reduced pulsatile behavior at higher frequencies. For consistency in subsequent experiments, a frequency of 24~Hz was selected for repeated pulse jet generation in the aerial jet (Fig.~\ref{sif_aerialjet}, C--H), boat (Fig.~\ref{sif_boat}), and mixing (Fig.~\ref{sif_mixing}) tests.

\section{Aerial jet experiment}\label{si:aerial}
\begin{figure*}[!htbp]
    \centering
    \includegraphics[width=\textwidth]{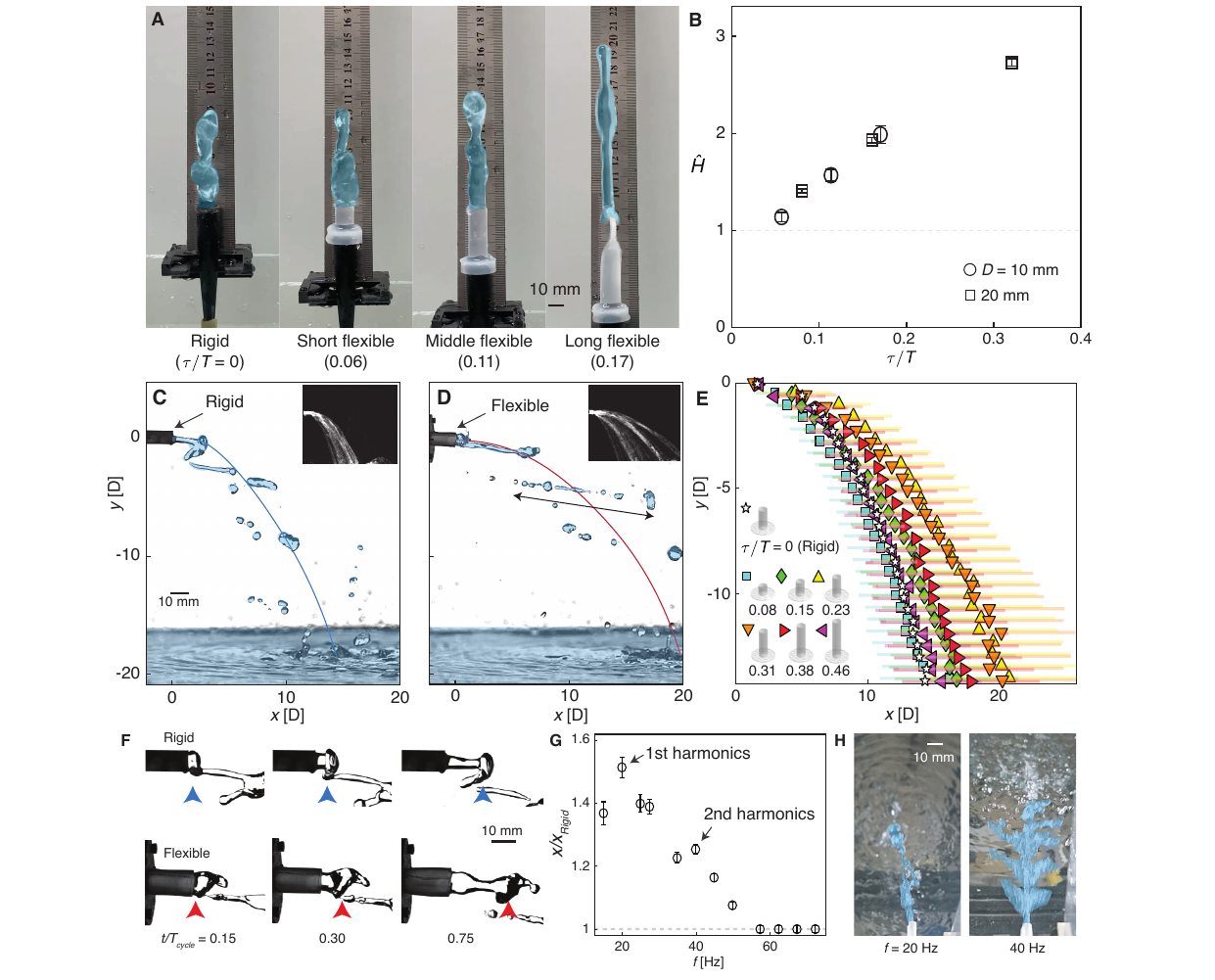}
    \caption{\textbf{Aerial jet experiments using flexible and rigid nozzles.}
    (\textbf{A})~Representative snapshots showing the maximum jet height for nozzles with varying $\tau/T$ values.
    (\textbf{B})~Normalized jet height, $\hat{H}=H_f/H_r$ (where $H_f$ is the jet height for the flexible nozzle and $H_r$ is for the rigid nozzle) as a function of the time ratio $\tau/T$. A total of N = 49 trials were conducted across n = 8 nozzle configurations \revred{(six flexible nozzles, plotted, and one rigid nozzle per diameter, which sets $\hat{H} = 1$)}. \revred{$H_f$ and $H_r$ are the means of the maximum jet heights over the trials of each nozzle. Error bars are the standard error over trials.}
    (\textbf{C}--\textbf{D})~Horizontal jet ejection from (\textbf{C}) a rigid nozzle ($\tau/T = 0$) and (\textbf{D}) a flexible nozzle ($\tau/T = 0.31$), respectively. The flexible nozzle produces a longer jet range; the double-headed arrow indicates the extended ligament and more significant formation of satellite droplets, compared to the rigid nozzle case. \revred{The insets show averaged images of the jets, from which the mean trajectories of (\textbf{E}) are obtained.}
    (\textbf{E})~Comparison of mean jet trajectories obtained via image processing for various $\tau/T$ values; the flexible nozzle ($\tau/T = 0.31$) achieves the greatest jet distance.
    (\textbf{F})~Time series illustrating nozzle deformation and corresponding formation of the liquid jet. Expansion is observed at the onset of the pulse ($t/T_{cycle} = 0.15$--0.75, where $T_{cycle}$ is the period of the pulsatile jet, which is different from the acceleration time of the actuator $T$), followed by contraction that releases stored elastic energy into the jet.
    (\textbf{G})~Horizontal jet distance (normalized by the rigid nozzle) as a function of input jet frequency, revealing resonance peaks at the fundamental ($\sim$20~Hz) and second harmonic ($\sim$40~Hz). Arrows indicate the resonance peaks.
    (\textbf{H})~Top-down views of horizontal jet development from the flexible nozzle at 20~Hz and 40~Hz. Above the resonant frequency, jet dispersion increases markedly due to deformation and nonlinear nozzle dynamics.
    In (\textbf{E}) and (\textbf{G}), error bars represent temporal standard deviations over the time duration of $t/T = 0$--$8{,}000$.
    }
    \label{sif_aerialjet}
\end{figure*}

The performance of the flexible nozzle in aerial jets was tested for both single pulsed jets (upward direction, Fig.~\ref{sif_aerialjet}, A--B) and repeated pulsed jets (horizontal direction, Fig.~\ref{sif_aerialjet}, C--H) across various nozzle conditions with $\tau/T$ values ranging from 0 (rigid) to \revred{0.32 for the single pulses and 0.46 for the repeated pulses}.

\begin{figure*}[!htbp]
    \centering
    \includegraphics[width=\textwidth]{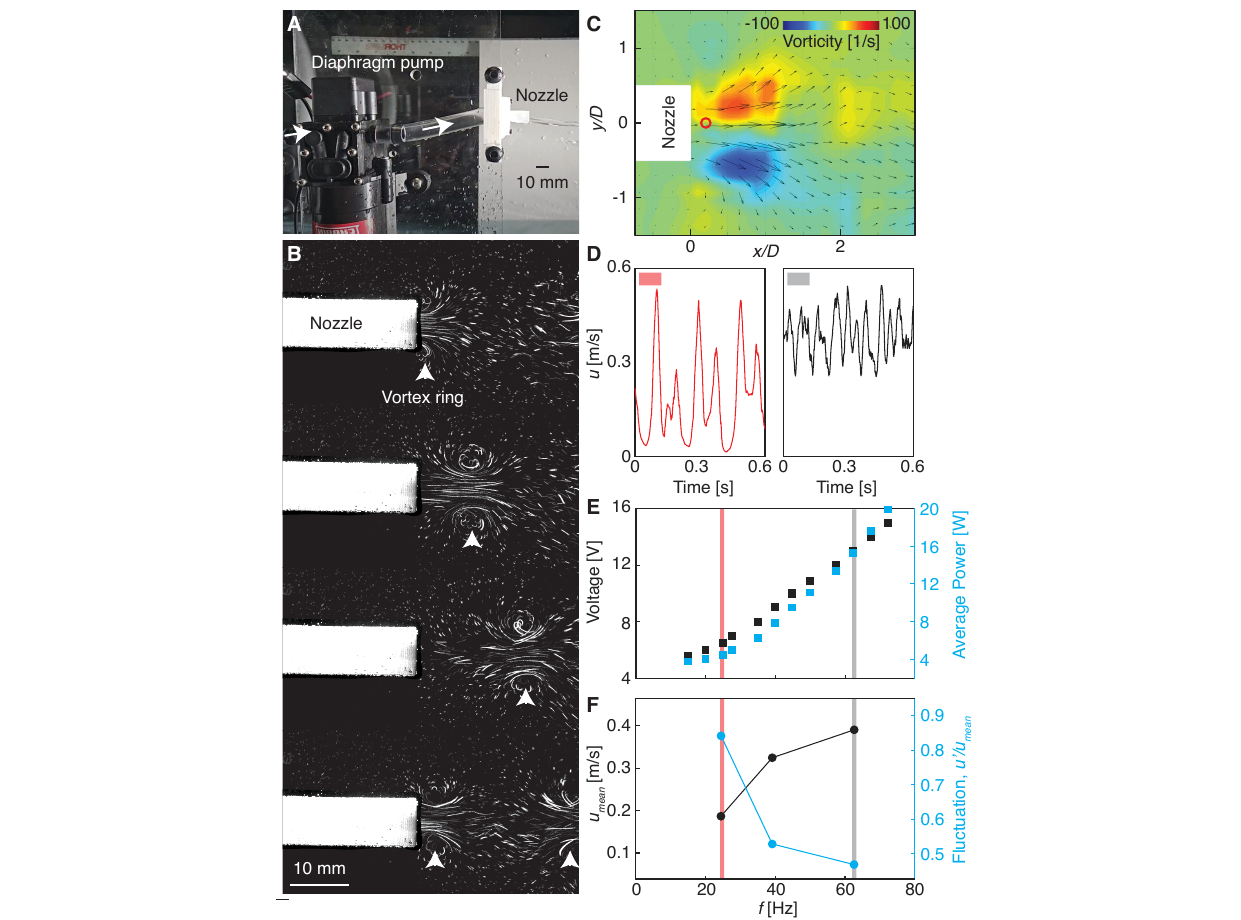}
    \caption{\textbf{Repeated pulse jet generation.}
    (\textbf{A})~Experimental setup illustrating the diaphragm pump, water tank, and nozzle secured with a flange-clamped mechanism (see Fig.~\ref{sif_nozzlefabrication}, A).
    (\textbf{B})~Visualization of sequential vortex ring ejection from the nozzle. Images are shown at 0.025~s intervals; the position of the vortex core (lower side) is marked by an arrowhead.
    (\textbf{C})~Vorticity field and velocity vectors at the nozzle tip, highlighting vortex ring formation. The probe location at $x = 0.1D$, $y = 0$, where the horizontal velocity is measured, is indicated by a red circle.
    (\textbf{D})~Temporal velocity profiles for two pump voltages (left: 6~V, right: 12~V).
    (\textbf{E})~Pump voltage and average power consumption, and (\textbf{F})~mean jet velocity and normalized velocity fluctuation as functions of the pump's dominant frequency. Results show that mean velocity increases while fluctuations decrease at higher frequencies, leading to diminished pulsatile flow characteristics.
    }
    \label{sif_repeated_pulse}
\end{figure*}

Single pulsed jets were generated using the jet generator described in Fig.~\ref{sif_piv} (PIV measurement), with a jet acceleration time of \revred{0.10}~s. \rev{The nozzles had} a \revred{10 or 20~mm} internal diameter, \revred{0.7~mm} nozzle thickness, \revred{15--60~mm} height, and a Young's modulus \rev{of} $1\times10^5$~Pa, corresponding to $\tau/T = 0$--\revred{0.32}. The nozzles were fixed onto the nozzle support with a press-fit mechanism (see Fig.~\ref{sif_nozzlefabrication}). Fig.~\ref{sif_aerialjet}, B shows the maximum jet height of both flexible nozzles with different heights and the rigid nozzle, with the maximum occurring at $\tau/T \approx \revred{0.32}$, matching the condition found in Fig.~\ref{f_mechanism}, E.

Repeated pulsed jets (Fig.~\ref{sif_repeated_pulse}) were generated using a diaphragm pump inside the water tank, with the time ratio $\tau/T$ varied from 0 to 0.46 and the input jet frequency $f$ varied from \revred{10 to 80}~Hz.
In these experiments, the nozzles were fixed with a flange-clamped mechanism (see Fig.~\ref{sif_nozzlefabrication}). Fig.~\ref{sif_aerialjet}, C--D show snapshots of horizontal jet ejection from rigid and flexible nozzles, respectively, at a jet input frequency of 24~Hz, demonstrating that the flexible nozzle ejects the jet \revred{44.6\%} farther compared to the rigid nozzle. Fig.~\ref{sif_aerialjet}, E compares the mean trajectory of the aerial jet from nozzles with different time ratios, obtained by image processing (see inset in Fig.~\ref{sif_aerialjet}, C--D), and shows that the flexible nozzle with $\tau/T \approx 0.3$ produces the farthest jet, consistent with theoretical predictions, $\tau/T = 0.2$--$0.4$. The time series of nozzle deformation is shown in Fig.~\ref{sif_aerialjet}, F. As the pulse begins ($t/T_{cycle} = 0.15$ in Fig.~\ref{sif_aerialjet}, F, where $T_{cycle}$ is the period of the pulsatile jet, which is different from the acceleration time of the actuator $T$), a faster-moving liquid blob is observed from the flexible nozzle compared to the rigid nozzle, due to the transfer of stored elastic energy into the kinetic energy of the jet, which results in superpropulsion. 
It should be noted that the horizontal variance of the jet trajectory from flexible nozzle is wider than that of the rigid nozzle (see the inset of Fig.~\ref{sif_aerialjet}, C and D). This is attributed to velocity differences within the liquid volume of a single pulse. The front part of the jet is faster, aided by power amplification from the soft nozzle, while the rear part is not accelerated as much due to early termination of the power amplification. This leads to an abrupt velocity gradient within one pulse, a longer ligament compared to the rigid nozzle (see the double-headed arrow in Fig.~\ref{sif_aerialjet}, D), and creates more satellites due to Rayleigh--Plateau instability \cite{choi2022atomization, driessen2014jet, martinez-calvo2020satellite}.

The aerial jet experiment captured both the harmonic response and nonlinear behavior of the flexible nozzle. Fig.~\ref{sif_aerialjet}, G shows the horizontal jet distance of the flexible nozzle, normalized by the rigid nozzle distance, as a function of input jet frequency from 10 to 80~Hz (see Fig.~\ref{sif_repeated_pulse} for frequency calculation). The plot reveals distinct first and second harmonic peaks at approximately 20~Hz and 40~Hz, although the overall jet distance decreases with increasing frequency due to reduced fluctuation, \rev{meaning that} the jet input becomes less pulsatile (see Fig.~\ref{sif_repeated_pulse}, D and F). 
The presence of the second harmonic suggests that the soft nozzle can be optimized under various conditions, not only at $\tau/T=0.3$ as proposed in this study.
This design flexibility, arising from multiple resonance points, is advantageous for industrial applications where constraints on nozzle aspect ratio, stiffness, and jet conditions may preclude operation at the primary resonance.

Fig.~\ref{sif_aerialjet}, H shows a non-linear response of the flexible nozzle at a frequency of 40~Hz, which is higher than the optimal frequency \rev{and therefore the optimal time ratio} $\tau/T$. Although the increase in jet distance is moderate (\revred{$\sim$25\%}), there is a significant increase in transverse jet dispersion due to nozzle deformation. This arises from the nonlinear interaction between the deforming nozzle and the internal pulsed input jet, which appears to have a notable effect on jet entrainment, mixing, and cooling \cite{gohil2015simulation, reynolds2003bifurcating}.

\begin{figure*}[!htbp]
    \centering
    \includegraphics[width=0.85\textwidth]{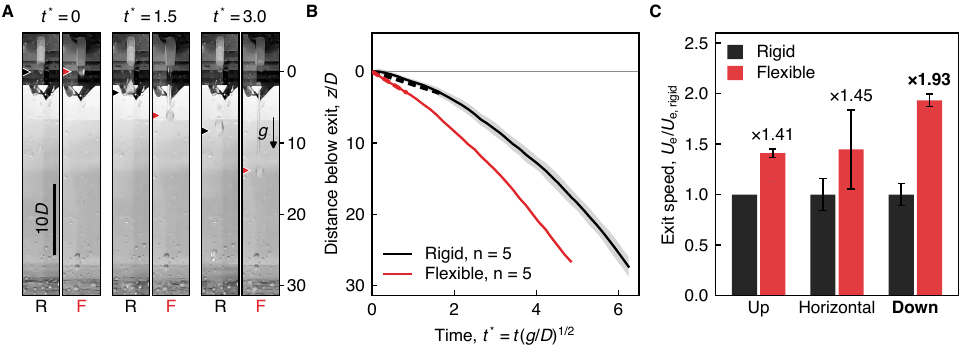}
    \caption{\rev{\textbf{Downward jets from flexible and rigid nozzles.}}
    \rev{Jets from the rigid nozzle (R, black) and the flexible nozzle (F, red) at $\tau/T = 0.23$, directed vertically downward. Lengths are scaled on the nozzle inner diameter $D = 7$~mm and time on $\sqrt{D/g} = 26.7$~ms, so that $t^* = t/\sqrt{D/g}$.}
    \rev{(\textbf{A})~One representative pulse per nozzle at $t^* = 0$, 1.5 and 3.0 after its front passes the exit, with the tracked front marked by triangles. The axis on the right gives $z/D$ on the scale of (\textbf{B}).}
    \rev{(\textbf{B})~Front position $z$ below the exit against $t^*$, mean $\pm$ SD of five pulses per nozzle, with $z = 0$ at the median first tracked position of each nozzle. Dashed lines run from the exit to $z = 3D$ at the exit speed $U_\mathrm{e}$ of each nozzle, which is the mean front speed over that distance.}
    \rev{(\textbf{C})~Exit speed relative to that of the rigid nozzle, $U_\mathrm{e}/U_\mathrm{e,rigid}$, at launch angles of $90^\circ$ (upward, Fig.~\ref{f_application}, B and Fig.~\ref{sif_aerialjet}, B), $0^\circ$ (horizontal, Fig.~\ref{f_application}, Q) and $-90^\circ$ (downward, panels \textbf{A} and \textbf{B}). Error bars are one SD, taken over trials upward, over time horizontally and over pulses downward. The actuation differs between the three tests.}
    }
    \label{sif_downwardjet}
\end{figure*}

\rev{The effect of the launch direction was tested in an additional experiment with the nozzle pointing vertically downward, so that gravity assists the jet (Fig.~\ref{sif_downwardjet}). Launch angles are measured from the horizontal, which gives $90^\circ$ for the upward jets, $0^\circ$ for the horizontal jets and $-90^\circ$ for the downward jets. Rigid and flexible nozzles of the same inner diameter, $D = 7$~mm, were driven by the same pump at a fixed 2.17~V ($\tau/T = 0.23$). The jet was filmed at 240~frames/s and its front was tracked over a fall of about 200~mm (29$D$), in five pulses per nozzle spread across each recording (Fig.~\ref{sif_downwardjet}, A).}

\rev{The gain in exit speed appears to increase gradually as the launch direction aligns with gravity. The exit speed of the flexible nozzle is \revred{1.41}, 1.45 and 1.93 times that of the rigid nozzle at $90^\circ$, $0^\circ$ and $-90^\circ$, in that order (Fig.~\ref{sif_downwardjet}, C). The upward value is the square root of the mean jet-height ratio of the nozzle of Fig.~\ref{f_application}, B (Fig.~\ref{sif_aerialjet}, B, \revred{$\tau/T = 0.17$}), and the horizontal value is the ratio of the mean jet distances of Fig.~\ref{f_application}, Q. The downward value is the mean speed of the jet front over the first 3$D$ below the exit. Farther from the exit, gravity accelerates both jets and the gain of the flexible nozzle decreases (Fig.~\ref{sif_downwardjet}, B).}

\rev{A scaling estimate gives the reason for the larger gain in the downward test. The nozzle has length $L$ and wave speed $c$ (Eq.~\ref{eq:lumped_io_main}), and the jet leaves it at velocity $V$ and at a launch angle $\theta$ from the horizontal, positive upward. Gravity changes the pressure that drives the liquid out of the nozzle during recoil by $-\rho g L \sin\theta$, and the elastic pressure of the wall scales as $\rho c V$. The ratio of the two pressures is $-\Gamma\sin\theta$, with $\Gamma = gL/(cV) = g\tau/V$. In the rigid nozzle the pump sets the flow rate, so gravity changes the pressure and not the exit speed. Relative to a horizontal jet, the gain of the flexible nozzle is therefore expected to change by $+\Gamma$ for the downward jet ($\theta = -90^\circ$) and by $-\Gamma$ for the upward jet ($\theta = 90^\circ$). The downward test has $\tau \approx \revred{5}$~ms and $V = 0.5$ to 0.9~m/s, which gives a change of \revred{$+5$ to $+10\%$}. The upward test has $\tau \approx \revred{17}$~ms and $V = 0.9$ to 1.4~m/s, which gives \revred{$-12$ to $-18\%$}. Together, the two estimates place the gain of the downward jet about \revred{15 to 30\%} above that of the upward jet, and the measured gains differ by about \revred{37\%} (1.93 against \revred{1.41}, Fig.~\ref{sif_downwardjet}, C). The two values cannot be compared directly, because the nozzle geometry and the jet conditions differ between the tests. The estimate nevertheless gives a reasonable explanation of the measured difference, with the same direction and the same order of magnitude.}

\FloatBarrier
\section{Boat experiment}\label{si:boat}
\begin{figure*}[!htbp]
    \centering
    \includegraphics[width=\textwidth]{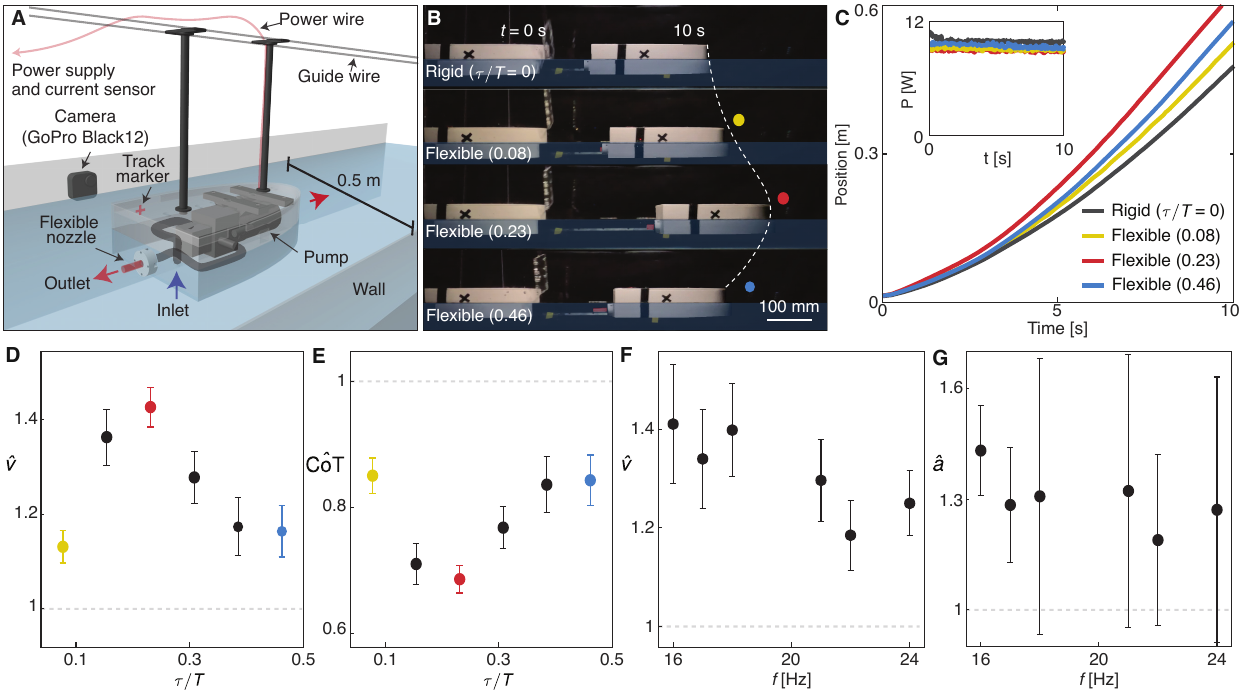}
    \caption{\textbf{Boat setup specifications and experimental results.}
    (\textbf{A})~Experimental setup showing the boat, water tank, guide wire, and nozzle with flange-clamped mechanism (see Fig.~\ref{sif_nozzlefabrication}, A).
    (\textbf{B})~Snapshots of the boat at 0 and 10 seconds, illustrating the difference in position for different nozzle conditions, $\tau/T$.
    (\textbf{C})~Horizontal position versus time comparison for each $\tau/T$. Power readings are shown in the inset, showing no significant difference in power.
    (\textbf{D-E})~Performance increase normalized to the rigid nozzle: (\textbf{D}) velocity, $\hat{v}=v_f/v_r$, and (\textbf{E}) cost of transport, $\hat{CoT}=CoT_f/CoT_r$, both as a function of time ratio, $\tau/T$, showing the highest improvement at $\tau/T = 0.23$.
    (\textbf{F-G})~Performance increase normalized to the rigid nozzle: (\textbf{F}) velocity, $\hat{v}=v_f/v_r$, and (\textbf{G}) acceleration, $\hat{a}=a_f/a_r$, both as a function of pump frequency, showing the highest improvement at lower frequencies due to \revred{the larger fluctuation (more pulsatile jet)} (see Fig.~\ref{sif_repeated_pulse}, F).
    }
    \label{sif_boat}
\end{figure*}

A boat was 3D printed using PLA filament to evaluate the transport performance of flexible nozzles (see Fig.~\ref{sif_boat}, A), with the size of 260~mm in length, 165~mm in width, and 68~mm in height, with a total mass of 1.2~kg including the pump and balancing weights. Two masts attached the boat to a guide wire to enforce 1D motion (see Fig.~\ref{sif_boat}, A). A 60-psi diaphragm pump (6-9206, Chapin) was installed inside the boat and hoses were connected to the inlet and outlet of the pump. The pump inlet hose was attached to a hole centered at the bottom of the boat, while the outlet hose was connected to a hole centered at the back of the boat, below the water surface, as seen in Fig.~\ref{f_application}, H. The nozzle attachment followed the flange-clamped mechanism as described in Fig.~\ref{sif_nozzlefabrication}, A. The pump was powered via a DC power supply (SPS-3010, JESVERTY), connected to a relay via an Arduino so that a mechanical switch could activate and deactivate the pump. 34 AWG wire with a diameter of 0.16~mm was used for the power supply to minimize any mechanical resistance due to wiring. A current sensor (INA219, Adafruit) with a 0.1-ohm shunt resistor and a maximum current of 3.2~A was used to read the bus and shunt voltage, and the current applied to the pump.

\begin{figure*}[!htbp]
    \centering
    \includegraphics[width=\textwidth]{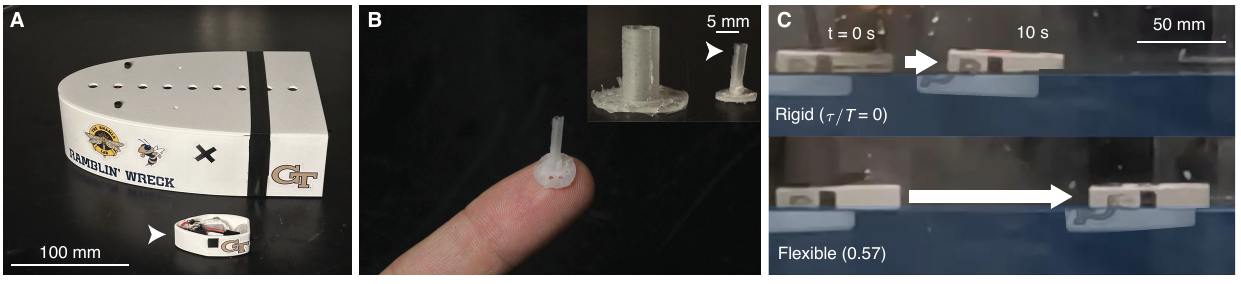}
    \caption{\textbf{Small boat setup and experimental results.}
    (\textbf{A})~Size comparison between the small-scale squid-boat (arrowheads) and the original boat.
    (\textbf{B})~Small nozzle on fingertip to demonstrate size. Inset shows size comparison between the flexible nozzle of small-scale squid-boat (arrowheads) and the original boat.
    (\textbf{C})~Snapshots of the small-scale squid-boat at 0 and 10 seconds, showing the flexible nozzle is $>$50\% faster than the rigid nozzle.
    }
    \label{sif_smallboat}
\end{figure*}

The boat was placed inside a water tank filled with tap water (see Fig.~\ref{sif_boat}, A). A camera (Black12, GoPro) captured the boat movement with a resolution of $1920\times1080$ pixels (in horizontal lock and linear mode view) at 30~fps. The boat was positioned in the tank with the back aligned to a plastic wand attached to a frame on the tank to ensure a consistent starting point, and the boat was released 3 seconds after the power switch was turned on. Six voltages ranging from 7.8--9.3~V with 3 trials at each voltage were applied to test the frequency dependency (see Fig.~\ref{sif_repeated_pulse}, E). 
Six nozzles with different heights ranging from 5 to 30~mm (corresponding to $\tau/T$ values from \revred{0.08} to \revred{0.46}), as well as one rigid nozzle with $\tau/T = 0$, were tested, with five trials conducted for each nozzle. A total of 18 trials were performed for the frequency tests ($f = 16$--24~Hz) and 35 trials for the nozzle geometry tests ($\tau/T = 0$--\revred{0.46}), respectively. Recorded videos were tracked using DLTdv8 in MATLAB, utilizing the Kalman algorithm location predictor with a search area width of 32 and a tracker threshold of 8. A black ``X'' was placed on the boat for tracking purposes. A third-order polynomial was fitted to the raw horizontal position curves over time (with $R^2 > 0.98$), and derivatives were computed to determine the velocity and acceleration at specific time points for each trial. \revred{Each position curve was fitted up to the frame at which the boat had travelled 2.4 body lengths (0.62~m), because after that point the boat slowed down and stopped at the end of the tank.} The velocities and accelerations at a target time of 5~s from the boat's start were calculated and averaged for each experiment, followed by comparisons between the flexible and rigid nozzles. A Savitzky-Golay smoothing method was used on the data to ensure no deviation from endpoint errors from the polynomial fitting. Velocity and cost of transport data from SG-smoothing were consistent with the fitting, however acceleration data was too noisy from the 2nd order derivative. Since the velocity was consistent across both it was concluded that the endpoint error on polynomial fitting does not affect our measurements and the acceleration estimates from polynomial fitting are still valid.

\rev{The small boat experiment was a proof-of-concept test of whether the advantage of the flexible nozzle persists at reduced scale. It spans two scales only, with nozzle diameter reduced from 7 to 2~mm, wall thickness to about \revred{0.25}~mm, actuation time from \revred{0.02} to 0.005~s, and vehicle mass and jet pulse volume by roughly a factor of 40, so scale invariance remains to be established. \revred{The miniature nozzle, with $\tau/T \approx 0.57$, was also not at the optimum of 0.2--0.4.}} The overall procedure closely followed that of the previous boat experiment, with only minor adjustments to accommodate the smaller vessel and nozzle. 
Specifically, the miniature nozzle was fabricated using the dipping method, similar to the previously described flange-embedded mechanism. This approach enabled the creation of a thin wall thickness of approximately \revred{0.25}~mm to ensure flexibility at the smaller scale. However, no embedded clamp was used for the miniature version; instead, a sandwich capping method was used.
The small-scale squid-boat and nozzle are compared with the original setup in Fig.~\ref{sif_smallboat}, A and B, respectively. Results showed that the flexible nozzle increased the boat's speed \revred{by more than 50\% compared with the rigid nozzle} (Fig.~\ref{sif_smallboat}, C), \rev{indicating that the enhancement persists at this reduced scale} (Total $N=6$; rigid N=3 and flexible N=3).

\FloatBarrier
\section{Cost-of-Transport interpretation}
\label{si:actuation_energetics}

The efficiency and cost-of-transport gains from superpropulsion depend on how the actuator is driven and on where losses sit in the power budget. In many laboratory demonstrations of superpropulsion (including ours), we prescribe actuator kinematics and operate with sufficient available input power that the actuator motion is only weakly affected by the jet load~\citep{prl119108001,challita2023superpropulsion,choi2024mechanism}. In that regime, a compliant element can change when energy reaches the jet and can also change how much mechanical work the actuator delivers over a cycle. The resulting performance gain should therefore be considered together with the actuator’s overhead losses and operating point.

Two limiting examples help frame the point. First, elastic ``one-shot'' launchers can exceed a rigid-body kinematic limit without increasing the initially stored energy by storing energy early and releasing it during the short interval when it most effectively increases directed kinetic energy~\citep{celestini2020contactlayer}. Second, in the piston--nozzle setting studied here, the compliant nozzle acts as a temporary volume and energy reservoir: it dilates during acceleration and returns stored elastic energy during recoil/collapse, reshaping the exit-flow history and increasing the momentum carried by the primary vortex ring relative to a rigid nozzle under the same prescribed stroke. We formalize how this affects transport efficiency using a gross cost-of-transport model.

\rev{Consider a self-propelled body of weight $W$ traveling at steady speed $v$. The power split and the cost-of-transport definition are given in the main text, as Eqs.~\eqref{eq:ptotal} and~\eqref{eq:gcot_def}, where $P_{\text{fixed}}$ is a speed-independent overhead \revred{(pump friction, valve and cam losses)}, the drag power scales as $P_{\text{prop}} = \tfrac{1}{2}\rho C_D \revred{A_{\mathrm{f}}}\,v^3 \equiv \revred{k}\,v^3$, and we write $\mathrm{GCOT}$ below for the same gross cost of transport. We continue the derivation from there.} The first term ($\propto v^{-1}$) reflects the dilution of fixed overhead over a greater distance traveled per unit time; the second ($\propto v^{2}$) captures the growing drag penalty. Their competition produces a U-shaped GCOT curve with a unique minimum at the optimal speed $v_{opt} = (P_{\text{fixed}}/2\revred{k})^{1/3}$ which satisfies $\partial\,\mathrm{GCOT}/\partial v = 0$ \rev{and therefore locates the minimum}. Introducing the dimensionless propulsive power $\revred{\Pi} = 2P_{\text{prop}}/P_{\text{fixed}}$ (so that $\revred{\Pi}=1$ at the GCOT minimum), the speed can be written as $v = \revred{\Pi}^{1/3}\,v_{\text{opt}}$ since $v=(P_{prop}/\revred{k})^{1/3}=(2P_{prop}/P_{fixed}\cdot P_{fixed}/(2\revred{k}))^{1/3}=(\revred{\Pi} \cdot v_{opt}^3)^{1/3}=\revred{\Pi}^{1/3} v_{opt}$. Therefore, each term of Eq.~\eqref{eq:gcot_def} can be expressed in terms of $\revred{\Pi}$ and $v_{\text{opt}}$ alone.

\begin{align}
  \mathrm{GCOT}
  &= \frac{P_{\text{fixed}}}{W\,v} + \frac{\revred{k}\,v^2}{W}
   = \frac{2\revred{k}\,v_{\text{opt}}^3}{W\,\revred{\Pi}^{1/3}\,v_{\text{opt}}} + \frac{\revred{k}\,\revred{\Pi}^{2/3}\,v_{\text{opt}}^2}{W}
   = \frac{2\revred{k}\,v_{\text{opt}}^2}{W}\,\revred{\Pi}^{-1/3} + \frac{\revred{k}\,v_{\text{opt}}^2}{W}\,\revred{\Pi}^{2/3}.
  \label{eq:gcot_expand}
\end{align}

\noindent Normalizing by $\mathrm{GCOT}_{\min} = 3\revred{k}\,v_{\text{opt}}^2/W$ \revred{yields} the relation
\begin{equation}
  \frac{\mathrm{GCOT}}{\mathrm{GCOT}_{\min}}
  = \frac{2}{3}\!\left(\revred{\Pi}^{-1/3} + \frac{\revred{\Pi}^{2/3}}{2}\right).
  \label{eq:gcot_universal}
\end{equation}

\noindent In the present pump-driven experiments, we estimate $\revred{\Pi} \sim O(10^{-4})$, given the electrical power input (${\sim}$5\,W, see Fig.~\ref{sif_repeated_pulse}, E) and the useful jet kinetic power ($\tfrac{1}{2}\rho A\,v_{\text{jet}}^3 \sim 0.5$\,mW). When $\revred{\Pi} \ll 1$, the first (overhead) term of Eq.~\eqref{eq:gcot_universal} dominates over the second (propulsive) term, and GCOT decreases with increasing $\revred{\Pi}$ (e.g., increasing $P_{\text{prop}}$ at constant $P_{\text{fixed}}$): channeling more energy into useful thrust, even at the expense of higher total consumption $P_{\text{total}} \,(=P_{fixed}+P_{prop})$, improves transport efficiency, which is responsible for the reduction in GCOT observed in boat experiment (Fig.~\ref{f_application}, T).

\FloatBarrier
\section{Mixing experiment}\label{si:mixing}

\begin{figure*}[!htbp]
    \centering
    \includegraphics[width=\textwidth]{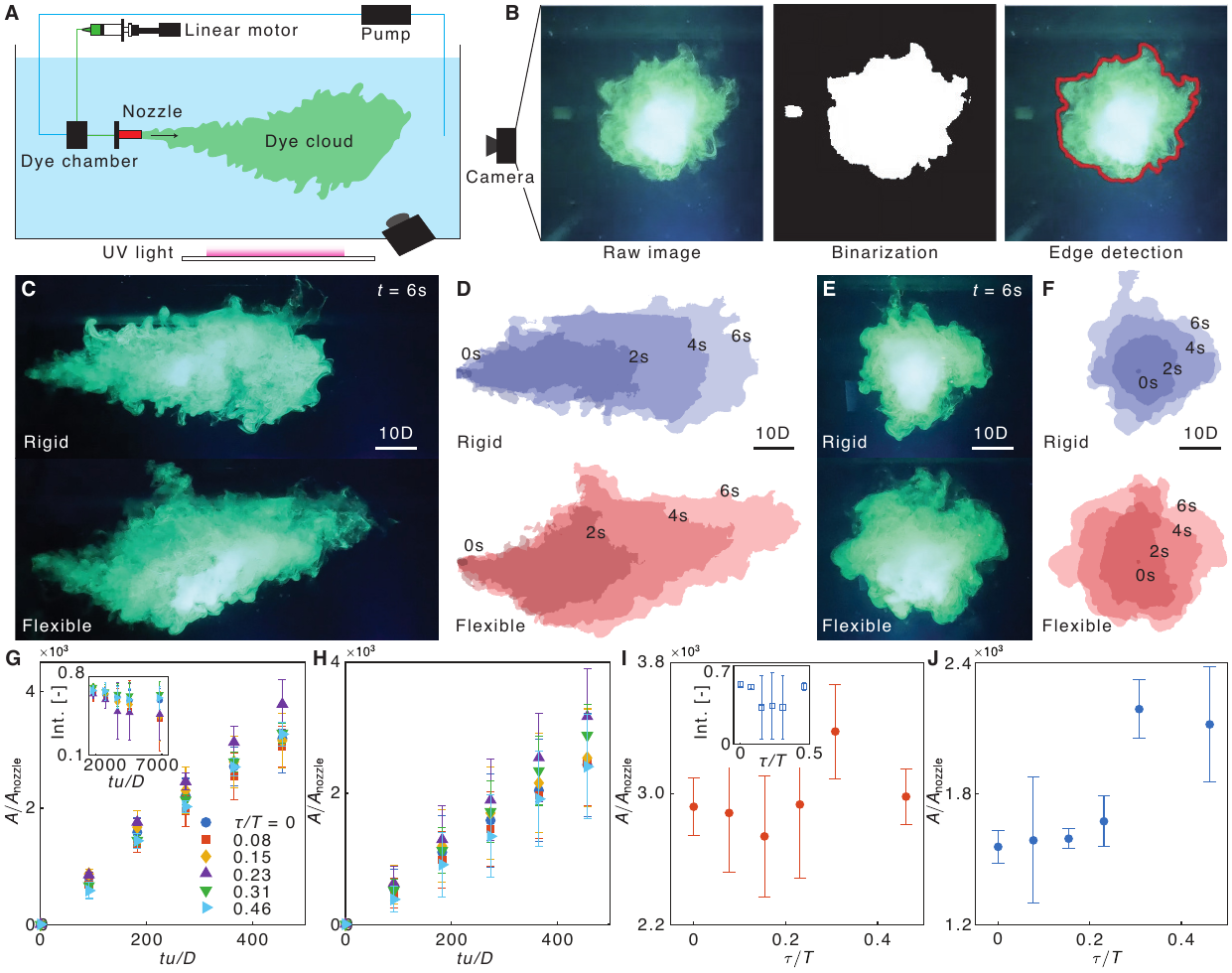}
    \caption{\textbf{Mixing experiment setup.}
    (\textbf{A})~Experimental setup with the water tank, diaphragm pump, automated dye injection system, and nozzle with flange-clamped mechanism (see Fig.~\ref{sif_nozzlefabrication}, A).
    (\textbf{B})~Image processing to capture the dye area from raw images, including binarization and edge detection of the largest connected region.
    (\textbf{C-F})~Visualization of dye cloud development from the (\textbf{C}) side view and (\textbf{E}) front view, shown at $t = 6\,\text{s}$ ($t^* = tu/D=274$), normalized by the jet velocity $u$ and internal diameter $D$ of the nozzle. (\textbf{D}) and (\textbf{F}) show overlaid edges of the dye cloud at 2~s intervals from $t = 0$ to 6~s.
    (\textbf{G-H})~Area development of the dye cloud as a function of normalized time ($t^* = tu/D$) from the (\textbf{G}) side and (\textbf{H}) front view. The inset in (\textbf{G}) shows the intensity of the dye cloud as a function of time. \revred{The inset in (\textbf{I}) shows the intensity at the end of the recording as a function of $\tau/T$.}  
    (\textbf{I-J})~Dye cloud area (at $t^* = 400$) depending on time ratio, $\tau/T$, from the (\textbf{I}) side and (\textbf{J}) front view, showing that the flexible nozzle produces a 40.5\% and \revred{15.7\%} wider dye cloud at $\tau/T \sim \revred{0.31}$ (for the front and side views, respectively) compared to the rigid nozzle.  
    }
    \label{sif_mixing}
\end{figure*}

Mixing performance of flexible nozzles was assessed using a water tank and automated dye injection system to ensure consistent dye dispersion (Fig.~\ref{sif_mixing}). 
A diaphragm pump (6-9206, Chapin) was fixed above the water tank and connected to a DC power supply (SPS-3010, JESVERTY) at 2.5~V. The pump inlet was placed far downstream from the nozzle ($>150D$, where $D$ is the internal diameter of the nozzle) to minimize its interference on jet development (Fig.~\ref{sif_mixing}, A). 
The pump outlet was connected to a dye injection chamber. In this chamber, dye from a syringe was naturally mixed with the incoming inlet flow before being expelled together through the flexible nozzle (Fig.~\ref{sif_mixing}, A). 
The dye solution consisted of 1 part fluorescent dye (Cool Green, Marblers) to 10 parts water (1:10 ratio). The dye was injected with a consistent volume (1~mL) and flow rate ($\sim$0.5~mL/s) through the syringe with a consistent piston movement guided by a linear actuator (LA-T8-12-50, DC House). Three 6W UV lights (Resin LED, ACADEEP) with a wavelength of 405~nm were used to illuminate the dye cloud located under the tank, and 2 semi-translucent plastic sheets were placed at the bottom of the tank to diffuse the light to evenly illuminate the dye cloud. Two cameras (GoPro Black 12) were positioned at the downstream and side locations, perpendicular to each other, to capture the side and front views of the dye cloud as it spread (Fig.~\ref{sif_mixing}, A). For each dye injection, $\sim$3~minutes were recorded to allow the dye to fully diffuse. After each trial, the tank and pump were completely drained and thoroughly cleaned to prevent any residual dye deposition. Three trials were performed for 5 different soft nozzles with different heights (5--30~mm, corresponding to $\tau/T$ values from \revred{0.08 to 0.46}) and one rigid nozzle with $\tau/T = 0$, totaling 18 trials. The nozzles were manufactured and installed with a flange-clamped mechanism (see Fig.~\ref{sif_nozzlefabrication}, A).

Dye dispersion was quantified using custom MATLAB (2025) code. To determine the onset of dye ejection, HSV color filtering was applied to the video frames to identify the emergence of green dye at the nozzle exit. The moment of initial dye appearance was defined as when the moving standard deviation of the green channel intensity surpassed a predetermined threshold.
Cropped image snapshots were then extracted at time intervals of 2~s after the initial green detection. For each snapshot, green dye regions were identified using the same HSV thresholding. Boundaries of the largest connected green region were extracted using the \texttt{bwboundaries} function to isolate the main dispersion area by excluding the small regions such as reflection. Two quantitative metrics were calculated to characterize mixing performance: (1) normalized area, defined as the area of the green dye region divided by the nozzle cross-sectional area ($A/A_{\text{nozzle}}$ where $A_{\text{nozzle}} = \pi D^2/4$ and $D$ is the internal diameter of the nozzle), and (2) intensity, measured as the average saturation value from the HSV color space across the entire image after the dye cloud filled the cross-section area of tank (after t=10~s). Time was normalized as $t^* = t \cdot u / D$, where $u$ is the jet velocity (u=319.5~mm/s) and $D$ is the internal diameter of the nozzle (D=7~mm). The temporal evolution of normalized area and intensity was analyzed to quantify mixing enhancement, with comparisons made at $t^* = 274$ and 400 for snapshots in Fig.~\ref{sif_mixing}, C--F and statistics in Fig.~\ref{sif_mixing}, G\textendash J, respectively, across different nozzle lengths.
Results showed that flexible nozzles with $\tau/T = \revred{0.31}$ produced normalized mixing areas that were 40.5\% and \revred{15.7\%} larger (for the front and side views, respectively) compared to rigid nozzles, indicating significantly enhanced mixing performance. Additionally, the flexible nozzle dissipated 35.6\% faster than the rigid nozzle, as shown in the insets of Fig.~\ref{sif_mixing}, G and I. Movies of the mixing experiment are provided in Supplementary Movie~\ref{mv_squid}.

\FloatBarrier
\clearpage
\section{Supplementary movie}

\refstepcounter{movie}\label{mv_squid}
\vspace{\baselineskip}
\textbf{Movie S1.} \revred{Squid funnels time their elastic recoil to amplify pulsed jets.} \newline
[00:05] Snapshot of squid showing its mantle and funnel \rev{(\textit{Sepioteuthis sepioidea})}.  \newline
[00:12] Histology microscopy of flexible funnel showing collagenous membrane \rev{(\textit{Doryteuthis opalescens})}.  \newline
[00:16] Squid escaping by jetting through a flexible funnel \rev{(\textit{Doryteuthis opalescens})}.  \newline
[00:27] Setup for measuring squid deformation \rev{(\textit{D. opalescens})}.  \newline
[00:33] Chromatophore tracking results showing funnel and mantle deformation \rev{(\textit{D. opalescens})}.  \newline
[00:51] Illustration of superpropulsion through elastic squid funnel.  \newline
[00:58] Fabrication of the flexible nozzle.  \newline
[01:01] Flow visualization via 2D particle image velocimetry, showing a faster vortex ring from the flexible nozzle.  \newline
[01:20] Wave-mediated nozzle deformation determines the response time and thrust amplification. \newline
[01:31] 3D FSI simulation of a flexible nozzle jet and comparison of $T/\tau$ between theory and measurement. \newline
[01:40] Optimal condition for impulse generation derived from theory and validated with experiments and CFD. \newline
[01:46] PIV visualization of the optimally flexible nozzle jet. \newline
[01:53] Photograph of a flying squid \rev{(\textit{Ommastrephes bartramii})}.  \newline
[01:56] Aerial jet performance, showing the more flexible nozzle generates a 110\% higher water column.  \newline
[02:05] Aerial jet performance, demonstrating the more flexible nozzle achieves a greater horizontal jet distance.  \newline
[02:20] Underwater propulsion of squid \rev{(\textit{Todarodes pacificus})}.  \newline
[02:26] Manufacturing of the squid-boat and attachment of the flexible nozzle.  \newline
[02:31] Deformation of the flexible nozzle attached to the squid-boat, responsible for superpropulsion.  \newline
[02:36] Squid-boat propulsion, with a \revred{43\%} speed increase for the flexible nozzle.  \newline
[02:50] Inking squid \rev{(\textit{Dosidicus gigas})} forming a thick, dispersed ink cloud.  \newline
[02:53] The faster vortex ring from the flexible nozzle disperses dye more quickly compared to the rigid nozzle.  \newline
[03:03] Dye dispersion experiment showing the flexible nozzle produces a 41\% wider dye cloud.

\newpage
\clearpage

\end{document}